\documentclass{article}%
\usepackage{amsmath}
\usepackage{geometry}
\usepackage{graphicx}%
\usepackage{amsfonts}%
\usepackage{amssymb}

\begin{document}

\title{A semiclassical theory of the chemical potential for the Atomic Elements}
\author{Bernard J. Laurenzi
\and Department of Chemistry
\and UAlbany, The State University of New York}
\date{Febuary 4, 2020}
\maketitle

\begin{abstract}
The chemical potiential for the ground states of the atomic elements have been
calculated within the semiclassical approximation The present work closely
follows Schwinger and Englert's semiclassical treatment of atomic structure.

\end{abstract}

\section{The Chemical Potential $\mu$ of the Atomic Elements}

For an atomic element containing $N$ electrons with nuclear charge $Z$, the
\textit{electronic} chemical potentials $\mu$ for neutral atomic species
(which are in their electronic ground states $\Psi$ with energy $E$) are
defined as
\[
\mu=\left(  \frac{\partial E}{\partial N}\right)  _{Z}\text{.}%
\]
Here $N$ and $Z$ will be regarded as continuous variables and $\mu$ can then
be calculated by noting that
\[
dE(Z,N)=\left(  \frac{\partial E}{\partial Z}\right)  _{N}dZ+\left(
\frac{\partial E}{\partial N}\right)  _{Z}dN,
\]
together with the well-known relation
\[
\left(  \frac{\partial E}{\partial Z}\right)  _{N}=\frac{\overline{V}_{ne}}%
{Z}.
\]
In the equation above, $\overline{V}_{ne}$ is the average value of the
nuclear-electronic potential energy (in atomic units) i.e.
\[
\frac{\overline{V}_{ne}}{Z}=-\sum_{i=1}^{N}\int\frac{\left|  \Psi_{0}\right|
^{2}}{r_{i}}d\tau,
\]
and $d\tau=d\mathbf{r}_{1}d\mathbf{r}_{2}d\mathbf{r}_{3}\ldots d\mathbf{r}%
_{N}.$ \ We have as a result%
\begin{equation}
\mu=\frac{d\,E(Z,Z)}{d\,Z}-\frac{\overline{V}_{ne}}{Z}, \label{1}%
\end{equation}
where $dE/dZ$ is the directional derivative of the electronic energy surface
$E(Z,N)$ along the curve $Z=N.$ \ In this work $E(Z,Z)$ is taken to be the
electronic energy which has been computed within the semiclassical
approximation by Schwinger and Englert.

\subsection{Properties and estimates of the chemical potential for the elements}

The physical interpretation or significance of the electronic chemical
potential $\mu$ is seen as a measure of the propensity of an electron to leave
an atom. \ In this context as the atomic number increases $\mu$ gives the
stability of an element relative to others in the periodic table.

An associated quantity $\eta$ defined by
\[
\eta=\frac{1}{2}\left(  \frac{\partial\mu}{\partial N}\right)  _{Z}\text{,}%
\]
has been called the \textit{hardness \cite{Makov} }and has been interpreted as
the resistance of an atom to the ingress of additional electrons. The higher
the hardness the the lower the polarizability of the atom's electron cloud and
the greater the resistance of that atom to add an electron. \ Various
estimates for the chemical potential and the hardness have been made. \ March
\cite{MARCH} has given an estimate of $\mu$ by assuming that the energy can be
written as a Taylor series in the variable $(N-Z)$ i.e.%
\[
E(Z,N)=E(Z,Z)+(N-Z)\left(  \frac{\partial E}{\partial N}\right)
|_{Z=N}+\frac{(N-Z)^{2}}{2}\left(  \frac{\partial^{2}E}{\partial N^{2}%
}\right)  |_{Z=N}+\cdots.
\]
In that work he has shown that to third order (from a fifth order polynomial
in \ $N-Z$ )\ that the chemical potential can be written as
\[
\mu=\mu_{2}+\mu_{3},
\]
with%
\begin{align*}
\mu_{2}  &  =-\tfrac{1}{2}(I_{1}+A),\\
\mu_{3}  &  =\tfrac{1}{60}\left(  3I_{4}-17I_{3}+43I_{2}-47I_{1}+18A\right)  ,
\end{align*}
and where $I_{n}$ is the n-th ionization potential of the atom and $A$ is it's
electron affinity. \ Using the empirical relation $I_{n}\cong nI_{1}$ the
chemical potential to third order $\mu_{3}=\frac{3}{10}A$ and we have two
estimates%
\[
\mu=-\tfrac{1}{2}(I_{1}+A),
\]
to lowest order and
\[
\mu=-\tfrac{1}{2}I_{1}-\tfrac{1}{5}A,
\]
which includes $\mu_{3}.$ \ It is interesting to note that Mulliken's
electronegativity function $\chi_{M}$ which is defined as
\[
\chi_{M}\equiv\tfrac{1}{2}(I_{1}+A)>0,
\]
and is interpreted as the ability of an atom to attract electrons is
approximately related to $-\mu$. \ Piris and March \cite{PIRIS} using natural
orbital functional (NOF) theory have estimated $\mu$ and compared it to
$-\,I_{1}$ for neutral atom (H-Kr) as seen in the Fig.(1) below.%

\begin{figure}
\begin{center}
\fbox{\includegraphics[
width=3.1635in
]%
{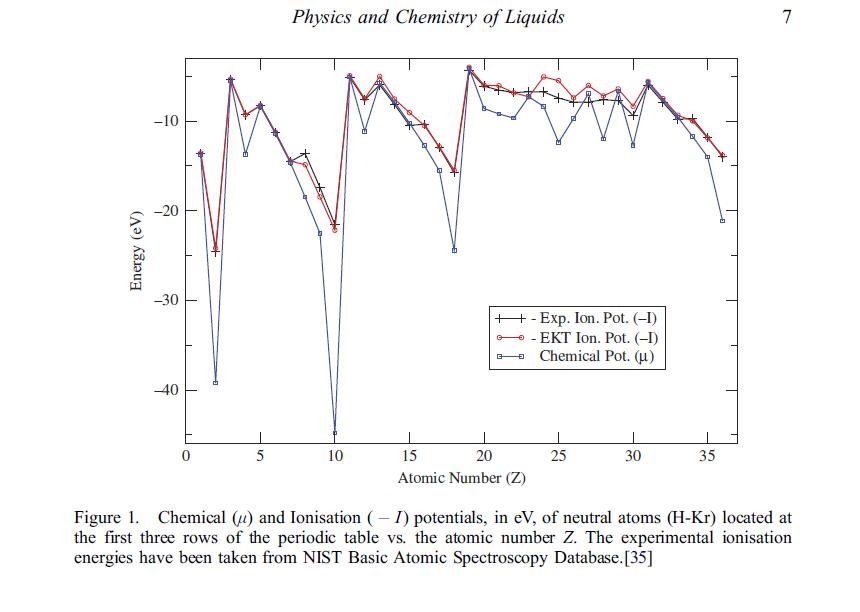}%
}\caption{The Chemical Potential $\mu$ vs. Z.}%
\end{center}
\end{figure}

\ 

Their chemical potential values parallel the oscillations in the experimental
ionization potential but deviate widely in magnitude from $-\,I_{1}$ in case
of the rare gases. If one wishes to interpret the electronic chemical
potential as the atomic analogue of the macroscopic thermodynamic chemical
potential, then $\mu$ as define above is an indication of the spontaneity of
the escaping tendency of an electron from an atom.

\subsection{A Semiclassical Approximation for the chemical potential $\mu$}

Within the ``semiclassical approximation,'' Schwinger and Englert \cite{SE}
(SE) have given an expression for $E(Z,Z)$. \ In that work, the authors have
shown that the total energy is made up of the semiclassial i.e. Thomas-Fermi
(TF) energy \cite{TF} and a quantum oscillating part i.e.%
\[
E(Z,Z)=E_{TF}(Z,Z)+E_{\,osc}(Z).
\]
Furthermore, the well-known value for $E_{TF}(Z,Z)$ is given by
\cite{engbook}
\begin{equation}
E_{TF}\,(Z,Z)=\tfrac{6}{7}\left(  \tfrac{4}{3\pi}\right)  ^{2/3}\Phi^{\prime
}(0)\,Z^{\,7/3}, \label{2}%
\end{equation}
with $\Phi^{\prime}(0)=-1.588071$ \cite{shengfeng} being the initial slope of
the TF function $\Phi(x)$ and where the TF potential $V_{TF}$ is
\[
V_{TF}=-Z\,\Phi(x)/r,
\]
and $x$ is the TF scaled distance $x=2^{7/3}r/(3\pi)^{2/3}$ The average value
of the TF potential energy is \cite{goldman}%
\begin{equation}
\overline{V}_{TF}=-Z\,\int_{0}^{\infty}\frac{\rho(\mathbf{r})}{r}%
d\mathbf{r=}2\left(  \tfrac{4}{3\pi}\right)  ^{2/3}\Phi^{\prime}%
(0)\,Z^{\,7/3}. \label{3}%
\end{equation}
\ \ 

We shall see below that the average value of the nuclear-electronic potential
$\overline{V}_{ne}$ like the total energy, can also be written as a sum of a
TF term and a quantum oscillating contribution i.e.%
\[
\overline{V}_{ne}=\overline{V}_{TF}+\overline{V}_{osc}.
\]
\ As a result of (2) and (3) the TF part to the chemical potential is seen to
vanishes,
\[
\mu_{TF}=0,
\]
there being no contribution of order $Z^{\,4/3}$and we have as a result%
\[
\mu=\frac{dE(Z,Z)_{osc}}{dZ}-\frac{\overline{V}_{ne,\,osc}}{Z}.
\]

The purpose of this work is to give the corresponding semiclassical expression
for $\overline{V}_{ne,osc}$ resulting in a semiclassical value for the
chemical potential for neutral atomic species. \ As will be seen below the
computation of that quantity unfortunately requires a rather elaborate
analysis. This investigation does not contain the effects due to the
antisymmetry \cite{SCH} of the system's wave functions or the effects of the
tightly bound electrons first taken into account by Scott \cite{SCOTT} nor the
quantum correction\ to the wave function due to the kinetic energy \cite{KE}.
\ Inclusion of these effects is problematic and beyond the scope of this work.

We begin the analysis of $\overline{V}_{ne}$ by recognizing that since the
terms within $\overline{V}_{ne\text{ }}$are single-particle operators,
integration over the $N-1$ coordinates of the $N$ particle wave function
reduces $\overline{V}_{ne}$ to
\[
-\frac{\overline{V}_{ne}}{Z}=\int\frac{\varrho(\mathbf{r})}{r}d\mathbf{r},
\]
where $\varrho$ is the single-particle electron density function defined as%
\[
\varrho(\mathbf{r})=N\int\ldots\int\left|  \Psi\right|  ^{2}d\mathbf{r}%
_{2}d\mathbf{r}_{3}\ldots d\mathbf{r}_{N}.
\]
This function could for example be taken to be the Thomas-Fermi electron
density. \ Instead, within the semi-classical (WKB) \cite{WKB}, and the
Hartree-Fock orbital approximations we take the density to be%
\begin{equation}
\varrho(\mathbf{r})=2\sum_{l=0}^{\infty}\sum_{n_{r}=0}^{\infty}(2l+1)\eta
(-E_{l,n_{r}}-\zeta)\,\,|u_{l,n_{r}}(r)/\sqrt{4\pi}r|^{2}. \label{4}%
\end{equation}
The single particle energies $E_{l,n_{r}}$ associated with the potential
$V(r)$ are labeled with the radial quantum number $n_{r}$ and the angular
quantum number $l$ respectively and the quantities $u\,_{l,\,n_{r}}%
(r)/\sqrt{4\pi}r$ are the WKB single-particle semiclassical wave functions
where $-\,\zeta$ is the single-particle energy of the highest occupied orbital
($\zeta\geq0$), and $\eta$ is the Heaviside function as shown in Fig. (2) below.%

\begin{figure}
\begin{center}
\includegraphics[
width=5.8686in
]%
{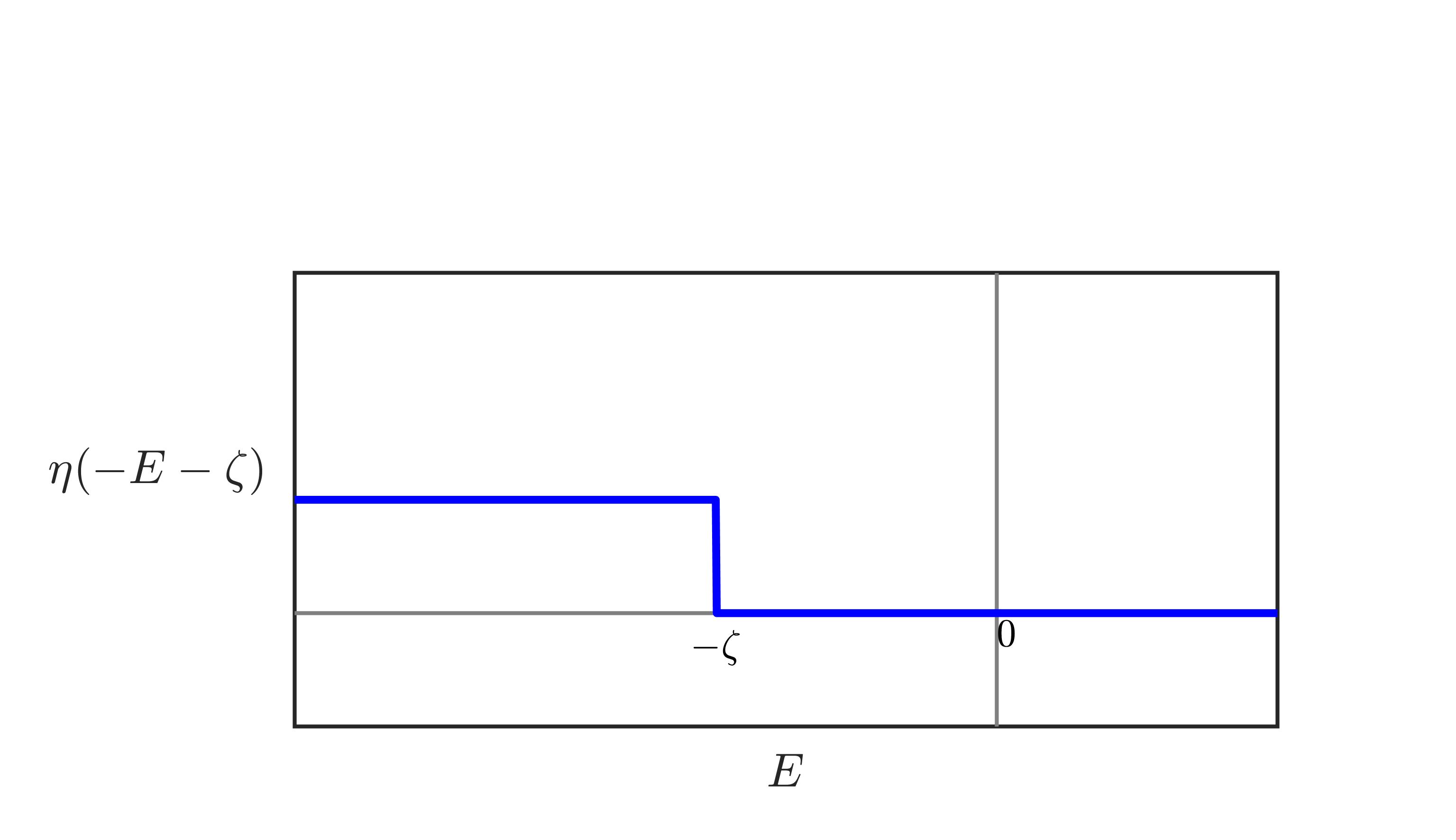}%
\caption{The Heaviside function vs. Energy.}%
\end{center}
\end{figure}

The latter function having been introduced in order to provide cutoffs in the
sums in (4) over the positive integer quantum numbers $l,$\thinspace$n_{r}$
thereby removing energies larger than $-\,\zeta$. \ The factor of 2 in Eq.(4)
has been included in the sum to account for the spin states. \ Using (4) we
have
\begin{equation}
\overline{V}_{ne}=2\sum_{l=0}^{\infty}\sum_{n_{r}=0}^{\infty}(2l+1)\,\eta
(-E_{l,\,n_{r}}-\varsigma)\overline{V}_{l,n_{r}}, \label{5}%
\end{equation}
with
\[
\overline{V}_{l\,,\,n_{r}}=-Z\int_{r_{l}}^{r_{u}}\frac{\left|  u_{l\,,\,n_{r}%
}(r)\right|  ^{2}}{r}dr,
\]
and $r_{l}$ and $r_{u}$ are the WKB lower and upper classical turning points
which define the classically allowed region. \ The WKB functions $u_{l,n_{r}%
}(r)$ being given by
\begin{equation}
u_{l,n_{r}}(r)=\frac{A_{l,n_{r}}}{\sqrt{2}\,\left[  2\{E_{l,n_{r}%
}-V(r)-{\small (}l{\small +}1/2{\small )}^{2}{\small /}2r^{2}\}\right]
^{1/4}}, \label{6}%
\end{equation}
where $A_{l,n_{r}}$ are normalization constants and $V(r)$ is a central but
not necessarily the Coulombic potential (The $r$ dependencies of the ``phase
factors'' \cite{PHASEFACTOR} of these functions are being ignored here). \ The
potential $V(r)$ represents the interaction of an electron with the nuclear
charge as well as with the other electrons in the atom, an approximate example
of\ which is the TF potential.\ 

Within the WKB approximation one also has the relation
\begin{equation}
n_{r}+1/2\,{\small =}\frac{1}{\pi}\int_{r_{l}}^{r_{u}}\sqrt{2\{E_{l,n_{r}%
}-V(r)-{\small (}l{\small +}1/2{\small )}^{2}{\small /}2r^{2}\}}\,dr,
\label{7}%
\end{equation}
where $r_{l}$ and $r_{u}$ referred to above are the roots of the quantity
within the square root of that expression. \ Following Schwinger and Englert
\cite{SE1} we define the quantities $\lambda,\nu,\varepsilon_{\lambda,\nu}%
,$and $\upsilon_{\lambda,\nu}$ as%
\begin{align*}
\lambda &  =l+1/2,\\
\nu &  =n_{r}+1/2,\\
\varepsilon_{\lambda,\nu}  &  =E_{l\,,\,n_{r}},\\
\upsilon_{\lambda,\nu}  &  =\overline{V}_{l\,,\,n_{r}},
\end{align*}
and regard them as continuous variables in the equations below. Then (7)
becomes \
\begin{equation}
\nu=\frac{1}{\pi}\int_{r_{l}}^{r_{u}}\sqrt{2\{\varepsilon_{\lambda,\nu
}-V(r)-\lambda^{2}{\small /}2r^{2}\}}\,dr. \label{8}%
\end{equation}
Rewriting (5), we note that the sums now extend over the negative as well as
the positive values of $l$ and $n_{r}$. \ The former values however, do not
contribute to the sums and we get
\[
\overline{V}_{ne}=4\int_{0}^{\infty}\lambda d\lambda\sum_{l=-\infty}^{\infty
}\delta(l+1/2-\lambda)\int_{0}^{\infty}d\nu\sum_{n_{r}=-\infty}^{\infty}%
\delta(n_{r}+1/2-\nu){\large \upsilon}_{\lambda,\nu}\eta(-\varepsilon
_{\lambda,\nu}-\varsigma),
\]
where $\delta(z)$ is the Dirac delta function. \ Using the Poisson identities
\begin{align*}
\sum_{l=-\infty}^{\infty}\delta(l+1/2-\lambda)  &  =\sum_{k=-\infty}^{\infty
}(-1)^{k}\exp(2\pi ik\lambda),\\
\sum_{n_{r}=-\infty}^{\infty}\delta(n_{r}+1/2-\nu)\,  &  =\sum_{j=-\infty
}^{\infty}(-1)^{j}\exp(2\pi ij\nu),
\end{align*}
the expression for $\overline{V}_{ne}$ becomes
\begin{equation}
\overline{V}_{ne}=4\sum_{k=-\infty}^{\infty}\;\sum_{j=-\infty}^{\infty
}(-1)^{k+j}\int_{0}^{\infty}\lambda\exp(2\pi ik\lambda)d\lambda\int
_{0}^{\infty}\exp(2\pi ij\nu)\,{\large \upsilon}_{\lambda,\nu}\eta
(-\varepsilon_{\lambda,\nu}-\varsigma)\,d\nu. \label{9}%
\end{equation}

\subsection{Interrelations Among The Variables $\lambda\,,\nu$ \thinspace And
$\varepsilon(\lambda\,,\nu)$}

Before proceeding, it is useful to examine the relations among the quantities
$\nu$, $\lambda$ and $\varepsilon(\lambda,\nu).$ \ For a given $V(r)$ and
energy $\varepsilon,$ and for a range of values of $\nu$ which will be
discussed below, the roots of the relation
\begin{equation}
2r^{2}[\varepsilon_{\lambda,\nu}-V(r)]-\lambda(r)^{2}=0, \label{10}%
\end{equation}
i.e. $r_{l}$ and $r_{u},$ define the classical turning points for the system.
\ We note that it follows from Eq. (4) that when $\nu=0$, that these roots
coalesce to the single value of $r_{\varepsilon}.$ \ This behavior can be seen
graphically in Fig. (3) where \ we have plotted the effective potential
$V^{\prime}=V(r)+\lambda^{2}/2r^{2}$ versus $r$.%
\begin{figure}
\begin{center}
\includegraphics[
width=5.6187in
]%
{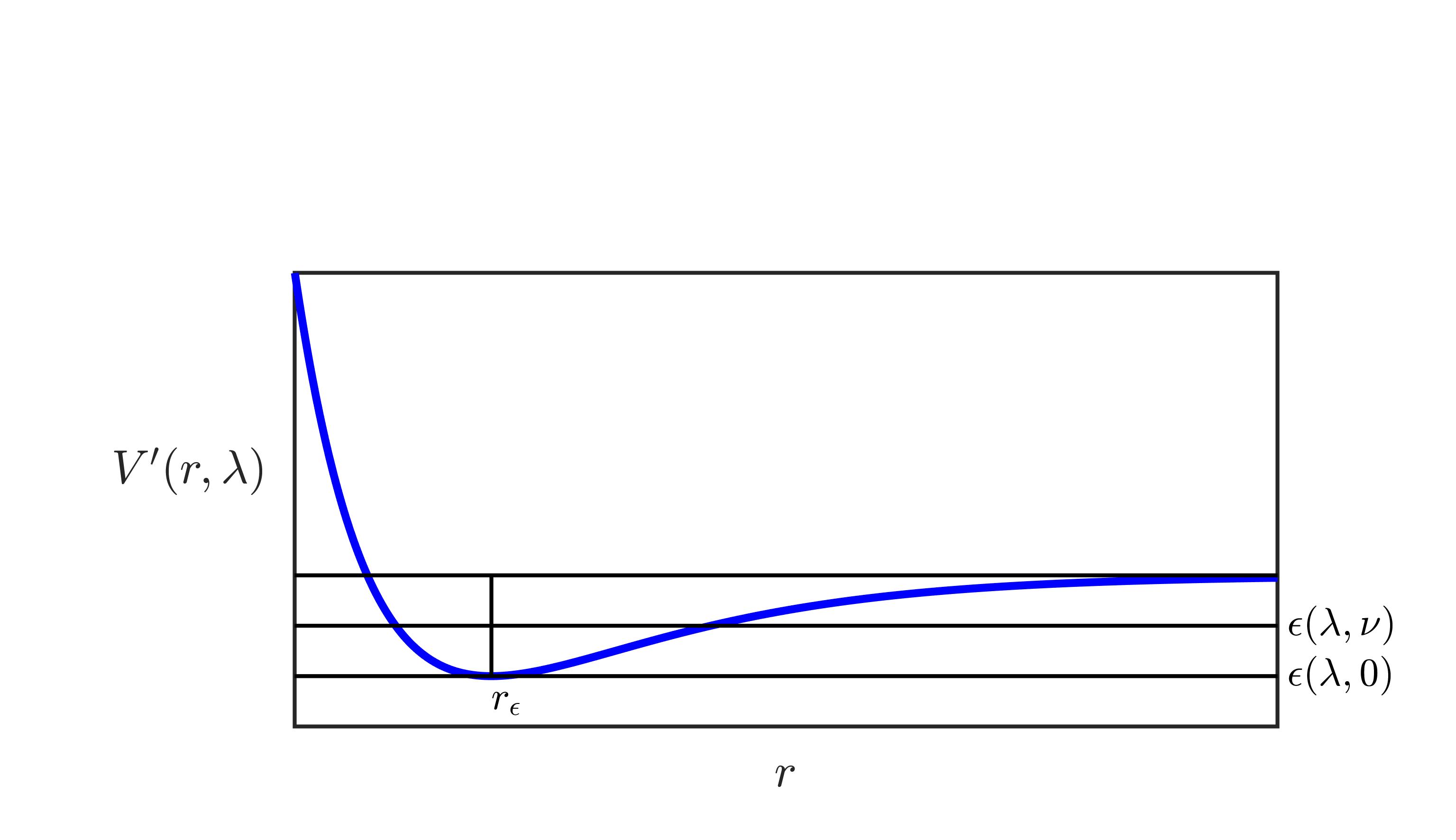}%
\caption{ The Effective potential vs. r}%
\end{center}
\end{figure}

For a given energy $\varepsilon(\lambda_{\varepsilon},0)$ the curve
$V^{\,\prime}(r)$ is seen to have a single turning point denoted by
$r_{\varepsilon}.$ \ However, the curve $V^{\,\prime}(r)$ for any
$\varepsilon(\lambda,\nu)>\varepsilon(\lambda_{\varepsilon},0)$ has two
turning points $r_{l},$ $r_{u}$\ which are less than and greater than
$r_{\varepsilon}$. \ Recalling that these turning points are the roots of Eq.
(8) we will see in what follows that $r_{\varepsilon}$ is the distance at
which $\lambda$ has it maximum value i.e. $\lambda_{\varepsilon}.$ \ This
value is given by%
\[
\lambda_{\varepsilon}(r_{\varepsilon})=\sqrt{2r_{\varepsilon}^{2}\left\{
\varepsilon(\lambda_{\varepsilon},0)-V(r_{\varepsilon})\right\}  }.
\]
Furthermore, in the case where the single-particle energy has its absolute
highest value referred to above as $-\zeta$ and hereafter as $\varepsilon
_{0\text{ }},$ the corresponding \textit{largest} of the maximum values of
$\lambda_{\varepsilon}$ is here denoted by $\lambda_{0}$ and satisfies the
relation
\[
\lambda_{0}(r_{0})=\sqrt{2r_{0}^{2}\left\{  \varepsilon_{0}-V(r_{0})\right\}
},
\]
where $r_{0}$ is the distance at which $\lambda$ has the largest maximum value
$\lambda_{0}$ and is the min point in the $V^{\,\prime}$ curve corresponding
to the energy $\varepsilon_{0}$.

To demonstrate the behavior of $\lambda$ we take as an example the case of the
classical turning points the TF potential \cite{ZHU} i.e. $-Z\Phi(x)/r$. \ In
Fig. (4) we have (for a given $\varepsilon$) plotted the scaled quantity
$\widetilde{\lambda}=\lambda(r)/\sqrt{a}Z^{1/3}$ in terms of the scaled energy
$\widetilde{\epsilon}=-a\,Z^{2/3}\varepsilon$ and the scaled distance
$x=Z^{1/3}r/a$ where $a=\frac{1}{2}(\frac{3\pi}{4})^{2/3}$ and $\left|
\widetilde{\epsilon}\right|  <$ $\left|  \widetilde{\epsilon}^{\,\prime
}\right|  ,$ then
\[
\widetilde{\lambda}=\sqrt{2\,x\,\left\{  \Phi(x)-x\,\widetilde{\epsilon
}\right\}  }.
\]

\ In Fig. (3) we see that at every energy $\varepsilon$ for a given $\lambda$
\ there are two turning points except at the maximum value of $\lambda,$ i.e.
$\lambda_{\varepsilon}(r_{\varepsilon})$ occurring at $r_{\varepsilon}.$ The
corresponding range of physical values of $x$ being $0\leq x\leq x_{\max}$
where $x_{\max}$ is determined by the roots of the equation $\Phi(x_{\max
})-x_{\max}\widetilde{\epsilon}=0.$ \ The quantity $r_{\varepsilon}$ which
allows calculation of $\lambda_{\varepsilon}(r_{\varepsilon})$ can be obtained
from the equation%
\[
\frac{d\,\widetilde{\lambda}}{d\,x}|_{x=x_{\varepsilon}}=0,
\]
or
\[
\frac{d\,\left\{  x\,\Phi(x)\right\}  }{dx}|_{x=x_{\varepsilon}}%
=2\,x_{\varepsilon}\,\widetilde{\epsilon}.
\]
This behavior is shown in Fig. (4)%

\begin{figure}
\begin{center}
\includegraphics[
width=4.6804in
]%
{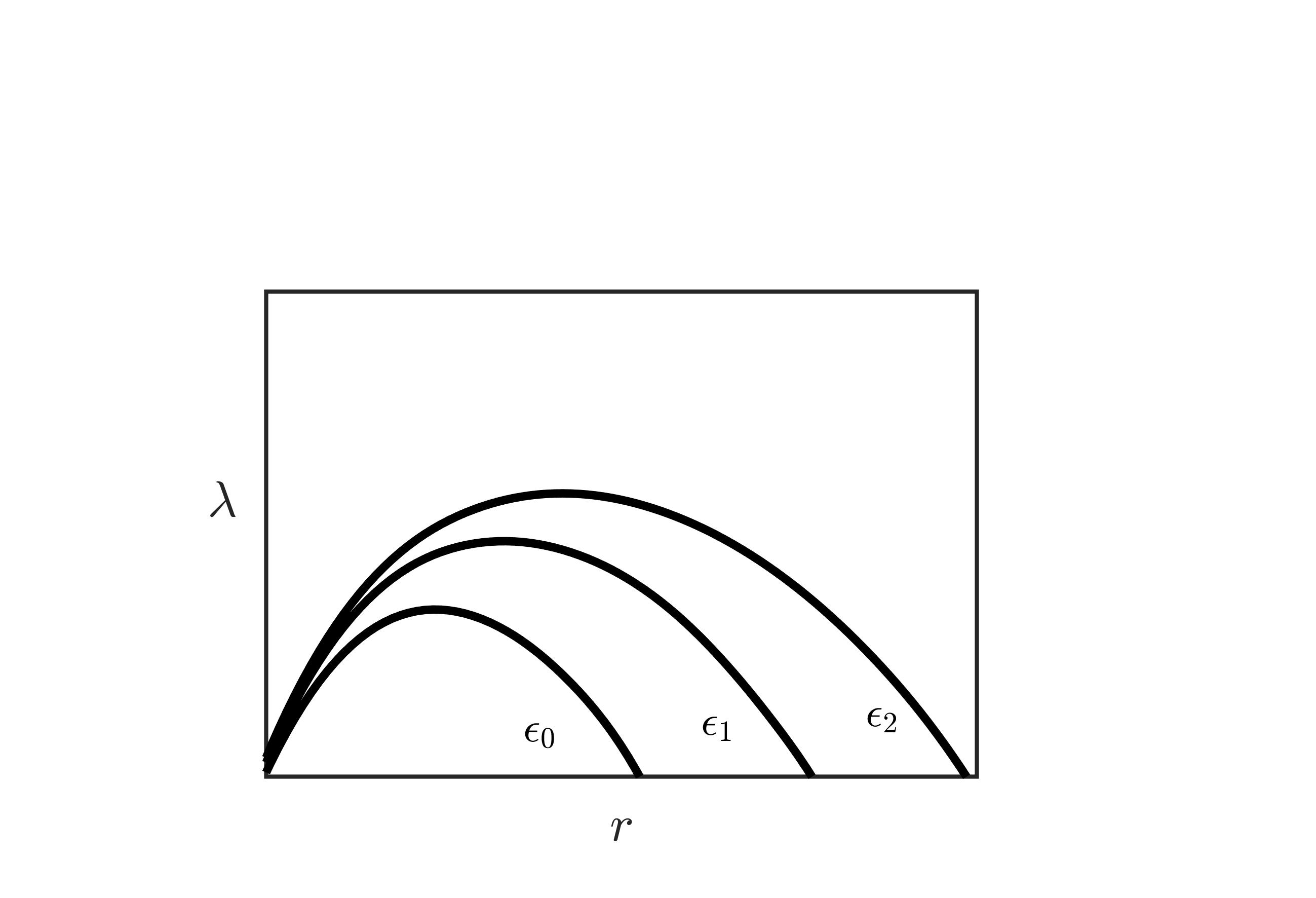}%
\caption{$\lambda$ vs $r$ for the TF potential.}%
\end{center}
\end{figure}

As a further example of the behavior of $\lambda,$ consider the case of the
Coulomb potential where we have
\[
Z\,r_{l}=\frac{1\pm\sqrt{1-2\,\epsilon\,\lambda^{2}}}{2\,\epsilon},\text{where
}\epsilon=-\varepsilon/Z^{2},
\]
which yield two turning points except when $\nu=0$ and where $\lambda
_{\varepsilon}(r)=\frac{Z}{\sqrt{2|\varepsilon|}}$ at which $r_{\varepsilon
}=\frac{Z}{2\left|  \varepsilon\right|  }.$ \ In Fig. (5) we have plotted
$\lambda_{\epsilon}=\sqrt{2\,\mathfrak{r}\,\left[  1-\mathfrak{r}%
\,\epsilon\right]  }$as a function of the scaled distance $\mathfrak{r}=Z\,r$
and the scaled energy $\epsilon$%

\begin{figure}
\begin{center}
\includegraphics[
width=5.3817in
]%
{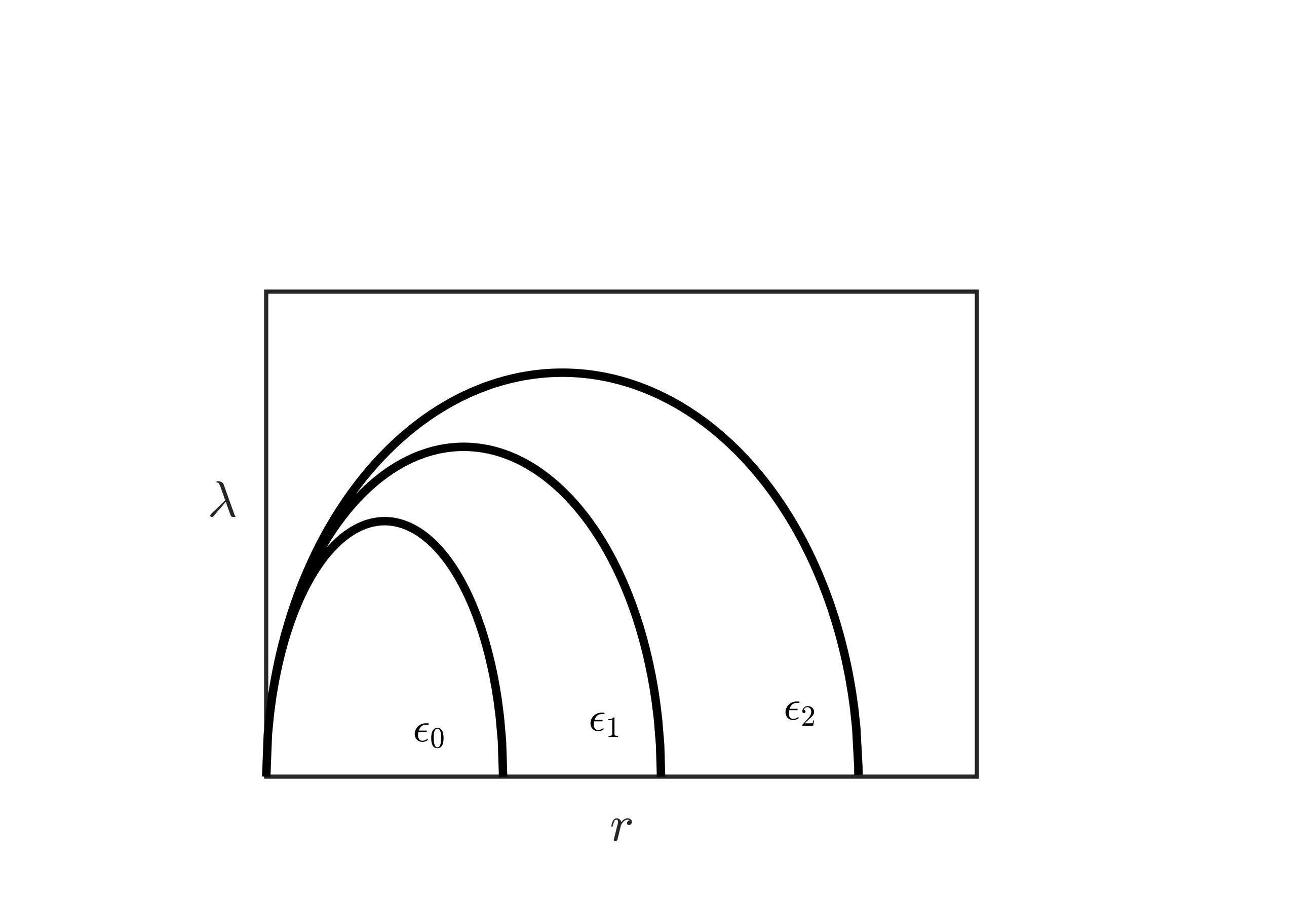}%
\caption{$\lambda$ vs $r$ for the Coulomb potential}%
\end{center}
\end{figure}

\ We see in the case of the Coulomb potential for a given $\varepsilon$ with
$\lambda_{\varepsilon}=0$, there is a set of classical turning points which
occur at $0$ and $\frac{Z}{\left|  \varepsilon\right|  },$whereas the maximum
values of $\lambda_{\varepsilon}$ are $\lambda_{\varepsilon,\max}%
(r)=\frac{Z}{\sqrt{2\left|  \varepsilon\right|  }}$, and the set of single
turning points occur at $r_{\varepsilon,\max}=\frac{Z}{2\left|  \varepsilon
\right|  }$. \ Furthermore, for a given $\varepsilon(\lambda,\nu)=const$ \ the
variables $\lambda$ and $\nu(\lambda|\varepsilon)$ are related as shown in the
Fig. (6) below.%

\begin{figure}
\begin{center}
\includegraphics[
width=4.7288in
]%
{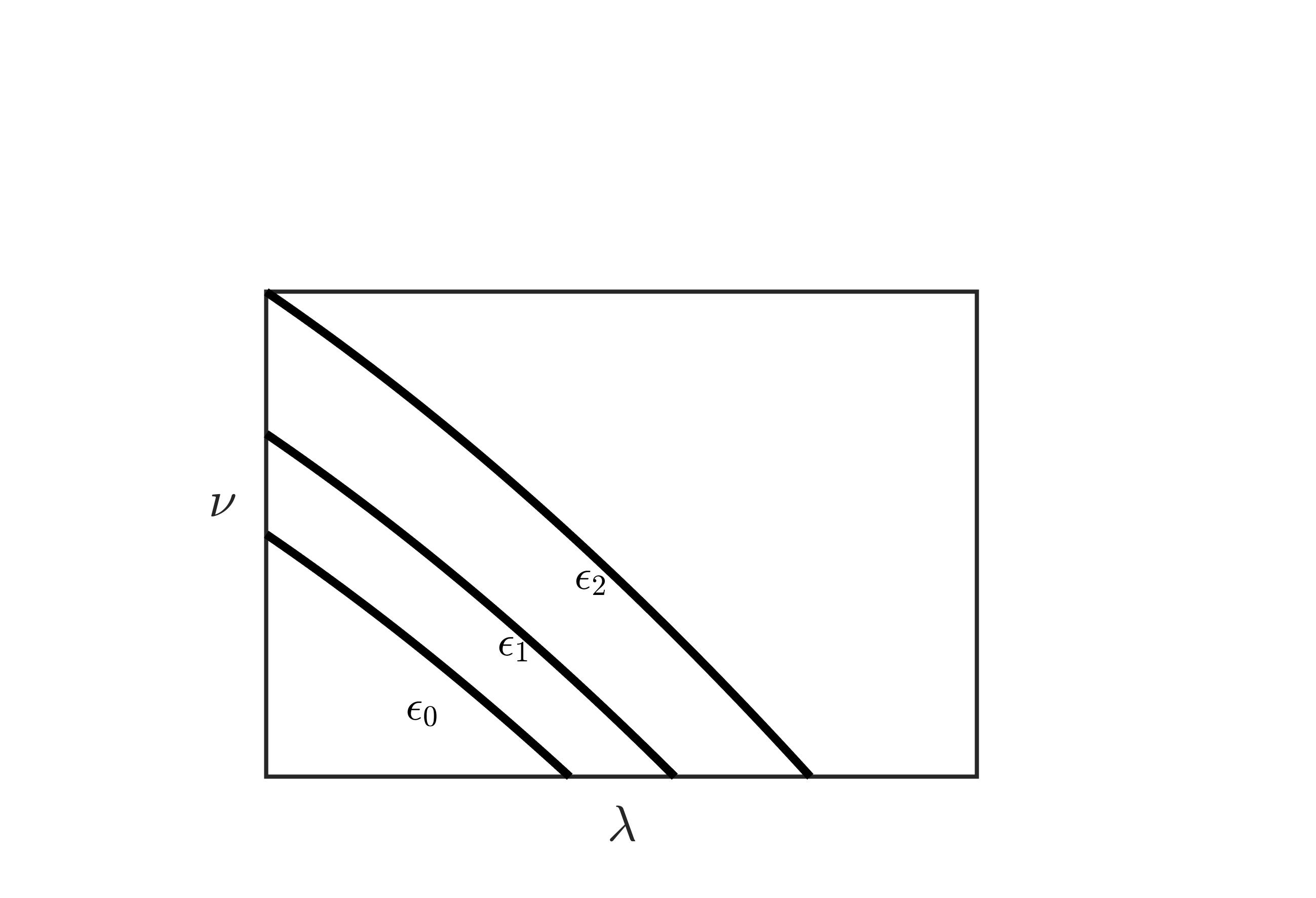}%
\caption{$\nu$ vs. $\lambda$ for the TF potential.}%
\end{center}
\end{figure}

We expect the shape of the curves in Fig. 6 to show concave curvature as is
the case of the TF potential. \ Along these curves the energy is constant
(\textit{curves of degeneracy)}.\ \ In the case of the Coulombic potential,
where the energy is $\varepsilon=-Z^{2}/2(\lambda+\nu)^{2}$ the curves in the
plot of $\nu(\lambda|\varepsilon)$ versus $\lambda$ for different
$\varepsilon$ consists of a family of \textit{straight lines} whereas in the
case of a general potential we expect these lines to be curved as shown above
\cite{PARAMETERNOTATION}. \ In addition we denote the maximum value of $\nu$
i.e. $\nu(0|\varepsilon)=\nu_{\varepsilon}.$ \ From this we see that for a
given $\varepsilon$ the quantities $\lambda$ and $\nu$ are restricted to the
ranges%
\[
0\leq\lambda\leq\lambda_{\varepsilon}(r),
\]%
\[
0\leq\nu\leq\nu(\lambda|\varepsilon).
\]

The domain of integration in $\lambda,\nu$ space is seen to consist of all
$\lambda,\nu$ values below the curves of degeneracy $\nu(\lambda|\varepsilon)$
corresponding to $\varepsilon=\varepsilon_{0}$. \ With this in mind we can for
a given $\varepsilon$ rewrite the average potential\text{ }$\overline{V}_{ne}$
as
\begin{equation}
\overline{V}_{ne}=4\lim_{\varepsilon\longrightarrow\varepsilon_{0}}%
\sum_{k=-\infty}^{\infty}\sum_{j=-\infty}^{\infty}(-1)^{k+j}\int_{0}%
^{\lambda_{\varepsilon}(r)}\lambda\exp(2\pi ik\lambda)\,d\lambda\int_{0}%
^{\nu(\lambda|\varepsilon)}\exp(2\pi ij\nu)\,{\large \upsilon}_{\lambda,\nu
}\,d\nu, \label{11}%
\end{equation}
with $\nu(\lambda|\varepsilon)$ denoting the\ \textit{curves of degeneracy.}

\ In the work to follow it is useful to define the integrals $\mathcal{N}%
_{j}\,\mathcal{(}\lambda_{\varepsilon},\varepsilon_{\lambda,\nu})$ as
\[
\mathcal{N}_{j}\mathcal{(}\lambda,\varepsilon)=\int_{0}^{\nu(\lambda
|\varepsilon)}\cos(2\pi j\,\nu)\,{\large \upsilon}_{\lambda,\nu}\,\,d\nu.
\]

\subsection{The Regions Of $k,\,j$ Space}

In order to make progress in evaluating the terms in the double sum in Eq.
(11) for $\overline{V}_{ne}$ it is useful to divide $k,\,j$ index space (note
that $\lambda$ and $\nu$ are associated with the indices $k,$ and $j$
respectively) into the regions shown in the diagram below. \ These regions
shown in Fig. (7) correspond roughly to those chosen by SE in their evaluation
the energy of the system.%

\begin{figure}
\begin{center}
\includegraphics[
width=4.5238in
]%
{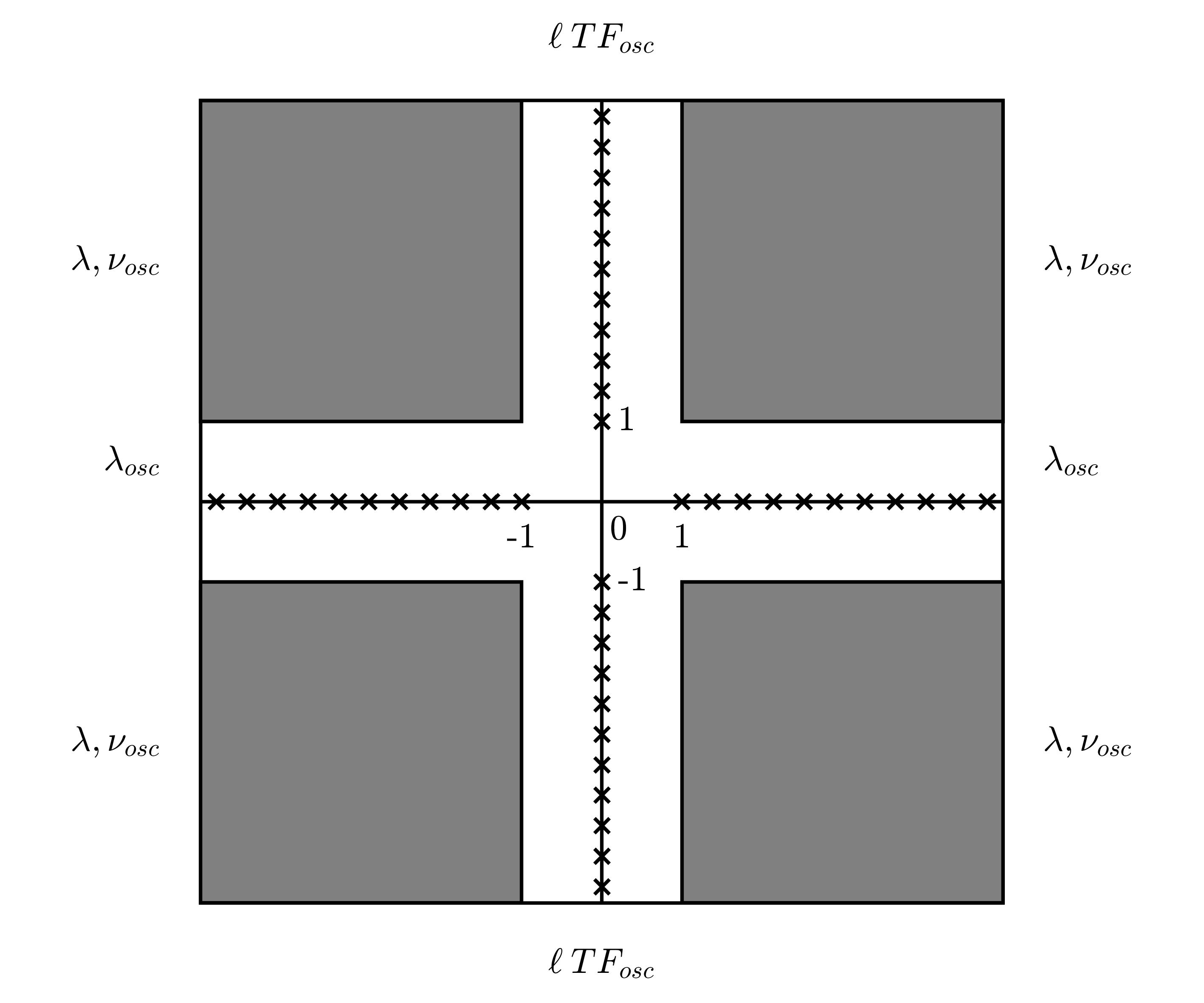}%
\caption{Regions within the $k,$ $j$ space.}%
\end{center}
\end{figure}

The TF region consists of the single point
\[
j=0,\hspace{0.15in}k=0
\]
the \textit{l\thinspace}TF region consists of the points on the vertical line%
\[
\text{\ }j=0,\hspace{0.25in}1\leq k\leq\infty,
\]
the $\lambda$ region is given by%
\[
k=0,\hspace{0.25in}1\leq j\leq\infty\text{,}%
\]
and the $\lambda,\nu$ region consists of the points covered by the ranges%
\[
1\leq j\leq\infty,\text{ \ \ }1\leq k\leq\infty.
\]
\ \ Rewriting $\overline{V}_{ne}$ in terms of these regions we have%
\begin{equation}
\overline{V}_{ne}=\lim_{\varepsilon\rightarrow\varepsilon_{0}}\left\{
\,\overline{V}_{ne,TF}(\varepsilon)+\overline{V}_{ne,lTF}(\varepsilon
)+\overline{V}_{ne,\,\lambda}(\varepsilon)+\overline{V}_{ne,\,\,\lambda
\mathbf{,}\nu}(\varepsilon)\,\right\}  , \label{12}%
\end{equation}
where
\begin{subequations}
\begin{align}
\overline{V}_{ne,TF}(\varepsilon)  &  =4\int_{0}^{\lambda_{\varepsilon}%
}\lambda\,\mathcal{N}_{0}\mathcal{(}\lambda,\varepsilon)\,d\lambda,\tag{13a}\\
\overline{V}_{ne,\,\mathit{l}TF}(\varepsilon)  &  =8\sum_{k=1}^{\infty
}(-1)^{k}\int_{0}^{\lambda_{\varepsilon}}\lambda\cos(2\pi k\lambda
)\,\mathcal{N}_{0}\mathcal{(}\lambda,\varepsilon)\,d\lambda,\tag{13b}\\
\overline{V}_{ne,\,\lambda}(\varepsilon)  &  =8\sum_{j=1}^{\infty}(-1)^{j}%
\int_{0}^{\lambda_{\varepsilon}}\lambda\,\mathcal{N}_{j}\mathcal{(}%
\lambda,\varepsilon)\,d\lambda\tag{13c}\\
\overline{V}_{ne,\,\,\lambda,\nu}(\varepsilon)  &  =16\sum_{k=1}^{\infty}%
\sum_{j=1}^{\infty}(-1)^{k+j}\int_{0}^{\lambda_{\varepsilon}}\lambda
\,\cos(2\pi k\lambda)\,\mathcal{N}_{j}\mathcal{(}\lambda,\varepsilon
)\,d\lambda\,. \tag{13d}%
\end{align}
The numerical factors appearing in Eqs (13) result from the contributions from
the second, third and fourth quadrants of Fig. (7). \ We will see below that
the first term shown in (13) i.e. $\overline{V}_{ne,TF}$ produces a
non-oscillatory, semiclassical expression for the average nuclear electronic
potential. \ The sum $\overline{V}_{ne,\,\mathit{l}TF,\,}$ over the second
region produces an oscillatory semiclassical expression which we will call the
`$l-$ quantized' semiclassical average potential. \ Oscillatory terms in the
remaining regions will be called the `$\lambda$, and the $\lambda,\nu$ -
quantized' \ contributions to the average potential respectively.

\subsection{Evaluation Of $\mathcal{N}_{j}\mathcal{(}\lambda_{\,}%
,\varepsilon_{\lambda,\nu})=\int_{0}^{\nu(\lambda|\varepsilon)}\cos(2\pi
\,j\nu)\,\,{\protect\Huge \upsilon}_{\lambda,\nu}\,\,d\nu$}

The program for the evaluation of the expression for $\overline{V}_{ne}$ in
Eq. (12), is best carried out by investigating its various parts in a stepwise
fashion in order to simplify the exposition of the work. \ We begin by noting
that (recall $\upsilon_{\lambda,\nu}=\overline{V}_{l\,,\,n_{r}},$)
\end{subequations}
\begin{equation}
-\frac{{\large \upsilon}_{\lambda,\nu}}{Z}=\int_{r_{l}}^{r_{u}}\frac{\left|
u_{l,n_{r}}(r)\right|  ^{2}}{r}dr=\frac{\left|  A_{\lambda,\nu}\right|  ^{2}%
}{2}\int_{r_{l}}^{r_{u}}\frac{dr}{r\sqrt{2\{\varepsilon_{\lambda,\nu
}-V(r)-{\small \lambda}^{2}{\small /2}r^{2}\}}}, \label{14}%
\end{equation}
and
\[
\int_{0}^{\infty}\left|  u_{l,n_{r}}(r)\right|  ^{2}dr=1=\frac{\left|
A_{\lambda,\nu}\right|  ^{2}}{2}\int_{r_{l}}^{r_{u}}\frac{dr}{\sqrt
{2\{\varepsilon_{\lambda,\nu}-V(r)-{\small \lambda}^{2}{\small /2}r^{2}\}}}.
\]
Differentiation of $\nu$ with respect to $\varepsilon$ in Eq. (8) gives
\[
\left(  \frac{\partial\nu}{\partial\varepsilon_{\lambda,\nu}}\right)
|_{\lambda}=\frac{1}{\pi}\int_{r_{l}}^{r_{u}}\frac{dr}{\sqrt{2\{\varepsilon
_{\lambda,\nu}-V(r)-{\small \lambda}^{2}{\small /2}r^{2}\}}},
\]
and thus
\[
\frac{\left|  A_{\lambda,\nu}\right|  ^{2}}{2}=\frac{1}{\pi}\left(
\frac{\partial\varepsilon_{\lambda,\nu}}{\partial\nu}\right)  |_{\lambda}.
\]
As a result we may rewrite Eq. (14) as (valid for all $\lambda$ and $\nu)$%
\[
-\frac{{\large \upsilon}_{\lambda,\nu}}{Z}=\frac{1}{\pi}\left(  \frac{\partial
\varepsilon_{\lambda,\nu}}{\partial\nu}\right)  _{\lambda}\int_{r_{l}}^{r_{u}%
}\frac{dr}{\sqrt{2r^{2}\{\varepsilon_{\lambda,\nu}-V(r)\}-{\small \lambda}%
^{2}}}.
\]
The integral $\mathcal{N}_{j}\mathcal{(}\lambda,\varepsilon)$ over the
variable $\nu$ then becomes (for a given $\lambda$ the energy $\nu$ can vary
with $\varepsilon$)
\begin{equation}
\mathcal{N}_{j}\mathcal{(}\lambda,\varepsilon)=-\frac{Z}{\pi}\int_{0}%
^{\nu(\lambda|\varepsilon)}d\nu\,\cos(2\pi j\nu)\,\left(  \frac{\partial
\varepsilon_{\lambda,\nu}}{\partial\nu}\right)  _{\lambda}\int_{r_{l}}^{r_{u}%
}\frac{dr}{\sqrt{2r^{2}\{\varepsilon_{\lambda,\nu}-V(r)\}-{\small \lambda}%
^{2}}}. \label{15}%
\end{equation}
Interchanging the order of integration we get%
\begin{equation}
\mathcal{N}_{j}\mathcal{(}\lambda,\varepsilon)=-\frac{Z}{\pi}\int_{r_{l}%
}^{r_{u}}dr\int_{0}^{\nu(\lambda|\varepsilon)}d\nu\,\,\left(  \frac{\partial
\varepsilon_{\lambda,\nu}}{\partial\nu}\right)  _{\lambda}\frac{\cos(2\pi
j\nu)}{\sqrt{2r^{2}\{\varepsilon_{\lambda,\nu}-V(r)\}-{\small \lambda}^{2}}}.
\label{16}%
\end{equation}
If the relation%
\begin{align*}
&  (\frac{\partial\varepsilon_{\lambda,\nu}}{\partial\nu})_{\lambda}%
\frac{\cos(2\pi j\nu)}{\sqrt{2r^{2}\{\varepsilon_{\lambda,\nu}%
-V(r)\}-{\small \lambda}^{2}}}\\
&  =2\pi j\sin(2\pi j\nu)\tfrac{\sqrt{2r^{2}\{\varepsilon_{\lambda,\nu
}-V(r)\}-{\small \lambda}^{2}}}{r^{2}}+\frac{\partial}{\partial\nu}[\cos(2\pi
j\nu)\tfrac{\sqrt{2r^{2}\{\varepsilon_{\lambda,\nu}-V(r)\}-{\small \lambda
}^{2}}}{r^{2}}]_{\lambda},
\end{align*}
is used, we get an expression for $\mathcal{N}_{j}\mathcal{(}\lambda
,\varepsilon)$ which is partially integrable i.e.%
\begin{align*}
&  \mathcal{N}_{j}\mathcal{(}\lambda,\varepsilon)\\
&  =-\frac{Z}{\pi}\int_{r_{l}}^{r_{u}}dr\int_{0}^{\nu(\lambda|\varepsilon
)}\{2\pi j\sin(2\pi j\nu)\tfrac{\sqrt{2r^{2}\{\varepsilon_{\lambda,\nu
}-V(r)\}-{\small \lambda}^{2}}}{r^{2}}+\frac{\partial}{\partial\nu}[\cos(2\pi
j\nu)\tfrac{\sqrt{2r^{2}\{\varepsilon_{\lambda,\nu}-V(r)\}-{\small \lambda
}^{2}}}{r^{2}}]_{\lambda}\}d\nu\,.
\end{align*}
And finally we obtain
\begin{align*}
\mathcal{N}_{j}\mathcal{(}\lambda,\varepsilon)  &  =-\frac{Z}{\pi}\int_{r_{l}%
}^{r_{u}}\frac{dr}{r^{2}}\cos(2\pi j\nu(\lambda|\varepsilon))\sqrt
{2r^{2}\{\varepsilon_{\lambda,\nu(\lambda|\varepsilon)}-V(r)\}-{\small \lambda
}^{2}}\\
&  -2Z\,j\int_{r_{l}}^{r_{u}}\frac{dr}{r^{2}}\int_{0}^{\nu(\lambda
|\varepsilon)}\sin(2\pi j\nu)\sqrt{2r^{2}\{\varepsilon_{\lambda,\nu
}-V(r)\}-{\small \lambda}^{2}}d\nu,
\end{align*}
an expression which will be useful in the evaluation of the integrals in Eq. (12).

\subsection{Thomas Fermi parameters $r_{0},$ $\lambda_{0},$ $\omega_{0},$
$\nu_{0}^{(1)},$ $\nu_{0}^{(2)},\mathcal{K}_{0}(r_{0})$}

At this juncture it is useful to review Thomas-Fermi (TF) theory and to
compute some of the parameters which will be needed in the final calculation
of the chemical potential. \ In TF theory the potential $V(r)$ is defined by
the equations
\[
\nabla^{2}V=-4\pi\varrho_{TF}(\mathbf{r}),
\]%
\begin{equation}
\varrho_{TF}(\mathbf{r})=\frac{1}{3\pi^{2}}\left(  2\{\varepsilon
_{0}-V(r)\}\right)  ^{3/2}. \label{17}%
\end{equation}
In the case of a spherically symmetric system this equation can be rewritten
as
\[
\frac{d^{2}\Phi}{d\,x^{2}}=\frac{\Phi(x)^{3/2}}{\sqrt{x}},\qquad\text{with
\ \thinspace}\Phi(0)=1,\text{ }\Phi(\infty)=0,
\]
\ As seen above in the case of the Coulombic potential the value of $r$ at
which $\lambda$ \ is a maximum for a given $\varepsilon$ had been given as
\[
r_{\varepsilon}=\frac{Z}{2\left|  \varepsilon\right|  }.
\]
Here using the TF potential we give the value of $r$ corresponding to the
maximum of $\lambda$ in the case where $\varepsilon\rightarrow\varepsilon
_{0}.$ \ Using Eq. (10) we have%
\[
\frac{d\,\lambda_{0}^{2}(r)}{d\,r}=0=\frac{d}{d\,r}\{-2r^{2}V(r)\},
\]
which is tantamount to
\[
\frac{d\{x\,\Phi(x)\}}{d\,x}=0.
\]
The maximum of the quantity $x\,\Phi(x)$ occurs at $x_{0}=2.10403$ where
$\Phi(x_{0})=0.231151$ the corresponding value of $r_{0}$ being
$2.10403\,a/Z^{1/3}$ \cite{xphivalue}. \ The largest of the maximum values of
$\lambda_{\varepsilon}$ i.e. $\lambda_{0\text{ }}$is then
\[
\lambda_{0\text{ }}(r_{0})=\sqrt{2a\,x_{0}\Phi(x_{0})}\,Z^{1/3}%
=0.927992\,Z^{1/3}.
\]
In the work that follows a collection of quantities (which have been computed
by Schwinger and Englert ) for the TF potential is given below and will prove
useful in the estimation of the $Z$ dependence of the remaining terms in
$\overline{V}_{ne}.$ We have

\begin{center}
$%
\begin{tabular}
[c]{l}%
$x_{0}=2.10403,$\\
\\
$r_{0}=1.86278\,Z^{-1/3},$\\
\\
$\omega_{\varepsilon}^{2}=-\frac{d^{2}\lambda_{\varepsilon}^{2}(r)}{d\,r^{2}%
}|_{r=r_{\varepsilon}}$\\
\\
$\omega_{0}=0.36359\,Z^{\,2/3},$\\
\\
$\lambda_{0}(r_{0})=0.92799\,Z^{\,1/3},$\\
\\
$\overset{\bullet}{\lambda}=\frac{\partial\lambda_{\varepsilon}(r_{0}%
)}{\partial\varepsilon}|_{\varepsilon=0}=r_{0}^{2}/\lambda_{0}(r_{0}%
)=3.73921Z$\\
\\
$\frac{\lambda_{0}(r_{0})}{\overset{\bullet}{\lambda}}=\left(  \frac{\lambda
_{0}(r_{0})}{r_{0}}\right)  ^{2}=0.24818Z^{4/3}$\\
\\
$\nu_{0}^{(1)}=-\frac{\partial\nu_{0}(\lambda)}{\partial\lambda}%
|_{\lambda=\lambda_{0}}=\sqrt{2}\frac{\lambda_{0}(r_{0})}{\omega_{0}r_{0}%
}=1.93768,$\\
\\
$\nu_{0}^{(2)}=-\frac{\partial^{2}\nu_{0}(\lambda)}{\partial\lambda^{2}%
}|_{\lambda=\lambda_{0}}=\frac{\nu_{0}^{(1)}}{24\lambda_{0}(r_{0})}[(\nu
_{0}^{(1)})^{2}-1][-15+23(\nu_{0}^{(1)})^{2}-5(\nu_{0}^{(1)})^{4}%
]=0.208673/Z^{1/3},$\\
\\
$\frac{\omega_{0}^{2}}{\lambda_{0}(r_{0})}=0.14246\,Z$\\
\\
$\mathcal{K}_{0}(r_{0})=\frac{\lambda_{0}^{2}(r_{0})}{Z\,r_{0}}=0.46230,$\\
\\
$\frac{\lambda_{0}^{3}(x_{0})}{\omega_{0}r_{0}^{3}\,}=0.34004\,Z^{\,4/3},$%
\end{tabular}
\ \ $
\end{center}

$\bigskip$

\subsection{The TF Term $\overline{V}_{ne,TF}$\bigskip}

In the case where\thinspace\ $j=0$ we have for $\mathcal{N}_{0}\mathcal{(}%
\lambda,\varepsilon_{\lambda,\nu(\lambda|\varepsilon)})$
\begin{equation}
\mathcal{N}_{0}\mathcal{(}\lambda,\varepsilon_{\lambda,\nu(\lambda
|\varepsilon)})=-\frac{Z}{\pi}\int_{r_{l}}^{r_{u}}\frac{dr}{r^{2}}\sqrt
{2r^{2}\{\varepsilon_{\lambda,\nu(\lambda|\varepsilon)}-V(r)\}-{\small \lambda
}^{2}}, \label{18}%
\end{equation}
the corresponding value of $\overline{V}_{ne,TF}$ then becomes%
\[
\overline{V}_{ne,TF}(\varepsilon)=-\frac{4Z}{\pi}\int_{0}^{\lambda
_{\varepsilon}}\lambda d\lambda\int_{r_{l}}^{r_{u}}\sqrt{2r^{2}\{\varepsilon
_{\lambda,\nu(\lambda|\varepsilon)}-V(r)\}-{\small \lambda}^{2}}dr/r^{2}.
\]
Interchanging the order of integration results in ($r_{l}\leq r\leq r_{u})$%
\[
\overline{V}_{ne,TF}(\varepsilon)=-\frac{4Z}{\pi}\int_{r_{l}}^{r_{u}}%
dr/r^{2}\int_{0}^{\lambda_{\varepsilon}}\lambda d\lambda\sqrt{2r^{2}%
\{\varepsilon_{\lambda,\nu(\lambda|\varepsilon)}-V(r)\}-\lambda^{2}},
\]
and we note that as $\lambda$ varies over its range from $0$ to $\lambda
_{\varepsilon}$ in the limit as $\varepsilon\rightarrow\varepsilon
_{0}\rightarrow0$ the turning points must vary from $r_{l}=0$ and
$r_{u}=\infty$ \ these being the roots of\ $2r^{2}\{\varepsilon_{0}%
-V(r)\}-\lambda_{0}^{2}=0.$ \ Integration over $\lambda$ gives the result
\begin{equation}
\overline{V}_{ne,TF}=-\frac{4Z}{3\pi}\int_{0}^{\infty}\left(  2r^{2}%
\{\varepsilon_{0}-V(r)\}\right)  ^{3/2}dr/r^{2}. \label{19}%
\end{equation}
Remarkably, if $V(r)$ is taken to be the Thomas-Fermi potential, the
corresponding particle density $\varrho_{TF}(\mathbf{r})$ with the form
\[
\varrho_{TF}(\mathbf{r})=\frac{1}{3\pi^{2}}\left(  2\{\varepsilon
_{0}-V(r)\}\right)  ^{3/2},
\]
then $\overline{V}_{ne,TF}$ in Eq. (19) rewritten as a integral over 3
dimensional space is just%
\[
\overline{V}_{ne,TF}=-Z\int\frac{\varrho_{TF}(\mathbf{r})}{r}\,d\mathbf{r}%
=\overline{V}_{TF}%
\]
We see that the leading term in the expression for $\overline{V}_{ne}$ is the
non-oscillatory Thomas-Fermi average value of the nuclear-electronic
interaction $\overline{V}_{TF}$. \ The remaining terms in the sums in Eq. (12)
represent the semi-classical and oscillatory contributions to the
nuclear-electronic interaction.

\subsection{\bigskip The Term $\overline{V}_{ne\,,\mathit{\,l}TF}\,$ The
$\mathit{l}$ TF Oscillations}

\bigskip

The sum representing $\overline{V}_{ne\,,\mathit{\,l}TF}\,$ can be rewritten
as
\begin{align*}
\overline{V}_{ne\,,\mathit{\,l}TF}\,(\varepsilon)  &  =8\sum_{k=1}^{\infty
}(-1)^{k}\int_{0}^{\lambda_{\varepsilon}}d\lambda\,\lambda\cos(2\pi
ik\lambda)\int_{0}^{\nu_{\varepsilon}(\lambda)}\,{\large \upsilon}%
_{\lambda\,,\nu}\,d\nu\\
\overline{V}_{ne\,,\mathit{\,l}TF}  &  =\lim_{\varepsilon\longrightarrow
\varepsilon_{0}}\,\overline{V}_{ne\,,\mathit{\,l}TF}\,(\varepsilon)
\end{align*}
\bigskip

Once again interchange of the order of integration in the integrals above and
use of the procedure to obtained $\overline{V}_{TF}$ (with the expression for
$\mathcal{N}_{0}\mathcal{(}\lambda,\varepsilon_{\lambda,\nu(\lambda
|\varepsilon)})$ in Eq. (18) $\overline{V}_{ne\,,\,lTF}\,(\varepsilon)$ can
then be rewritten as
\[
\overline{V}_{ne\,,\mathit{\,l}TF}\,(\varepsilon)=-\,\frac{8Z}{\pi}\sum
_{k=1}^{\infty}(-1)^{k}\int_{r_{l}}^{r_{u}}\frac{dr}{r^{2}}\int_{0}%
^{\lambda_{\varepsilon}}d\lambda\cdot\lambda\cos(2\pi ik\lambda)\sqrt
{2r^{2}\{\varepsilon_{\lambda_{\varepsilon}},_{{\large \nu}(\lambda
|\varepsilon)}-V(r)\}-\lambda^{2}}.
\]
Now we write
\[
\lambda_{\varepsilon}^{2}=2r^{2}\{\varepsilon_{\lambda_{\varepsilon}}%
,_{\nu(\lambda|\varepsilon)}-V(r)\},
\]
and $\lambda$ as
\[
\lambda=\lambda_{\varepsilon}(r)\cos\theta,\text{ \hspace{0.25in}}0\leq
\theta\leq\pi/2.
\]
The expression for\ $\overline{V}_{ne\,,\mathit{\,l}TF}\,(\varepsilon)$ with
this change of variable becomes%
\[
\overline{V}_{ne\,,\mathit{\,l}TF}\,(\varepsilon)=-\frac{8Z}{\pi}\sum
_{k=1}^{\infty}(-1)^{k}\int_{r_{l}}^{r_{u}}\frac{\lambda_{\varepsilon}^{3}%
(r)}{r^{2}}dr\int_{0}^{\pi/2}\sin^{2}\theta\cos\theta\cos(2\pi k\lambda
_{\varepsilon}(r)\cos\theta)\,d\theta.
\]
The angular integral appearing in the equation above is well-known and we have%
\[
\int_{0}^{\pi/2}\sin^{2}\theta\cos\theta\cos(2\pi k\lambda_{\varepsilon}%
\cos\theta)\,d\theta=\frac{1}{3}-\frac{\pi\mathbf{H}_{2}(2\pi k\lambda
_{\varepsilon})}{4\pi\,k\,\lambda_{\varepsilon}},
\]
where $\mathbf{H}_{2}(z)$ are the Struve functions \cite{STRUVE} of order $2$.
\ The required sum is then%
\[
\overline{V}_{ne\,,\mathit{\,l}TF}\,(\varepsilon)=-\frac{8Z}{\pi}\sum
_{k=1}^{\infty}(-1)^{k}\int_{r_{l}}^{r_{u}}\frac{\lambda_{\varepsilon}^{3}%
(r)}{r^{2}}\left[  \frac{1}{3}-\frac{\pi\mathbf{H}_{2}(2\pi k\lambda
_{\varepsilon}(r))}{4\pi\,k\,\lambda_{\varepsilon}(r)}\right]  \,dr.
\]
In the case of integer order, the Struve functions $\mathbf{H}_{k}(z)$ are
related to the Weber functions $\mathbf{E}_{k}(z)$ \cite{WEBER} . \ In this
case%
\[
\frac{1}{3}-\frac{\pi}{2z}\mathbf{H}_{2}(z)=\frac{\pi}{2z}\mathbf{E}_{2}(z),
\]
and we can write within the semiclassical approximation an exact expression
for$\overline{V}_{ne\,,\mathit{\,l}TF}\,(\varepsilon)$ as
\[
\overline{V}_{ne\,,\mathit{\,l}TF}\,(\varepsilon)=-\frac{2Z}{\pi}\sum
_{k=1}^{\infty}\frac{(-1)^{k}}{k}\int_{r_{l}}^{r_{u}}\frac{\lambda
_{\varepsilon}^{2}(r)}{\,\,\,r^{2}}\mathbf{E}_{2}(2\pi k\lambda_{\varepsilon
}(r))dr.
\]

For the Thomas-Fermi function $\Phi(x)$ where $\varepsilon=-\,\varsigma=0$ we
note that
\[
\lambda_{0}(r)=Z^{1/3}\sqrt{2a\,x_{0}\,\Phi(x_{0})},
\]
and see that for large $Z,$ that $\lambda_{0}(r)$ is large. \ For large $Z,$
using the asymptotic expansion for the leading and next to leading terms
\cite{ASYMPT} for $\mathbf{E}_{2}(z)$ i.e.%
\[
\frac{\pi}{2z}\mathbf{E}_{2}(z)\thicksim\sqrt{\frac{\pi}{2}}\frac{1}{z^{3/2}%
}\left\{  -\cos(z+\pi/4)+\frac{15}{8z}\sin(z+\pi/4)\right\}  ,
\]
the integral to be evaluated is%
\[
\int_{r_{l}}^{r_{u}}\frac{\lambda_{\varepsilon}^{2}(r)}{\,\,\,r^{2}}%
\mathbf{E}_{2}(2\pi k\lambda_{\varepsilon}(r))dr=
\]%
\[
-\frac{1}{\pi k^{1/2}}\int_{0}^{\infty}\frac{\lambda_{\varepsilon}^{3/2}%
(r)}{\,\,\,r^{2}}\left\{  \cos(2\pi k\lambda_{\varepsilon}(r)+\pi
/4)-\frac{15}{16\pi k\lambda_{\varepsilon}(r)}\sin(2\pi k\lambda_{\varepsilon
}(r)+\pi/4)\right\}  dr.
\]
As stated in the SE paper we are not interested in the detailed content in
this quantity, but instead only in the leading oscillatory contributions to
it. \ Evaluation of this integral for large $\lambda_{\varepsilon}$ can be
obtained using the `\textit{stationary phase approximation}.'\cite{STATIONARY}
\ Recalling that $\lambda_{_{\varepsilon}}(r)$ has a maximum at the point
$r_{_{\varepsilon}}$ expansion of that function around $r_{_{\varepsilon}}$
gives
\[
\lambda_{_{\varepsilon}}(r)=\lambda_{_{\varepsilon}}(r_{\varepsilon}%
)-\tfrac{1}{4}\tfrac{\omega_{_{\varepsilon}}^{2}}{\lambda_{_{\varepsilon}%
}(r_{_{\varepsilon}})}(r-r_{_{\varepsilon}})^{2}+\ldots,
\]
where $\omega_{0}^{2}/\lambda_{0}(r_{0})$ is proportional to $Z$ and is large.
\ In the limit as $\varepsilon\rightarrow\varepsilon_{0}$ and within that
approximation the leading oscillatory terms for the average potential
$\overline{V}_{ne,\,\mathit{l}TF}$ is \
\[
-\frac{\overline{V}_{ne\,,\mathit{\,l}TF}}{{\small Z}}\,={\small 2}%
\sqrt{{\small 2}}\frac{\,\lambda_{0}^{2}(r_{0})}{\omega_{0\,}r_{0}^{2}}\left[
\sum_{k=1}^{\infty}\frac{(-1)^{k}}{\left(  \pi k\right)  ^{2}}\cos
({\small 2\pi}k\lambda_{0}{\small (r}_{0}{\small )})-\frac{15}{16\lambda_{0}%
}\sum_{k=1}^{\infty}\frac{(-1)^{k}}{\left(  \pi k\right)  ^{3}}\sin
({\small 2\pi}k\lambda_{0}{\small (r}_{0}{\small )})\right]  .
\]
In the equation appearing above, sums of the kind
\begin{align*}
S_{n}(z)  &  =\sum_{k=1}^{\infty}\frac{(-1)^{k}}{(\pi k)^{2n+1}}\sin(2\pi
k\,z),\\
C_{n}(z)  &  =\sum_{k=1}^{\infty}\frac{(-1)^{k}}{(\pi k)^{2n}}\cos(2\pi k\,z),
\end{align*}
occur. \ These infinite sums can be rewritten \cite{SE2} in closed-form in
terms of the periodic function $\left\langle z\right\rangle $ def\thinspace
ined by
\[
\left\langle z\right\rangle =z-\lfloor z+1/2\rfloor\,,\hspace{0.5in}%
-\frac{1}{2}\leq\left\langle z\right\rangle <\frac{1}{2},
\]
where $\lfloor z\rfloor\,$\ is the floor function. \ The f\thinspace irst few
of these sums are given here as%
\begin{align*}
S_{0}(z)  &  =-\,\left\langle z\right\rangle ,\\
C_{1}(z)  &  =\left\langle z\right\rangle ^{2}-\tfrac{1}{12},\\
S_{1}(z)  &  =\tfrac{2}{3}\left\langle z\right\rangle [\left\langle
z\right\rangle ^{2}-\tfrac{1}{4}],\\
C_{2}(z)  &  =\tfrac{1}{90}-\tfrac{1}{3}[\left\langle z\right\rangle
^{2}-\tfrac{1}{4}]^{2}.
\end{align*}

The leading terms in the potential energy $\overline{V}_{ne,\,\mathit{l}%
TF}/{\small Z}$ written in terms of the closed-form expressions becomes
\begin{equation}
-\frac{\overline{V}_{ne,\,\mathit{l}TF}}{{\small Z}}={\small 2}\sqrt
{{\small 2}}\frac{\,\lambda_{0}^{2}(r_{0})}{\omega_{0\,}r_{0}^{2}}\left[
C_{1}(\lambda_{0}{\small )}-\frac{15}{16\lambda_{0}(r_{0})}S_{1}(\lambda
_{0}{\small )}\right]  , \label{20}%
\end{equation}
with%
\[
\frac{\,\lambda_{0}^{2}(r_{0})}{\omega_{0\,}r_{0}^{2}}=0.68258Z^{\,2/3}.
\]
The $\mathit{l}TF$ contribution is seen to contain terms of orders $Z^{2/3}$
and $Z^{1/3}$.

\subsection{\bigskip The $\overline{V}_{ne,\lambda}$ and $\overline
{V}_{ne,\lambda\,,\nu\,}$ \ Oscillations}

The terms $\overline{V}_{ne,\lambda\,}$ and $\overline{V}_{ne,\lambda,\nu\,}$
are more complex in nature. \ In those cases the integrals $\mathcal{N}%
_{j}\mathcal{(}\lambda,\varepsilon_{\lambda,\nu})$ with $j>0$ are complicated
by the presence of the variable $\nu$ and trigonometric terms thereby
requiring a more elaborate analysis. \ Using the bounds defined by the
appropriate regions of integration for $\lambda,\,\nu$ we have
\begin{subequations}
\begin{align}
\overline{V}_{ne,\,\lambda}(\varepsilon)  &  =8\sum_{j=1}^{\infty}(-1)^{j}%
\int_{0}^{\lambda_{\varepsilon}}\lambda\,\mathcal{N}_{j}\mathcal{(}%
\lambda,\varepsilon_{\lambda,\nu})\,d\lambda\tag{21a}\\
\overline{V}_{ne,\lambda,\nu\,}(\varepsilon)  &  =16\sum_{k=1}^{\infty}%
\sum_{j=1}^{\infty}(-1)^{k+j}\int_{0}^{\lambda_{\varepsilon}}\lambda
\,\cos(2\pi k\lambda)\mathcal{N}_{j}\mathcal{(}\lambda,\varepsilon
_{\lambda,\nu})\,d\lambda\,\,, \tag{21b}%
\end{align}
where the $\lambda,\,$and the $\lambda,\nu$ region has been divided into the
two, subregions def\thinspace ined by $k=0,$ $1\leq j\leq\infty$ and $1\leq
j\leq\infty,$ $\ \ 1\leq k\leq\infty$, respectively.%

\begin{figure}
\begin{center}
\includegraphics[
width=3.5725in
]%
{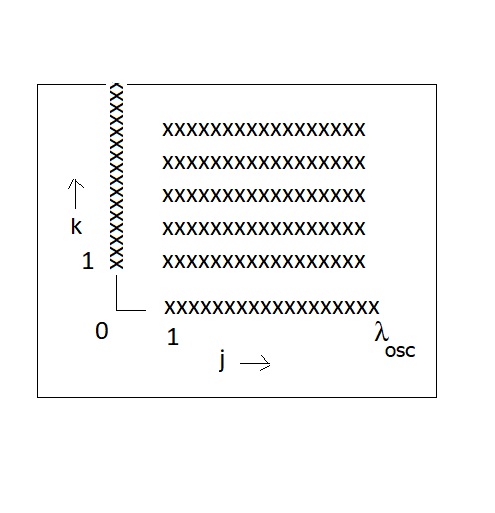}%
\caption{The $k$ vs. $j$ regions of summation}%
\end{center}
\end{figure}

\subsubsection{The Term $\mathcal{N}_{j}\mathcal{(}\lambda,\varepsilon
_{\lambda,\nu})$ For $j\geq1$ The $\lambda$\thinspace Oscillations}

We have seen that $\mathcal{N}_{j}\mathcal{(}\lambda,\varepsilon_{\lambda,\nu
})$ has been partially integrated with respect to $\nu$ i.e.
\end{subequations}
\begin{align*}
\mathcal{N}_{j}\mathcal{(}\lambda,\varepsilon_{\lambda,\nu})  &
=-\frac{Z}{\pi}\int_{r_{l}}^{r_{u}}\frac{dr}{r^{2}}\cos(2\pi j\nu
(\lambda|\varepsilon))I_{0}(\nu(\lambda|\varepsilon),r,\lambda)\\
-  &  2Z\,j\int_{r_{l}}^{r_{u}}\frac{dr}{r^{2}}\int_{0}^{\nu(\lambda
|\varepsilon)}\sin(2\pi j\nu)\,I_{0}(\nu,r,\lambda)\,d\nu{\small ,}%
\end{align*}
where%
\[
I_{0}(\nu,r,\lambda)=\sqrt{2r^{2}\{\varepsilon_{\lambda,\nu}%
-V(r)\}-{\small \lambda}^{2}}.
\]
The quantity $I_{0}(\nu,\lambda,r)$ evaluated at $\nu(\lambda|\varepsilon)$
simplifies and we get
\begin{align*}
I_{0}(\nu(\lambda|\varepsilon),\lambda,r)  &  =\sqrt{2r^{2}\{\varepsilon
_{\lambda,\nu(\lambda|\varepsilon)}-V(r)\}-\lambda^{2}}=\sqrt{\lambda
_{\varepsilon}^{2}(r)-\lambda^{2}},\\
I_{0}(\nu(\lambda|\varepsilon),\lambda,r)  &  =\lambda_{\varepsilon}%
(r)\sin\theta.
\end{align*}
Integration by parts of the integral with respect to $\nu$ gives
\begin{equation}
\mathcal{N}_{j}\mathcal{(}\lambda,\varepsilon_{\lambda,\nu})=-\frac{Z}{\pi
}\{\int_{r_{l}}^{r_{u}}\frac{dr}{r^{2}}\cos(2\pi j\nu(\lambda|\varepsilon
))\,I_{0}(\nu(\lambda|\varepsilon),r,\lambda)\nonumber
\end{equation}%
\[
+2\pi j\int_{r_{l}}^{r_{u}}\frac{dr}{r^{2}}[\sin(2\pi j\nu(\lambda
|\varepsilon)I_{1}(\nu(\lambda|\varepsilon),\lambda,r)-2\pi j\int_{0}%
^{\nu(\lambda|\varepsilon)}\cos(2\pi j\nu(\lambda|\varepsilon))I_{1}%
(\nu(\lambda|\varepsilon),\lambda,r)d\nu]\},
\]
where%
\[
I_{1}(\nu(\lambda|\varepsilon),\lambda,r)=\int_{0}^{\nu(\lambda|\varepsilon
)}I_{0}(\nu(\lambda|\varepsilon),\lambda,r)\,d\nu.
\]
This process can be continued and the integral with respect to $\nu$ in the
equation above can be integrated by parts once more to yield
\begin{align*}
\mathcal{N}_{j}\mathcal{(}\lambda,\varepsilon_{\lambda,\nu})  &
=-\frac{Z}{\pi}\{\int_{r_{l}}^{r_{u}}\frac{dr}{r^{2}}\cos(2\pi j\nu
(\lambda|\varepsilon))I_{0}(\nu(\lambda|\varepsilon),\lambda,r)\\
&  +\left(  2\pi j\right)  \int_{r_{l}}^{r_{u}}\frac{dr}{r^{2}}\sin(2\pi
j\nu(\lambda|\varepsilon))I_{1}(\nu(\lambda|\varepsilon),\lambda,r)\\
&  -(2\pi j)^{2}\int_{r_{l}}^{r_{u}}\frac{dr}{r^{2}}\cos(2\pi j\nu
(\lambda|\varepsilon))I_{2}(\nu(\lambda|\varepsilon),\lambda,r)\\
&  -(2\pi j)^{3}\int_{r_{l}}^{r_{u}}\frac{dr}{r^{2}}\int_{0}^{\nu
(\lambda|\varepsilon)}\sin(2\pi j\nu)I_{2}(\nu,\lambda,r)d\nu\},
\end{align*}
where
\[
I_{2}(\nu(\lambda|\varepsilon),\lambda,r)=\int_{0}^{\nu(\lambda|\varepsilon
)}I_{1}(\nu(\lambda|\varepsilon),\lambda,r)\,d\nu.
\]
The process introduced above can in principle be continued indefinitely
however, it suff\thinspace ices to terminate the expression for $\mathcal{N}%
_{j}\mathcal{(}\lambda,\varepsilon_{\lambda,\nu})$ at order $j^{\,2}$. \ 

The integrals $I_{1}(\nu(\lambda|\varepsilon),\lambda,r)$ and $I_{2}%
(\nu(\lambda|\varepsilon),\lambda,r)$ have been approximately evaluated in
appendix A and are given here by%
\[
\frac{I_{1}(\nu(\lambda|\varepsilon),\lambda,r)}{Zr}=\frac{\lambda
_{\varepsilon}(r)}{Zr}\sqrt{\lambda_{\varepsilon}^{2}-\lambda^{2}}%
-\arctan(\frac{\lambda_{\varepsilon}(r)}{Zr}\sqrt{\lambda_{\varepsilon
}(r)-\lambda^{2}}),
\]
or in terms of the angular variable $\theta$%
\begin{align*}
\frac{I_{1}(\nu(\lambda|\varepsilon),\lambda,r)}{Zr}  &  =\mathcal{K}%
_{\varepsilon}\sin\theta-\arctan(\mathcal{K}_{\varepsilon}\sin\theta),\\
&  =\left[  \mathcal{K}_{\varepsilon}\sin\theta\right]  ^{3}\,\mathcal{I}%
_{1}(\mathcal{K}_{\varepsilon}\sin\theta),
\end{align*}
where $\mathcal{I}_{1}(z)$ is
\begin{align*}
\mathcal{I}_{1}(z)  &  =\sum_{\kappa=0}^{\infty}\frac{(-1)^{\kappa
}\,z^{2\kappa}}{\left(  2\kappa+3\right)  },\hspace{0.25in}\left|  z\right|
\leq1\\
\mathcal{I}_{1}(z)  &  =\frac{1}{3}-\frac{z^{2}}{5}+\frac{z^{4}}{7}+\cdots,
\end{align*}
and $\mathcal{K}_{0}$ is the unit less constant
\[
\mathcal{K}_{0}=\frac{\lambda_{0}^{2}(r_{0})}{Z\,r_{0}}=0.46230.
\]
Then
\[
\frac{I_{1}(\nu(\lambda|\varepsilon),\lambda,r)}{Zr}=\frac{1}{3}\left[
\mathcal{K}_{\varepsilon}\sin\theta\right]  ^{3}-\frac{1}{5}[\mathcal{K}%
_{\varepsilon}\sin\theta]^{5}+\cdots
\]
Similarly we have%
\begin{align*}
\frac{I_{2}(\nu(\lambda|\varepsilon),\lambda,r)}{2Zr\lambda_{\varepsilon}(r)}
&  =\frac{\lambda_{\varepsilon}(r)}{Zr}\sqrt{\lambda_{\varepsilon}%
(r)^{2}-\lambda^{2}}-2\arctan(\frac{\lambda_{\varepsilon}(r)}{Zr}\sqrt
{\lambda_{\varepsilon}(r)^{2}-\lambda^{2}})\\
&  +\frac{\arcsin h(\frac{\lambda_{\varepsilon}(r)}{Zr}\sqrt{\lambda
_{\varepsilon}(r)^{2}-\lambda^{2}})}{\sqrt{1+\left(  \frac{\lambda
_{\varepsilon}(r)}{Zr}\right)  ^{2}(\lambda_{\varepsilon}(r)^{2}-\lambda^{2}%
)}},
\end{align*}
or in terms of $\theta$ we have%
\begin{align*}
\frac{I_{2}(\nu(\lambda|\varepsilon),\lambda,r)}{2rZ\,\lambda_{\varepsilon
}(r)}  &  =\mathcal{K}_{\varepsilon}\sin(\theta)-2\arctan(\mathcal{K}%
_{\varepsilon}\sin\theta)+\frac{\arcsin h(\mathcal{K}_{\varepsilon}\sin
\theta)}{\sqrt{1+\mathcal{K}_{\varepsilon}^{2}\sin^{2}\theta}},\\
& \\
&  =\left[  \mathcal{K}_{\varepsilon}\sin\theta\right]  ^{5}\,\mathcal{I}%
_{2}(\mathcal{K}_{\varepsilon}\sin\theta),
\end{align*}
where $\mathcal{I}_{2}(z)$ is
\begin{align*}
\mathcal{I}_{2}(z)  &  =\sum_{\kappa=0}^{\infty}{\Large [}\Gamma
(\kappa+5/2)-\tfrac{\sqrt{\pi}}{2}\Gamma(\kappa+3){\Large ]}\tfrac
{(-1)^{k}\,z^{2\kappa}}{\Gamma(\kappa+7/2)},\hspace{0.25in}\left|  z\right|
\leq1\\
\,\mathcal{I}_{2}(z)  &  =\tfrac{2}{15}-\tfrac{6}{35}z^{2}+\tfrac{58}%
{315}z^{4}-\tfrac{130}{693}z^{6}+\,\cdots.
\end{align*}
Finally
\[
\frac{I_{2}(\nu(\lambda|\varepsilon),\lambda,r)}{2rZ\lambda_{\varepsilon}%
(r)}=\tfrac{2}{15}\left[  \mathcal{K}_{\varepsilon}(r_{\varepsilon})\sin
\theta\right]  ^{5}-\tfrac{6}{35}\left[  \mathcal{K}_{\varepsilon
}(r_{\varepsilon})\sin\theta\right]  ^{7}+\cdots.
\]
We note that the integral $I_{2}$ is small compared to $I_{1}$ and will be dropped.

The integrals in $\overline{V}_{ne,\,\lambda}(\varepsilon)$ and $\overline
{V}_{ne,\,\lambda,\nu}(\varepsilon)$ can then be written as%

\begin{align}
\overline{V}_{ne\lambda;\lambda,\nu\,}  &  =\lim_{\varepsilon\rightarrow
0}\,\left[  \overline{V}_{ne,\,\lambda}(\varepsilon)+\overline{V}%
_{ne,\,\lambda,\nu}(\varepsilon)\right] \nonumber\\
&  =\left\{  8\sum_{j=1}^{\infty}(-1)^{\,j}\,\mathcal{V}_{0}(\,j)+16\sum
_{k=1}^{\infty}\sum_{j=1}^{\infty}(-1)^{k+j}\mathcal{V}_{1}(\,j,k)\,\right\}
. \label{23}%
\end{align}
The $\mathcal{V}_{0}(\,j)$ integrals contain the $\lambda$ oscillation$\ $%
terms and the $\mathcal{V}_{1}(\,j,k)$ terms contain the mixed $\lambda,\nu$
oscillation terms. \ Now we get the expressions
\begin{align*}
\mathcal{V}_{0}(\,j)  &  =-\frac{Z}{\pi}\{\int_{r_{l}}^{r_{u}}\frac{dr}{r^{2}%
}\int_{0}^{\lambda_{\varepsilon}}\lambda\,I_{0}(\nu(\lambda|\varepsilon
))\cos(2\pi j\,\nu(\lambda|\varepsilon))d\lambda\\
&  +2\pi j\,\int_{r_{l}}^{r_{u}}\frac{dr}{r^{2}}\int_{0}^{\lambda
_{\varepsilon}}\lambda\,I_{1}(\nu(\lambda|\varepsilon))\sin(2\pi
j\,\nu(\lambda|\varepsilon))d\lambda\},\\
& \\
\mathcal{V}_{1}(\,j,k)  &  =-\frac{Z}{\pi}\{\int_{r_{l}}^{r_{u}}%
\frac{dr}{r^{2}}\int_{0}^{\lambda_{\varepsilon}}\lambda\,I_{0}(\nu
(\lambda|\varepsilon))\cos(2\pi k\lambda)\cos(2\pi j\,\nu(\lambda
|\varepsilon))d\lambda\,\\
&  +2\pi j\int_{r_{l}}^{r_{u}}\frac{dr}{r^{2}}\int_{0}^{\lambda_{\varepsilon}%
}\,\lambda I_{1}(\nu(\lambda|\varepsilon))\cos(2\pi k\lambda)\sin(2\pi
j\,\nu(\lambda|\varepsilon))\,d\lambda\}.
\end{align*}
In the equations above we have interchange the order of integration over $r$
and $\lambda$ and used $\lambda=\lambda_{\varepsilon}\cos\theta$ where
$0\leq\theta$ $\leq\pi/2$. to give%
\begin{align}
\mathcal{V}_{0}(\,j)  &  =-\frac{Z}{\pi}\{\int_{r_{l}}^{r_{u}}\frac{\lambda
_{\varepsilon}^{3}(r)}{r^{2}}\int_{0}^{\pi/2}\sin^{2}\theta\cos\theta\cos(2\pi
j\,\nu(\lambda|\varepsilon))d\theta dr\label{24a}\\
&  +\frac{2\pi j}{Z^{2}}\,\int_{r_{l}}^{r_{u}}\frac{\lambda_{\varepsilon}%
^{8}(r)}{r^{4}}\int_{0}^{\pi/2}\,\sin^{4}\theta\cos\theta\sin(2\pi
j\,\nu(\lambda|\varepsilon))\,\mathcal{I}_{1}(\mathcal{K}_{\varepsilon}%
\sin\theta)d\theta dr\},\nonumber\\
& \nonumber\\
\mathcal{V}_{1}(\,j,k)  &  =-\frac{Z}{\pi}\{\int_{r_{l}}^{r_{u}}%
\frac{\lambda_{\varepsilon}^{3}(r)}{r^{2}}\int_{0}^{\pi/2}\sin^{2}\theta
\cos\theta\cos(2\pi k\lambda_{\varepsilon}\cos\theta)\cos(2\pi j\,\nu
(\lambda|\varepsilon))d\theta dr\,\tag{24b}\\
&  +\frac{2\pi j}{Z^{2}}\int_{r_{l}}^{r_{u}}\frac{\lambda_{\varepsilon}%
^{8}(r)}{r^{4}}\int_{0}^{\pi/2}\sin^{4}\theta\cos\theta\,\cos(2\pi
k\lambda_{\varepsilon}\cos\theta)\sin(2\pi j\,\nu(\lambda|\varepsilon
))\mathcal{I}_{1}(\mathcal{K}_{\varepsilon}\sin\theta)d\theta dr\,\}.\nonumber
\end{align}
Written in more compact form the Eqs. (24) became
\begin{align*}
\mathcal{V}_{0}(\,j)  &  =-\frac{Z}{\pi}\left\{  \int_{r_{l}}^{r_{u}%
}\frac{\lambda_{\varepsilon}^{3}(r)\,}{r^{2}}\mathcal{C}_{0}^{\,\left(
1\right)  }(b,z,\varrho)\,dr+\frac{2\pi j}{Z^{2}}\int_{r_{l}}^{r_{u}%
}\frac{\lambda_{\varepsilon}^{8}(r)}{r^{4}}\mathcal{C}_{0}^{\,\left(
2\right)  }(\mathcal{K}_{\varepsilon},b,z,\varrho)\,dr\right\}  ,\\
& \\
\mathcal{V}_{1}(\,j,k)  &  =-\frac{Z}{\pi}\left\{  \int_{r_{l}}^{r_{u}%
}\frac{\lambda_{\varepsilon}^{3}(r)}{r^{2}}\mathcal{C}_{1}^{\,\left(
1\right)  }(b,z,z^{\prime},\varrho)\,dr+\frac{2\pi j}{Z^{2}}\int_{r_{l}%
}^{r_{u}}\frac{\lambda_{\varepsilon}^{8}(r)}{r^{4}}\mathcal{C}_{1}^{\,\left(
\,2\right)  }(\mathcal{K}_{\varepsilon},b,z,z^{\prime},\varrho))\,dr\right\}
,
\end{align*}
where the angular integrals are defined by%
\begin{align*}
C_{0}^{\,\,\left(  1\right)  }(b,z,\varrho)  &  =\int_{0}^{\pi/2}\sin
^{2}\theta\cos\theta\cos(2\pi j\,\nu(\lambda|\varepsilon))\,d\theta,\\
C_{1}^{\,\,\left(  1\right)  }(b,z,z^{\prime},\varrho)  &  =\int_{0}^{\pi
/2}\sin^{2}\theta\cos\theta\cos(2\pi k\lambda_{\varepsilon}\cos\theta
)\cos(2\pi j\,\nu(\lambda|\varepsilon))\,d\theta,\\
C_{0}^{\,\,\left(  2\right)  }(\mathcal{K}_{\varepsilon},b,z,\varrho)  &
=\int_{0}^{\pi/2}\sin^{4}\theta\cos\theta\,\sin(2\pi j\,\nu(\lambda
|\varepsilon))\,\mathcal{I}_{1}(\mathcal{K}_{\varepsilon}\sin\theta
)\,d\theta,\\
C_{1}^{\,\left(  \,2\right)  }(\mathcal{K}_{\varepsilon},b,z,z^{\prime
},\varrho)  &  =\int_{0}^{\pi/2}\sin^{4}\theta\cos\theta\,\cos(2\pi
k\lambda_{\varepsilon}\cos\theta)\sin(2\pi j\,\nu(\lambda|\varepsilon
))\,\mathcal{I}_{1}(\mathcal{K}_{\varepsilon}\sin\theta)\,d\theta,
\end{align*}

\bigskip

As will be seen below these angular integrals are complicated in that they
contain the functions $\cos(2\pi j\,\nu(\lambda|\varepsilon))$, and $\cos(2\pi
k\lambda_{\varepsilon}\cos\theta)\cos(2\pi j\,\nu(\lambda|\varepsilon))$ as
well as their $\sin$ counterparts, quantities which are functions\ of $\theta$
as well as $\lambda_{\varepsilon}(r)$. \ In order to make progress in
evaluating these integrals which contain the term $\nu(\lambda|\varepsilon)$,
we will expand that quantity as follows. Taking into account the fact that the
$\nu(\lambda|\varepsilon)$ versus $\lambda$ curves show curvature, the Thomas
Fermi lines of nonlinear degeneracy will be replaced by a \textit{quadratic}
polynomial with the choice of parameters used by SE (in SE's notation
$\nu_{\varepsilon}^{\,(1)}=\nu_{\varepsilon}^{\prime\,\text{\ }},$ and
$\nu_{\varepsilon}^{\,(2)}=\nu_{\varepsilon}^{\,^{\prime\prime}})$ that is we
write%
\[
\nu(\lambda|\varepsilon)=\nu_{\varepsilon}^{\,(1)}\,[\lambda_{\varepsilon
}(r)-\lambda]-\tfrac{1}{2}\nu_{\varepsilon}^{\,(2)}\,[\lambda_{\varepsilon
}(r)-\lambda]^{2},
\]
with (we use SE's values for $\nu_{\varepsilon}^{\,(1)},\nu_{\varepsilon
}^{\,(2)}$ and assume that they are constants independent of $r$)
\begin{align*}
\nu_{\varepsilon}^{(1)}  &  =\sqrt{2}\lambda_{\varepsilon}/\omega
_{\varepsilon}r_{\varepsilon},\\
\nu_{\varepsilon}^{(2)}  &  =(\nu_{\varepsilon}^{(1)}-1)/\lambda_{\varepsilon
}(r_{\varepsilon}).
\end{align*}
Expressing $\lambda$ in terms of the angle $\theta$\ we write%
\begin{equation}
2\pi j\nu(\lambda|\varepsilon)=b+z\cos\theta+\varrho\cos^{2}\theta, \label{25}%
\end{equation}
where using the values given above we have%

\begin{align*}
b  &  =2\pi j\lambda_{\varepsilon}(r)\,[\nu_{\varepsilon}^{(1)}-\tfrac{1}%
{2}\lambda_{\varepsilon}(r)\,\nu_{\varepsilon}^{(2)}]\\
b  &  =\pi j\,\lambda_{\varepsilon}(r)\,[\nu_{\varepsilon}^{\left(  1\right)
}+1],\\
& \\
\varrho &  =-\pi j\lambda_{\varepsilon}^{2}(r)\,\nu_{\varepsilon}^{(2)}\\
\varrho &  =-\pi j\lambda_{\varepsilon}(r)\,[\nu_{\varepsilon}^{\left(
1\right)  }-1],\\
& \\
z  &  =-2\pi j\lambda_{\varepsilon}(r)\,[\nu_{\varepsilon}^{(1)}%
-\lambda_{\varepsilon}(r)\,\nu_{\varepsilon}^{(2)}]\\
z  &  =-2\pi j\lambda_{\varepsilon}(r),\\
& \\
z^{\prime}  &  =2\pi k\lambda_{\varepsilon}(r).
\end{align*}
Recalling that $\,\lambda_{\varepsilon}(r)=\lambda_{_{\varepsilon}%
}(r_{\varepsilon})-\tfrac{1}{4}\tfrac{\omega_{_{\varepsilon}}^{2}}%
{\lambda_{_{\varepsilon}}(r_{_{\varepsilon}})}(r-r_{_{\varepsilon}}%
)^{2}+\ldots,$ and within the limit as $\varepsilon$ approaches zero%
\begin{align*}
b_{0}  &  =\lim_{\varepsilon\rightarrow0}\,\pi j\,\lambda_{\varepsilon
}(r_{\varepsilon})\,[\nu_{\varepsilon}^{\left(  1\right)  }+1]=2.72614\,\pi
\,j\,Z^{1/3},\\
\left|  \,z\,\right|  _{0}  &  =\lim_{\varepsilon\rightarrow0}\,2\pi
j\,\lambda_{\varepsilon}(r_{\varepsilon})=1.85598\pi\,j\,Z^{1/3},\\
z_{0}^{\prime}  &  =\lim_{\varepsilon\rightarrow0}\,2\pi k\,\lambda
_{\varepsilon}(r_{\varepsilon})=1.85598\pi\,k\,Z^{1/3},\\
\left|  \,\varrho\,\right|  _{0}  &  =\lim_{\varepsilon\rightarrow0}\,\pi
j\,\lambda_{\varepsilon}(r_{\varepsilon})\,[\nu_{\varepsilon}^{\left(
1\right)  }-1]=0.87358\,\pi\,j\,Z^{1/3}.
\end{align*}
With $z=-2\pi j\lambda_{\varepsilon}(r_{\varepsilon}),$ $z^{\prime}=2\pi
k\lambda_{\varepsilon}(r_{\varepsilon})$ and $\varrho=-\pi j\lambda
_{\varepsilon}(r_{\varepsilon})[v_{0}^{(1)}-1]$ we have upon expanding the
$\cos(2\pi j\,\nu(\lambda|\varepsilon))$ and sin$(2\pi j\,\nu(\lambda
|\varepsilon))$ terms and obtained the trigonometric expressions%

\begin{align}
\cos(2\pi j\,\nu(\lambda|\varepsilon))  &  =\;\;\cos(b)\left\{  \cos
(z\cos\theta)\cos(\varrho\cos^{2}\theta)-\sin(z\cos\theta)\sin(\varrho\cos
^{2}\theta)\right\} \tag{T1}\\
&  -\sin(b)\left\{  \sin(z\cos\theta)\cos(\varrho\cos^{2}\theta)+\cos
(z\cos\theta)\sin(\varrho\cos^{2}\theta)\right\}  ,\nonumber
\end{align}
and%
\begin{align}
\sin(2\pi j\,\nu(\lambda|\varepsilon))  &  =\;\;\sin(b)\left\{  \cos
(z\cos\theta)\cos(\varrho\cos^{2}\theta)-\sin(z\cos\theta)\sin(\varrho\cos
^{2}\theta)\right\} \tag{T2}\\
&  +\cos(b)\left\{  \sin(z\cos\theta)\cos(\varrho\cos^{2}\theta)+\cos
(z\cos\theta)\sin(\varrho\cos^{2}\theta)\right\}  .\nonumber
\end{align}
Similarly we have for the trigonometric expressions which contain the terms
$\lambda_{\varepsilon}(r_{\varepsilon})$, and $\nu(\lambda|\varepsilon)$
\begin{align}
\cos(2\pi k\lambda_{\varepsilon}(r_{\varepsilon}))\cos(2\pi j\,\nu
(\lambda|\varepsilon))  &  =\tfrac{1}{2}\cos(b)[\cos([z-z^{\prime}]\cos
\theta)\cos(\varrho\cos^{2}\theta)\tag{T3}\\
&  +\cos([z+z^{\prime}]\cos\theta)\cos(\varrho\cos^{2}\theta)\nonumber\\
&  -\sin([z-z^{\prime}]\cos\theta)\sin(\varrho\cos^{2}\theta)\nonumber\\
&  -\sin([z+z^{\prime}]\cos\theta)\sin(\varrho\cos^{2}\theta)]\nonumber\\
& \nonumber\\
&  -\tfrac{1}{2}\sin(b)\{\sin([z-z^{\prime}]\cos\theta)\cos(\varrho\cos
^{2}\theta)\nonumber\\
&  +\sin([z+z^{\prime}]\cos\theta)\cos(\varrho\cos^{2}\theta)\nonumber\\
&  +\cos([z-z^{\prime}]\cos\theta)\sin(\varrho\cos^{2}\theta)\nonumber\\
&  +\cos([z+z^{\prime}]\cos\theta)\sin(\varrho\cos^{2}\theta)\},\nonumber
\end{align}
and%
\begin{align}
\cos(2\pi k\lambda_{\varepsilon}(r_{\varepsilon}))\sin(2\pi j\,\nu
(\lambda|\varepsilon))  &  =\tfrac{1}{2}\cos(b)[\sin([z-z^{\prime}]\cos
\theta)\cos(\varrho\cos^{2}\theta)\tag{T4}\\
&  +\sin([z+z^{\prime}]\cos\theta)\cos(\varrho\cos^{2}\theta)\nonumber\\
&  +\cos([z-z^{\prime}]\cos\theta)\sin(\varrho\cos^{2}\theta)\nonumber\\
&  +\cos([z+z^{\prime}]\cos\theta)\sin(\varrho\cos^{2}\theta)\nonumber\\
&  \tfrac{1}{2}\sin(b)[\cos([z-z^{\prime}]\cos\theta)\cos(\varrho\cos
^{2}\theta)\nonumber\\
&  +\cos([z+z^{\prime}]\cos\theta)\cos(\varrho\cos^{2}\theta)\nonumber\\
&  -\sin([z-z^{\prime}]\cos\theta)\sin(\varrho\cos^{2}\theta)\nonumber\\
&  -\sin([z+z^{\prime}]\cos\theta)\sin(\varrho\cos^{2}\theta)],\nonumber
\end{align}
Using the $\nu(\lambda|\varepsilon)$ expressions and (T1) the integrals
$C_{0}^{\,\,\left(  1\right)  }$ becomes
\[
C_{0}^{\,\,\left(  1\right)  }(b,\,z,\varrho)=\cos(b)\{\mathcal{C}%
_{cc}\mathcal{(}0,z,\varrho)-\mathcal{S}_{ss}\mathcal{(}0,z,\varrho
)\}-\sin(b)\{\mathcal{C}_{cs}(0,z,\varrho)+\mathcal{S}_{sc}(0,z,\varrho)\}.
\]
then using (T3) $C_{1}^{\,(1)}$ can be written as
\begin{align*}
C_{1}^{\,(1)}(b,\,z,z^{\prime},\varrho)  &  =\tfrac{1}{2}\cos(b)[C_{cc}%
(0,z-z^{\prime},\varrho)-S_{ss}(0,z-z^{\prime},\varrho)+C_{cc}(0,z+z^{\prime
},\varrho)-S_{ss}(0,z+z^{\prime},\varrho)]\\
&  -\tfrac{1}{2}\sin(b)[C_{cs}(0,z-z^{\prime},\varrho)+S_{sc}(0,z-z^{\prime
},\varrho)+C_{cs}(0,z+z^{\prime},\varrho)+S_{sc}(0,z+z^{\prime},\varrho)],
\end{align*}
where the \textit{primitive} angular integrals $\mathcal{C}_{cc}%
\mathcal{(}\kappa,z,\zeta),$ $\mathcal{S}_{ss}\mathcal{(}\kappa,z,\zeta),$
$\mathcal{C}_{cs}\mathcal{(}\kappa,z,\zeta),$ $\mathcal{S}_{sc}\mathcal{(}%
\kappa,z,\zeta)$ are defined as
\begin{subequations}
\begin{align}
\mathcal{C}_{cc}\mathcal{(}\kappa,z,\varrho)  &  =\int_{0}^{\pi/2}%
\sin^{2\kappa+2\,}\theta\cos\theta\cos(z\cos\theta)\cos(\varrho\cos^{2}%
\theta)d\theta,\tag{26a}\\
\mathcal{S}_{ss}\mathcal{(}\kappa,z,\varrho)  &  =\int_{0}^{\pi/2}%
\sin^{2\kappa+2\,}\theta\cos\theta\sin(z\cos\theta)\sin(\varrho\cos^{2}%
\theta)d\theta,\tag{26b}\\
\mathcal{C}_{cs}\mathcal{(}\kappa,z,\varrho)  &  =\int_{0}^{\pi/2}%
\sin^{2\kappa+2\,}\theta\cos\theta\cos(z\cos\theta)\sin(\varrho\cos^{2}%
\theta)d\theta,\tag{26c}\\
\mathcal{S}_{sc}\mathcal{(}\kappa,z,\varrho)  &  =\int_{0}^{\pi/2}%
\sin^{2\kappa+2\,}\theta\cos\theta\sin(z\cos\theta)\cos(\varrho\cos^{2}%
\theta)d\theta. \tag{26d}%
\end{align}
Collecting terms in $C_{1}^{\,(1)}(b,\,z,z^{\prime},\varrho)$ with argument
$z+z^{\prime}$ and $z-z^{\prime}$ we get%

\end{subequations}
\[
C_{1}^{\,\,\left(  1\right)  }(\,b,z,z^{\prime},\varrho)=\tfrac{1}{2}%
C_{0}^{\,\,\left(  1\right)  }(b,\,z+z^{\prime},\varrho)+\tfrac{1}{2}%
C_{0}^{\,\,\left(  1\right)  }(b,\,z-z^{\prime},\varrho),
\]
The integrals $\mathcal{C}_{0}^{\,\,\left(  2\right)  }$ becomes with (T2)
\[
\mathcal{C}_{0}^{\,\,\left(  2\right)  }(\mathcal{K}_{\varepsilon}%
,b,z,\varrho)=\cos(b)\{T_{cs}(\mathcal{K}_{\varepsilon},z,\varrho
)+T_{sc}(\mathcal{K}_{\varepsilon},z,\varrho)\}+\sin(b)\{T_{cc}(\mathcal{K}%
_{\varepsilon},z,\varrho)-T_{ss}(\mathcal{K}_{\varepsilon},z,\varrho)\},
\]
then we can also write for $\mathcal{C}_{1}^{\,\,\left(  2\right)
}(\mathcal{K}_{\varepsilon},b,z,z^{\prime},\varrho)$ using (T4)%
\begin{align*}
\mathcal{C}_{1}^{\,\,\left(  2\right)  }(\mathcal{K}_{\varepsilon
},b,z,z^{\prime},\varrho)  &  =\tfrac{1}{2}\cos(b)[T_{cs}(\mathcal{K}%
_{\varepsilon},z-z^{\prime},\varrho)+T_{sc}(\mathcal{K}_{\varepsilon
},z-z^{\prime},\varrho)]\\
&  +\tfrac{1}{2}\sin(b)[T_{cc}(\mathcal{K}_{\varepsilon},z-z^{\prime}%
,\varrho)-T_{cs}(\mathcal{K}_{\varepsilon},z-z^{\prime},\varrho)]\\
&  +\tfrac{1}{2}\cos(b)[T_{sc}(\mathcal{K}_{\varepsilon},z+z^{\prime}%
,\varrho)+T_{cs}(\mathcal{K}_{\varepsilon},z+z^{\prime},\varrho)]\\
&  +\tfrac{1}{2}\sin(b)[T_{cc}(\mathcal{K}_{\varepsilon},z+z^{\prime}%
,\varrho)-T_{cs}(\mathcal{K}_{\varepsilon},z+z^{\prime},\varrho)],
\end{align*}
where the angular integrals $T_{cc}(\mathcal{K}_{\varepsilon},z,\zeta
),T_{ss}(\mathcal{K}_{\varepsilon},z,\zeta),T_{cs}(\mathcal{K}_{\varepsilon
},z,\zeta),T_{sc}(\mathcal{K}_{\varepsilon},z,\zeta)$ have been defined as
\begin{subequations}
\begin{align}
T_{cc}(\mathcal{K}_{\varepsilon},z,\varrho)  &  =\int_{0}^{\pi/2}%
\mathcal{I}_{1}(\mathcal{K}_{\varepsilon}\sin\theta)\sin^{4}\theta\cos
\theta\cos(z\cos\theta)\cos(\varrho\cos^{2}\theta)d\theta,\tag{27a}\\
T_{ss}(\mathcal{K}_{\varepsilon},z,\varrho)  &  =\int_{0}^{\pi/2}%
\mathcal{I}_{1}(\mathcal{K}_{\varepsilon}\sin\theta)\sin^{4}\theta\cos
\theta\sin(z\cos\theta)\sin(\varrho\cos^{2}\theta)d\theta,\tag{27b}\\
T_{cs}(\mathcal{K}_{\varepsilon},z,\varrho)  &  =\int_{0}^{\pi/2}%
\mathcal{I}_{1}(\mathcal{K}_{\varepsilon}\sin\theta)\sin^{4}\theta\cos
\theta\cos(z\cos\theta)\sin(\varrho\cos^{2}\theta)d\theta,\tag{27c}\\
T_{sc}(\mathcal{K}_{\varepsilon},z,\varrho)  &  =\int_{0}^{\pi/2}%
\mathcal{I}_{1}(\mathcal{K}_{\varepsilon}\sin\theta)\sin^{4}\theta\cos
\theta\sin(z\cos\theta)\cos(\varrho\cos^{2}\theta)d\theta. \tag{27d}%
\end{align}
Using the expressions above we get
\end{subequations}
\[
\mathcal{C}_{1}^{\,\,\left(  2\right)  }(\mathcal{K}_{\varepsilon
},b,z,z^{\prime},\varrho)=\tfrac{1}{2}C_{0}^{\,\,\left(  2\right)
}(\mathcal{K}_{\varepsilon},b,z+z^{\prime},\varrho)+\tfrac{1}{2}%
C_{0}^{\,\,\left(  2\right)  }(\mathcal{K}_{\varepsilon},b,z-z^{\prime
},\varrho).
\]
\ \ 

\bigskip

We note that all of the $C_{i}^{(k)}$ integrals appearing above can be written
in terms of the single quantity $\mathcal{F}(b,\kappa,z,\varrho)$ defined by
\[
\mathcal{F}(b,\kappa,z,\varrho)=\cos(b)\{C_{cc}(\kappa,z,\varrho
)-S_{ss}(\kappa,z,\varrho)\}-\sin(b)\{C_{cs}(\kappa,z,\varrho)+S_{sc}%
(\kappa,z,\varrho)\}.
\]
Observing that
\begin{equation}
\mathcal{F}(b+\pi/2,\kappa,z,\varrho)=-\sin(b)\{C_{cc}(\kappa,z,\varrho
)-S_{ss}(\kappa,z,\varrho)\}-\cos(b)\{C_{cs}(\kappa,z,\varrho)+S_{sc}%
(\kappa,z,\varrho)\}, \label{28}%
\end{equation}
the $C_{i}^{(k)}$ integrals can be rewritten in the compact forms \
\begin{subequations}
\begin{align}
C_{0}^{\,\,\left(  1\right)  }(b,z,\varrho)  &  =\mathcal{F}(b,0,z,\varrho
),\tag{29a}\\
C_{1}^{\,\,\left(  1\right)  }(\,b,z,z^{\prime},\varrho)  &  =\tfrac{1}%
{2}\{\mathcal{F}(b,0,z+z^{\prime},\varrho)+\mathcal{F}(b,0,z-z^{\prime
},\varrho)\},\tag{29b}\\
C_{0}^{\,\,\left(  2\right)  }(\mathcal{K}_{\varepsilon},b,z,\varrho)  &
=\sum_{\kappa=1}^{\infty}\frac{(-1)^{\kappa}\,\mathcal{K}_{\varepsilon
}^{\,2(\kappa-1)}}{(2\kappa+1)}\mathcal{F}(b+\pi/2,\kappa,z,\varrho
),\tag{29c}\\
C_{1}^{\,\,\left(  2\right)  }(\mathcal{K}_{\varepsilon},b,z,z^{\prime
},\varrho)  &  =\tfrac{1}{2}\sum_{\kappa=1}^{\infty}\frac{(-1)^{\kappa
}\,\mathcal{K}_{\varepsilon}^{\,2(\kappa-1)}}{(2\kappa+1)}\{\mathcal{F}%
(b+\pi/2,\kappa,z+z^{\prime},\varrho)+\mathcal{F}(b+\pi/2,\kappa,z-z^{\prime
},\varrho)\}. \tag{29d}%
\end{align}
Recalling that $z<0$ the quantities $z\pm z^{\prime}$ must also be less than
zero in (29b) and (29d). \ The latter restrictions imply that $j>k$ when sums
containing $z\pm z^{\prime}$ are evaluated.

For the radial integrals in (24a) and (24b) i.e. those performed over the
region $0\leq$ $r\leq\infty$ we define the integrals $\mathcal{V}%
_{i}^{\,\,\left(  k\right)  }$
\end{subequations}
\begin{subequations}
\begin{align}
\mathcal{V}_{0}^{\,\,\left(  1\right)  }(j)  &  =\int_{0}^{\infty
}\frac{\lambda_{0}^{3}(r)\,}{r^{2}}C_{0}^{\,\,\left(  1\right)  }%
(b,\,z,\varrho)\,dr,\tag{30a}\\
\mathcal{V}_{1}^{\,\,\left(  1\right)  }(j,k)  &  =\int_{0}^{\infty
}\frac{\lambda_{0}^{3}(r)\,}{r^{2}}C_{1}^{\,\,\left(  1\right)  }%
(\,b,z,z^{\prime},\varrho)\,dr,\tag{30b}\\
\mathcal{V}_{0}^{\,\left(  2\right)  }(j)  &  =\int_{0}^{\infty}%
\frac{\lambda_{0}^{8}(r)\,}{r^{4}}C_{0}^{\,\,\left(  2\right)  }%
(\,\mathcal{K}_{\varepsilon},b,z,\varrho)\,dr,,\tag{30c}\\
\mathcal{V}_{1}^{\,\,\left(  2\right)  }(j,k)  &  =\int_{0}^{\infty
}\frac{\lambda_{0}^{8}(r)\,}{r^{4}}C_{1}^{\,\,\left(  \,2\right)
}(\,\mathcal{K}_{\varepsilon},b,z,z^{\prime},\varrho)\,dr, \tag{30d}%
\end{align}
or written in terms of the $\mathcal{F}$ functions
\end{subequations}
\begin{subequations}
\begin{align}
\mathcal{V}_{0}^{\,\,\left(  1\right)  }(j)  &  =\int_{0}^{\infty
}\frac{\lambda_{0}^{3}(r)\,\mathcal{F}(b,0,z,\varrho)}{r^{2}}dr,\tag{31a}\\
\mathcal{V}_{1}^{\,\,\left(  1\right)  }(j,k)  &  =\tfrac{1}{2}\int
_{0}^{\infty}\frac{\lambda_{0}^{3}(r)\,}{r^{2}}\{\mathcal{F}(b,0,z+z^{\prime
},\varrho)+\mathcal{F}(b,0,z-z^{\prime},\varrho)\}dr,\nonumber\\
& \nonumber\\
\mathcal{V}_{0}^{\,\left(  2\right)  }(j)  &  =\sum_{\kappa=1}^{\infty
}\frac{(-1)^{\kappa}\,\mathcal{K}_{\varepsilon}^{\,2(\kappa-1)}}{(2\kappa
+1)}\int_{0}^{\infty}\frac{\lambda_{0}^{8}(r)\,}{r^{4}}\mathcal{F}%
(b+\pi/2,\kappa,z,\varrho)dr,\tag{31b}\\
\mathcal{V}_{1}^{\,\,\left(  2\right)  }(j,k)  &  =\tfrac{1}{2}\sum_{\kappa
=1}^{\infty}\frac{(-1)^{\kappa}\,\mathcal{K}_{\varepsilon}^{\,2(\kappa-1)}%
}{(2\kappa+1)}\cdot\nonumber\\
&  \cdot\int_{0}^{\infty}\frac{\lambda_{0}^{8}(r)\,}{r^{4}}\{\mathcal{F}%
(b+\pi/2,\kappa,z+z^{\prime},\varrho)+\mathcal{F}(b+\pi/2,\kappa,z-z^{\prime
},\varrho)\}.\nonumber
\end{align}
Then the terms in Eqs. (24a) and (24b) become
\end{subequations}
\begin{subequations}
\begin{align}
\mathcal{V}_{0}^{\,}(\,j)  &  =-\frac{Z}{\pi}[\mathcal{V}_{0}^{\,\,\left(
1\right)  }(j)+\left(  \frac{2\pi j}{Z^{2}}\right)  \,\mathcal{V}%
_{0}^{\,\left(  2\right)  }(j)],\tag{32a}\\
\mathcal{V}_{1}^{\,}(\,j,k)  &  =-\frac{Z}{\pi}[\mathcal{V}_{1}^{\,\,\left(
1\right)  }(j,k)+\left(  \frac{2\pi j}{Z^{2}}\right)  \,\mathcal{V}%
_{1}^{\,\,\left(  2\right)  }(j,k)]. \tag{32b}%
\end{align}

Because of the rapidly oscillating terms contained in the $\cos(b)$ and
$\sin(b)$ factors in the $\mathcal{V}_{0}$ and $\mathcal{V}_{1}$ integrals,
the stationary phase approximation \cite{STATIONARY}\ will be used to evaluate
these integrals. \ Since $z$ and $\varrho$ are negative quantities and are
arguments of the Bessel and Struve functions which are involved in the
calculations indicated above, the required asymptotic expansions for those
functions are ($z<0$) taken to be
\end{subequations}
\begin{align}
J_{0}(z)  &  \thicksim\sqrt{\frac{2}{\pi\left|  z\right|  }}\cos(\left|
z\right|  -\pi/4)+\cdots,\label{33}\\
J_{1}(z)  &  \thicksim-\sqrt{\frac{2}{\pi\left|  z\right|  }}\sin(\left|
z\right|  -\pi/4)+\cdots,\nonumber\\
H_{0}(z)  &  \thicksim-\sqrt{\frac{2}{\pi\left|  z\right|  }}\sin(\left|
z\right|  -\pi/4)-\frac{2}{\pi\left|  z\right|  }\cdots,\nonumber\\
H_{1}(z)  &  \thicksim-\sqrt{\frac{2}{\pi\left|  z\right|  }}\cos(\left|
z\right|  -\pi/4)+\frac{2}{\pi}+\cdots,\nonumber
\end{align}
as a result we have%
\begin{align}
\mathcal{F(}b,\kappa,z,\varrho)  &  =\frac{1}{\sqrt{\left|  z\,\varrho
\,\right|  }}\left\{
\begin{array}
[c]{c}%
\sin(b+\left|  \varrho\right|  -\left|  z\right|  )[\mathcal{F}_{1}%
(\kappa,z,\varrho)-\mathcal{F}_{2}(\kappa,z,\varrho)]\\
+\cos(b+\left|  \varrho\right|  -\left|  z\right|  )[\mathcal{F}_{7}%
^{(+)}(\kappa,z,\varrho)+\mathcal{F}_{8}^{(+)}(\kappa,z,\varrho)]\\
-\sin(b-\left|  \varrho\right|  -\left|  z\right|  )[\mathcal{F}_{7}%
^{(-)}(\kappa,z,\varrho)-\mathcal{F}_{8}^{(-)}(\kappa,z,\varrho)]\\
-\cos(b-\left|  \varrho\right|  -\left|  z\right|  )[\mathcal{F}_{5}%
^{(+)}(\kappa,z,\varrho)+\mathcal{F}_{9}(\kappa,z,\varrho)]
\end{array}
\right\} \label{34}\\
&  +\sqrt{\frac{2}{\pi\left|  \,\varrho\,\right|  }}\left\{
\begin{array}
[c]{c}%
\cos(b+\left|  \varrho\right|  -\pi/4)\left[  \mathcal{F}_{4}^{\,(+)}%
(\kappa,z,\varrho)+\dfrac{\mathcal{F}_{5}^{\,(-)}(\kappa,z,\varrho)}%
{2\pi\,j\,\lambda_{\varepsilon}(r)}\right] \\
\\
+\sin(b+\left|  \varrho\right|  -\pi/4)\left[  \mathcal{F}_{6}^{\,(-)}%
(\kappa,z,\varrho)+\dfrac{\mathcal{F}_{7}^{\,(+)}(\kappa,z,\varrho)}%
{2\pi\,j\,\lambda_{\varepsilon}(r)}\right] \\
\\
-\sin(b-\left|  \varrho\right|  -\pi/4)\left[  \mathcal{F}_{4}^{\,(-)}%
(\kappa,z,\varrho)+\dfrac{\mathcal{F}_{5}^{\,(+)}(\kappa,z,\varrho)}%
{2\pi\,j\,\lambda_{\varepsilon}(r)}\right] \\
\\
+\cos(b-\left|  \varrho\right|  -\pi/4)\left[  \mathcal{F}_{6}^{\,(+)}%
(\kappa,z,\varrho)+\dfrac{\mathcal{F}_{7}^{\,(-)}(\kappa,z,\varrho)}%
{2\pi\,j\,\lambda_{\varepsilon}(r)}\right]
\end{array}
\right\}  ,\nonumber
\end{align}
where the $\mathcal{F}_{i}(\kappa,z,\varrho)$ quantities appearing in (34) are
defined as
\begin{align}
\mathcal{F}_{1}(\kappa,z,\varrho)  &  =-\,\widehat{\mathbf{P}}_{c}%
^{(2)}(\kappa,0,z,\varrho)+\mathbf{Q}_{s}^{(2)}(\kappa,1,z,\varrho
),\label{35}\\
\mathcal{F}_{2}(\kappa,z,\varrho)  &  =-\,\mathbf{P}_{s}^{(1)}(\kappa
,1,z,\varrho)-\widehat{\mathbf{Q}}_{c}^{(1)}(\kappa,0,z,\varrho),\nonumber\\
\mathcal{F}_{4}^{(\pm)}(\kappa,z,\varrho)  &  =-\,[\widehat{\mathbf{Q}}%
_{c}^{(2)}(\kappa,0,z,\varrho)+\widehat{\mathbf{R}}_{c}^{(2)}(\kappa
,0,z,\varrho)]\pm\lbrack\mathbf{Q}_{s}^{(1)}(\kappa,1,z,\varrho)+\mathbf{R}%
_{s}^{(1)}(\kappa,1,z,\varrho)],\nonumber\\
\mathcal{F}_{5}^{(\pm)}(\kappa,z,\varrho)  &  =\widehat{\mathbf{P}}_{c}%
^{(2)}(\kappa,0,z,\varrho)\pm\mathbf{P}_{s}^{(1)}(\kappa,1,z,\varrho
),\nonumber\\
\mathcal{F}_{6}^{(\pm)}(\kappa,z,\varrho)  &  =\pm\,[\widehat{\,\mathbf{Q}%
}_{c}^{(1)}(\kappa,0,z,\varrho)+\widehat{\mathbf{R}}_{c}^{(1)}(\kappa
,0,z,\varrho)]-[\mathbf{Q}_{s}^{(2)}(\kappa,1,z,\varrho)+\mathbf{R}_{s}%
^{(2)}(\kappa,1,z,\varrho)],\nonumber\\
\mathcal{F}_{7}^{(\pm)}(\kappa,z,\varrho)  &  =\pm\,\widehat{\mathbf{P}}%
_{c}^{(1)}(\kappa,0,z,\varrho)+\mathbf{P}_{s}^{(2)}(\kappa,1,z,\varrho
),\nonumber\\
\mathcal{F}_{8}^{(\pm)}(\kappa,z,\varrho)  &  =\pm\widehat{\,\mathbf{Q}}%
_{c}^{(2)}(\kappa,0,z,\varrho)-\mathbf{Q}_{s}^{(1)}(\kappa,1,z,\varrho
),\nonumber\\
\mathcal{F}_{9}(\kappa,z,\varrho)  &  =\widehat{\,\mathbf{Q}}_{c}^{(1)}%
(\kappa,0,z,\varrho)-\mathbf{Q}_{s}^{(2)}(\kappa,1,z,\varrho),\nonumber
\end{align}
and where the $\mathbf{P}$, $\mathbf{Q,}$ and $\mathbf{R,}$ quantities are
polynomials in $1/z$ and $1/\varrho$ and have been given in Appendix B. \ 

The leading terms for the $\mathcal{F}_{i}(\kappa,z,\varrho)$ appearing above
are (Cf. Appendix B)%
\begin{equation}%
\begin{tabular}
[c]{ll}%
$\mathcal{F}_{1}(2\kappa,z,\varrho)\thicksim(-1)^{\kappa}\frac{9}%
{2z^{2\kappa+1}}(4\kappa+1)!!,$ & $\mathcal{F}_{1}(2\kappa+1,z,\varrho
)\thicksim O(1/z^{2\kappa+3}),$\\
$\mathcal{F}_{2}(2\kappa,z,\varrho)\thicksim(-1)^{\kappa}\frac{5}%
{z^{2\kappa+1}}(4\kappa+1)!!,$ & $\mathcal{F}_{2}(2\kappa+1,z,\varrho
)\thicksim O(1/z^{2\kappa+3}),$\\
$\mathcal{F}_{4}^{(\pm)}(2\kappa,z,\varrho)\thicksim O(1/z^{2\kappa+2}),$ &
$\mathcal{F}_{4}^{(\pm)}(2\kappa+1,z,\varrho)\thicksim O(1/z^{2\kappa+3}),$\\
$\mathcal{F}_{5}^{(\pm)}(2\kappa,z,\varrho)\thicksim(-1)^{\kappa}%
\frac{1}{2z^{2\kappa+1}}(4\kappa+1)!!,$ & $\mathcal{F}_{5}^{(\pm)}%
(2\kappa+1,z,\varrho)\thicksim O(1/z^{2\kappa+3}),$\\
$\mathcal{F}_{6}^{(\pm)}(2\kappa,z,\varrho)\thicksim O(1/\varrho z^{2\kappa
+2}),$ & $\mathcal{F}_{6}^{(\pm)}(2\kappa+1,z,\varrho)\thicksim O(1/\varrho
z^{2\kappa+2}),$\\
$\mathcal{F}_{7}^{(\pm)}(2\kappa,z,\varrho)\thicksim O(1/\varrho z^{2\kappa
+1}),$ & $\mathcal{F}_{7}^{(\pm)}(2\kappa+1,z,\varrho)\thicksim O(1/\varrho
z^{2\kappa+3}),$\\
$\mathcal{F}_{8}^{(\pm)}(2\kappa,z,\varrho)\thicksim O(1/z^{2\kappa+2}),$ &
$\mathcal{F}_{8}^{(\pm)}(2\kappa+1,z,\varrho)\thicksim O(1/z^{2\kappa+2})$\\
$\mathcal{F}_{9}(2\kappa,z,\varrho)\thicksim O(1/\varrho z^{2\kappa+2}),$ &
$\mathcal{F}_{9}(2\kappa+1,z,\varrho)\thicksim O(1/\varrho z^{2\kappa+2}).$%
\end{tabular}
\label{36}%
\end{equation}

The argument $b-\left|  \varrho\right|  -\left|  z\right|  =0$ in the
trigonometric expressions above causes the $sin$ of that argument to vanish
and the $cos$ of the same argument to produce a non-oscillating terms which
are of no interest here and has been dropped. \ 

The arguments of the trigonometric functions in the remaining forms occurring
in (34) within the $\mathcal{F(}b,\kappa,z,\varrho)$ function when written in
explicit terms are
\begin{align*}
b+\left|  \varrho\right|  -\left|  z\right|  +\delta\pi/2  &  =2\pi
j\,[\nu_{\varepsilon}^{\left(  1\right)  }-1]\lambda_{\varepsilon}%
(r)+\delta\pi/2,\\
b\pm\left|  \varrho\right|  -\pi/4+\delta\pi/2  &  =2\pi j\,N_{\pm}%
\lambda_{\varepsilon}(r)+(\delta-1/2)\pi//2,
\end{align*}
where
\begin{align*}
\delta &  =0\text{ or }1,\\
N_{+}  &  =\nu_{\varepsilon}^{\left(  1\right)  }\text{,}\\
N_{-}  &  =1\text{.}%
\end{align*}
We have%
\begin{align}
&  \mathcal{F(}b\mathcal{+}\delta\pi/2,\kappa,z,\varrho)=\label{37}\\
&  \tfrac{1}{\pi\,j\,\lambda_{\varepsilon}(r)}\sqrt{\tfrac{1}{2[\nu
_{\varepsilon}^{\left(  1\right)  }-1]}}\left\{
\begin{array}
[c]{c}%
\sin(2\pi j[\nu_{\varepsilon}^{\left(  1\right)  }-1]\lambda_{\varepsilon
}(r)+\delta\pi/2)\\
\cdot\lbrack\mathcal{F}_{1}(\kappa,z,\varrho)-\mathcal{F}_{2}(\kappa
,z,\varrho)]\\
\\
+\cos(2\pi j[\nu_{\varepsilon}^{\left(  1\right)  }-1]\lambda_{\varepsilon
}(r)+\delta\pi/2)\\
\cdot\lbrack\mathcal{F}_{7}^{(+)}(\kappa,z,\varrho)+\mathcal{F}_{8}%
^{(+)}(\kappa,z,\varrho)]
\end{array}
\right\} \nonumber\\
&  +\tfrac{1}{\pi}\sqrt{\tfrac{2}{\,j\,\lambda_{\varepsilon}(r)[\nu
_{\varepsilon}^{\left(  1\right)  }-1]}}\left\{
\begin{array}
[c]{c}%
\cos(2\pi\,j\,\nu_{\varepsilon}^{\left(  1\right)  }\lambda_{\varepsilon
}(r)+(\delta-1/2)\pi/2)\\
\cdot\left[  \mathcal{F}_{4}^{(+)}(\kappa,z,\varrho)+\mathcal{F}_{5}%
^{(-)}(\kappa,z,\varrho)\,/\,2\pi\,j\,\lambda_{\varepsilon}(r)\right] \\
\\
+\sin(2\pi\,j\,\nu_{\varepsilon}^{\left(  1\right)  }\lambda_{\varepsilon
}(r)+(\delta-1/2)\pi/2)\\
\cdot\left[  \mathcal{F}_{6}^{(-)}(\kappa,z,\varrho)+\mathcal{F}_{7}%
^{(+)}(\kappa,z,\varrho)\,/\,2\pi\,j\,\lambda_{\varepsilon}(r)\right] \\
\\
-\sin(2\pi\,j\,\lambda_{\varepsilon}(r)+(\delta-1/2)\pi/2)\\
\cdot\left[  \mathcal{F}_{4}^{(-)}(\kappa,z,\varrho)+\mathcal{F}_{5}%
^{(+)}(\kappa,z,\varrho)\,/\,2\pi\,j\,\lambda_{\varepsilon}(r)\right] \\
\\
+\cos(2\pi\,j\,\lambda_{\varepsilon}(r)+(\delta-1/2)\pi/2)\\
\cdot\left[  \mathcal{F}_{6}^{(+)}(\kappa,z,\varrho)+\mathcal{F}_{7}%
^{(-)}(\kappa,z,\varrho)\,/\,2\pi\,j\,\lambda_{\varepsilon}(r)\right]
\end{array}
\right\}  ,\nonumber
\end{align}
an expression which will occur in its most general form within the terms
$\mathcal{V}_{i}^{\,\,\left(  2\right)  }$ as will be seen below.

\subsection{The radial integrals $\mathfrak{F}\mathcal{(}b\mathcal{+}\delta
\pi/2,\kappa,z,\varrho)$ and $\mathfrak{G\mathcal{(}}b+\delta\pi
/2,\kappa,z,\varrho\mathfrak{)}$}

\ The radial integrals $\mathfrak{F}\mathcal{(}b\mathcal{+}\delta\pi
/2,\kappa,z,\varrho)$ are a generalized form of the $\mathcal{V}%
_{i}^{\,\,\left(  1\right)  }$ integrals shown in Eqs. (31a) i.e.
\begin{equation}
\fbox{$\mathfrak{F}\mathcal{(}b\mathcal{+}\delta\pi/2,\kappa,z,\varrho
)=\int_{0}^{\infty}\dfrac{\lambda_{0}^{3}(r)}{r^{2}}\,\mathcal{F(}b+\delta
\pi/2,\kappa,z,\varrho)\,dr.$} \label{38a}%
\end{equation}

Integrals containing a smooth integrand $F(r)$ such as the cases occuring
above, are given within the stationary state approximation by%

\begin{align}
&  \int_{0}^{\infty}\frac{\lambda_{0}^{3}(r)}{r^{2}}F(r)\left\{
\begin{tabular}
[c]{c}%
$\cos(2\pi j[\nu_{\varepsilon}^{\left(  1\right)  }-1]\lambda_{0}(r)+\delta
\pi/2)$\\
$\sin(2\pi j[\nu_{\varepsilon}^{\left(  1\right)  }-1]\lambda_{0}(r)+\delta
\pi/2)$%
\end{tabular}
\ \ \ \ \ \ \ \ \right.  dr\label{39a}\\
&  =\sqrt{\tfrac{2}{j[\nu_{\varepsilon}^{\left(  1\right)  }-1]}}%
\tfrac{\lambda_{0}^{7/2}(r_{0})}{\omega r_{0}^{2}}F(r_{0})\left\{
\begin{tabular}
[c]{c}%
$\cos(2\pi j[\nu_{\varepsilon}^{\left(  1\right)  }-1]\lambda_{0}%
(r_{0})+(\delta-1/2)\pi/2)$\\
$\sin(2\pi j[\nu_{\varepsilon}^{\left(  1\right)  }-1]\lambda_{0}%
(r_{0})+(\delta-1/2)\pi/2)$%
\end{tabular}
\ \ \ \ \ \ \ \ \ \right.  ,\nonumber
\end{align}
and%
\begin{align}
&  \int_{0}^{\infty}\frac{\lambda_{0}^{3}(r)}{r^{2}}F(r)\left\{
\begin{tabular}
[c]{c}%
$\cos(2\pi jN_{\pm}\lambda_{0}(r)+(\delta-1/2)\pi/2)$\\
$\sin(2\pi jN_{\pm}\lambda_{0}(r)+(\delta-1/2)\pi/2)$%
\end{tabular}
\ \ \ \ \ \ \ \ \ \right.  dr\label{39sb}\\
&  =\sqrt{\tfrac{2}{jN_{\pm}}}\tfrac{\lambda_{0}^{7/2}(r_{0})}{\omega
r_{0}^{2}}F(r_{0})\left\{
\begin{tabular}
[c]{c}%
$\cos(2\pi jN_{\pm}\lambda_{0}(r_{0})+\delta\pi/2)$\\
$\sin(2\pi jN_{\pm}\lambda_{0}(r_{0})+\delta\pi/2)$%
\end{tabular}
\ \ \ \ \ \ \ \ \ \right.  ,\nonumber
\end{align}
where $\lambda_{0}(r)$ is taken to be
\[
\lambda_{0}(r)=\lambda_{_{0}}(r_{0})-\tfrac{1}{4}\tfrac{\omega_{_{0}}^{2}%
}{\lambda_{_{0}}(r_{0})}(r-r_{_{0}})^{2}.
\]
The radial integrals $\mathfrak{F}\mathcal{(}b\mathcal{+}\delta\pi
/2,\kappa,z,\varrho)$ are a generalized form of the $\mathcal{V}%
_{i}^{\,\,\left(  1\right)  }$ integrals shown in Eqs. (31a) i.e.
\begin{equation}
\fbox{$\mathfrak{F}\mathcal{(}b\mathcal{+}\delta\pi/2,\kappa,z,\varrho
)=\int_{0}^{\infty}\dfrac{\lambda_{0}^{3}(r)}{r^{2}}\,\mathcal{F(}b+\delta
\pi/2,\kappa,z,\varrho)\,dr.$} \label{40}%
\end{equation}
When the integration over $r$ is performed all of the quantities contained in
(40) are evaluated with the constants $r_{0},$ $\lambda_{0},$ $\omega_{0}%
,\nu_{0}^{\left(  1\right)  },$ $\nu_{0}^{(2)}$and with $\mathcal{K}_{0}%
(r_{0})=\lambda_{0}^{2}(r_{0})/Zr_{0}=0.462303.$ We have%
\[
\mathfrak{F}\mathcal{(}b\mathcal{+}\delta\pi/2,\kappa,z,\varrho)=
\]%
\begin{align}
&  \tfrac{\lambda_{0}^{5/2}(r_{0})}{\pi j^{3/2}\omega_{0}\,r_{0}^{2}\,[\nu
_{0}^{\left(  1\right)  }-1]}\left\{
\begin{array}
[c]{c}%
\sin(2\pi j[\nu_{0}^{\left(  1\right)  }-1]\lambda_{0}(r_{0})+(\delta
-1/2)\pi/2)[\mathcal{F}_{1}(\kappa,z,\varrho)-\mathcal{F}_{2}(\kappa
,z,\varrho)\,]\\
\cos(2\pi j[\nu_{0}^{\left(  1\right)  }-1]\lambda_{0}(r_{0})+(\delta
-1/2)\pi/2)+[\mathcal{F}_{7}^{(+)}(\kappa,z,\varrho)+\mathcal{F}_{8}%
^{(+)}(\kappa,z,\varrho)]
\end{array}
\right\} \label{41s}\\
&  +\tfrac{\lambda_{0}^{3}(r_{0})}{\pi j\omega_{0}\,r_{0}^{2}\sqrt{N_{+}%
[\nu_{0}^{\left(  1\right)  }-1]}}\left\{
\begin{array}
[c]{c}%
\cos(2\pi jN_{+}\lambda_{0}(r_{0})+\delta\pi/2)[\mathcal{F}_{4}^{(+)}%
(\kappa,z,\varrho)+\mathcal{F}_{5}^{(-)}(\kappa,z,\varrho)\,/\,2\pi
j\lambda_{0}(r_{0})]\\
+\sin(2\pi jN_{+}\lambda_{0}(r_{0})+\delta\pi/2)[\mathcal{F}_{6}^{(-)}%
(\kappa,z,\varrho)+\mathcal{F}_{7}^{(+)}(\kappa,z,\varrho)\,/\,2\pi
j\lambda_{0}(r_{0})]
\end{array}
\right\} \nonumber\\
&  +\tfrac{\lambda_{0}^{3}(r_{0})}{\pi j\omega_{0}\,r_{0}^{2}\sqrt{N_{-}%
[\nu_{0}^{\left(  1\right)  }-1]}}\left\{
\begin{array}
[c]{c}%
\cos(2\pi jN_{-}\lambda_{0}(r_{0})+\delta\pi/2)[\mathcal{F}_{4}^{(-)}%
(\kappa,z,\varrho)+\mathcal{F}_{5}^{(+)}(\kappa,z,\varrho)\,/\,2\pi
j\lambda_{0}(r_{0})]\\
-\sin(2\pi jN_{-}\lambda_{0}(r_{0})+\delta\pi/2)[\mathcal{F}_{6}^{(+)}%
(\kappa,z,\varrho)+\mathcal{F}_{7}^{(-)}(\kappa,z,\varrho)\,/\,2\pi
j\lambda_{0}(r_{0})]
\end{array}
\right\}  .\nonumber
\end{align}
The radial integrals which contain higher-order powers of $\lambda_{0}(r)$
i.e. $\mathfrak{G\mathcal{(}}b+\delta\pi/2,\kappa,z,\varrho\mathfrak{)}$ are a
generalized form of the $\mathcal{V}_{i}^{\,\,\left(  2\right)  }$ integrals
occurring in Eqs. (31b) and are defined by
\begin{equation}
\fbox{$\mathfrak{G\mathcal{(}}b+\delta\pi/2,\kappa,z,\varrho\mathfrak{)=}%
\int_{0}^{\infty}\dfrac{\lambda_{0}^{8}(r)}{r^{4}}$\thinspace$\mathcal{F(}%
b+\delta\pi/2,\kappa,z,\varrho)\,dr$.} \label{42}%
\end{equation}
Then we have%
\begin{align*}
&  \mathfrak{G\mathcal{(}}b+\delta\pi/2,\kappa,z,\varrho\mathfrak{)}%
\mathfrak{=}\\
&  \tfrac{\lambda_{0}^{15/2}(r_{0})}{\pi j^{3/2}\omega_{0}\,r_{0}^{4}\sqrt
{\nu_{0}^{\left(  1\right)  }-1}}\left\{
\begin{array}
[c]{c}%
\sin(2\pi j[\nu_{0}^{\left(  1\right)  }-1]\lambda_{0}(r_{0})+\delta
\pi/2)[\mathcal{F}_{1}(\kappa,z,\varrho)-\mathcal{F}_{2}(\kappa,z,\varrho
)\,]\\
\cos(2\pi j[\nu_{0}^{\left(  1\right)  }-1]\lambda_{0}(r_{0})+\delta
\pi/2)[\mathcal{F}_{7}^{(+)}(\kappa,z,\varrho)+\mathcal{F}_{8}^{(+)}%
(\kappa,z,\varrho)\,]
\end{array}
\right\} \\
&  +\tfrac{\lambda_{0}^{8}(r_{0})}{\pi j\omega_{0}\,r_{0}^{4}\sqrt{N_{+}%
[\nu_{0}^{\left(  1\right)  }-1]}}\left\{
\begin{array}
[c]{c}%
\cos(2\pi jN_{+}\lambda_{0}(r_{0})+\delta\pi/2)[\mathcal{F}_{4}^{(+)}%
(\kappa,z,\varrho)+\mathcal{F}_{5}^{(-)}(\kappa,z,\varrho)\,/\,2\pi
j\lambda_{0}(r_{0})]\\
+\sin(2\pi jN_{+}\lambda_{0}(r_{0})+\delta\pi/2)[\mathcal{F}_{6}^{(-)}%
(\kappa,z,\varrho)+\mathcal{F}_{7}^{(+)}(\kappa,z,\varrho)\,/\,2\pi
j\lambda_{0}(r_{0})]
\end{array}
\right\} \\
&  +\tfrac{\lambda_{0}^{8}(r_{0})}{\pi j\omega_{0}\,r_{0}^{4}\sqrt{N_{-}%
[\nu_{0}^{\left(  1\right)  }-1]}}\left\{
\begin{array}
[c]{c}%
\cos(2\pi jN_{-}\lambda_{0}(r_{0})+\delta\pi/2)[\mathcal{F}_{4}^{(-)}%
(\kappa,z,\varrho)+\mathcal{F}_{5}^{(+)}(\kappa,z,\varrho)\,/\,2\pi
j\lambda_{0}(r_{0})]\\
-\sin(2\pi jN_{-}\lambda_{0}(r_{0})+\delta\pi/2)[\mathcal{F}_{6}^{(+)}%
(\kappa,z,\varrho)+\mathcal{F}_{7}^{(-)}(\kappa,z,\varrho)\,/\,2\pi
j\lambda_{0}(r_{0})]
\end{array}
\right\}  .
\end{align*}
Finally we write%
\begin{align}
\mathcal{V}_{0}^{\,\,\left(  1\right)  }(j)  &  =\mathfrak{F}\mathcal{(}%
b,0,z,\varrho),\nonumber\\
\mathcal{V}_{1}^{\,\,\left(  1\right)  }(j,k)  &  =\tfrac{1}{2}\{\mathfrak{F}%
\mathcal{(}b,0,z+z^{\prime},\varrho)+\mathfrak{F}\mathcal{(}b,0,z-z^{\prime
},\varrho)\},\nonumber\\
\mathcal{V}_{0}^{\,\left(  2\right)  }(j)  &  =\sum_{\kappa=1}^{\infty
}\frac{(-1)^{\kappa}\,\mathcal{K}_{\varepsilon}^{\,2(\kappa-1)}}{(2\kappa
+1)}\mathfrak{G}\mathcal{(}b+\pi/2,\kappa,z,\varrho),\nonumber\\
\mathcal{V}_{1}^{\,\,\left(  2\right)  }(j,k)  &  =\tfrac{1}{2}\sum_{\kappa
=1}^{\infty}\frac{(-1)^{\kappa}\,\mathcal{K}_{\varepsilon}^{\,2(\kappa-1)}%
}{(2\kappa+1)}\{\mathfrak{G}\mathcal{(}b+\pi/2,\kappa,z+z^{\prime}%
,\varrho)+\mathfrak{G}\mathcal{(}b+\pi/2,\kappa,z-z^{\prime},\varrho)\}.
\label{43}%
\end{align}
The terms needed in the average value $\overline{V}_{ne,\,}$ being
\begin{subequations}
\begin{align}
\mathcal{V}_{0}^{\,}(\,j)  &  =-\frac{Z}{\pi}\mathfrak{F}\mathcal{(}%
b,0,z,\varrho)-\frac{2j}{Z}\sum_{\kappa=1}^{\infty}\frac{(-1)^{\kappa
}\,\mathcal{K}_{\varepsilon}^{\,2(\kappa-1)}}{(2\kappa+1)}\mathfrak{G}%
\mathcal{(}b+\pi/2,\kappa,z,\varrho),\tag{44a}\\
& \nonumber\\
\mathcal{V}_{1}^{\,}(\,j,k)  &  =-\frac{Z}{2\pi}\{\mathfrak{F}\mathcal{(}%
b,0,z+z^{\prime},\varrho)+\mathfrak{F}\mathcal{(}b,0,z-z^{\prime}%
,\varrho)\}\nonumber\\
&  -\frac{j}{Z}\sum_{\kappa=1}^{\infty}\frac{(-1)^{\kappa}\,\mathcal{K}%
_{\varepsilon}^{\,2(\kappa-1)}}{(2\kappa+1)}\{\mathfrak{G}\mathcal{(}%
b+\pi/2,\kappa,z+z^{\prime},\varrho)+\mathfrak{G}\mathcal{(}b+\pi
/2,\kappa,z-z^{\prime},\varrho)\}. \tag{44b}%
\end{align}
Retaining only the leading terms in the quantities $\mathcal{F}_{i}%
(\kappa,z,\varrho)$ we find that only the difference
\end{subequations}
\begin{equation}
\mathcal{F}_{1}(\kappa,z,\varrho)-\mathcal{F}_{2}(\kappa,z,\varrho
)=\frac{(-1)^{\kappa}}{2[2\pi j\lambda_{\varepsilon}(r)]^{2\kappa+1}},
\label{45}%
\end{equation}
survives whereas all of the others terms are small and have been dropped.
\ Then%
\begin{equation}
Z\,\mathfrak{F}\mathcal{(}b,0,z,\varrho)=-\frac{Z\lambda_{0}(r_{0})^{5/2}%
\sin(2\pi j\,[\nu_{\varepsilon}^{\left(  1\right)  }-1]\lambda_{\varepsilon
}(r)-\pi/4)}{2\pi^{2}[\nu_{0}^{\left(  1\right)  }-1]\,\omega_{0}\,r_{0}%
^{2}\,\,j^{3/2}\,z}, \label{46}%
\end{equation}
then%
\[
Z\,\mathfrak{F}\mathcal{(}b,0,z,\varrho)=\frac{Z\lambda_{0}(r_{0})^{3/2}%
\sin(2\pi j\,[\nu_{\varepsilon}^{\left(  1\right)  }-1]\lambda_{\varepsilon
}(r)-\pi/4)}{4\pi^{2}[\nu_{0}^{\left(  1\right)  }-1]\,\omega_{0}\,r_{0}%
^{2}\,\,j^{5/2}\,},
\]
and%
\begin{equation}
\tfrac{1}{2}Z\{\mathfrak{F}\mathcal{(}b,0,z+z^{\prime},\varrho)+\mathfrak{F}%
\mathcal{(}b,0,z-z^{\prime},\varrho)\}=\frac{Z\lambda_{0}(r_{0})^{3/2}%
\sin(2\pi j\,[\nu_{\varepsilon}^{\left(  1\right)  }-1]\lambda_{\varepsilon
}(r)-\pi/4)}{2\pi^{2}[\nu_{0}^{\left(  1\right)  }-1]\omega_{0}\,r_{0}%
^{2}\,\,j^{1/2}[j^{2}-k^{2}]}, \label{47}%
\end{equation}
terms which are on the order of $Z^{\,3/2}$. \ Similarly one find that
\[
\mathfrak{G}\mathcal{(}b+\pi/2,\kappa,z,\varrho)/Z\thicksim O(Z^{-3/2}),
\]
a quantity which is small and has been dropped.

\subsection{ The sum over $\mathcal{V}_{0}^{\,}(\,j)$}

The first part of the average $\overline{V}_{ne,\lambda}$ \ i.e.%
\begin{equation}
\sum_{j=1}^{\infty}(-1)^{\,j}\mathcal{V}_{0}^{\,}(\,j)=\sum_{j=1}^{\infty
}(-1)^{\,j+1}\{\frac{Z}{\pi}\mathfrak{F}\mathcal{(}b,0,z,\varrho)+\frac{2j}%
{Z}\sum_{\kappa=1}^{\infty}\frac{(-1)^{\kappa}\,\mathcal{K}_{\varepsilon
}^{\,2(\kappa-1)}}{(2\kappa+1)}\mathfrak{G}\mathcal{(}b+\pi/2,\kappa
,z,\varrho)\}, \label{48}%
\end{equation}
reduces to
\begin{equation}
\sum_{j=1}^{\infty}(-1)^{\,j}\mathcal{V}_{0}^{\,}(\,j)=\frac{1}{4\pi^{3}%
}\frac{Z\,\lambda_{0}(r_{0})^{3/2}}{\sqrt{[\nu_{0}^{\left(  1\right)  }%
-1]}\omega_{0}\,r_{0}^{2}}\sum_{j=1}^{\infty}(-1)^{\,j}\frac{\sin(2\pi
j[\nu_{0}^{\left(  1\right)  }-1]\lambda_{0}(r_{0})-\pi/4)}{j^{5/2}},
\label{49}%
\end{equation}
when only the leading terms in $Z$ are kept. \ We have after performing the
sum in (49)%

\begin{align}
-\frac{\overline{V}_{ne,\lambda}}{Z}  &  =-\frac{8}{Z}\sum_{j=1}^{\infty
}(-1)^{\,j}\mathcal{V}_{0}^{\,}(\,j)\label{50}\\
&  =\frac{\sqrt{2}}{\pi^{3}}\frac{\,\lambda_{0}(r_{0})^{3/2}}{[\nu
_{0}^{\left(  1\right)  }-1]\omega_{0}\,r_{0}^{2}}\left\{
\begin{array}
[c]{c}%
\operatorname{Re}[Li_{5/2}(-\exp(-2\pi i[\nu_{0}^{\left(  1\right)
}-1]\lambda_{0}(r_{0}))]\\
+\operatorname{Im}[Li_{5/2}(-\exp(-2\pi i[\nu_{0}^{\left(  1\right)
}-1]\lambda_{0}(r_{0}))]
\end{array}
\right\}  ,\nonumber
\end{align}
a quantity which is of order $Z^{1/2}$.

\subsection{The sum over $\mathcal{V}_{1}^{\,}(\,j,k)$}

The second part of the average $\overline{V}_{ne,\,\lambda\nu}$\ i.e.%
\begin{align}
&  \sum_{j=1}^{\infty}\sum_{k=1}^{\infty}(-1)^{\,j+k}\mathcal{V}_{1}%
^{\,}(\,j,k)\label{51s}\\
&  =-\frac{Z}{2\pi}\sum_{j=1}^{\infty}\sum_{k=1}^{\infty}(-1)^{\,j+k}%
\{\mathfrak{F}\mathcal{(}b,0,z+z^{\prime},\varrho)+\mathfrak{F}\mathcal{(}%
b,0,z-z^{\prime},\varrho)\}\nonumber\\
&  -\frac{j}{Z}\sum_{j=1}^{\infty}\sum_{k=1}^{\infty}(-1)^{\,j+k}\sum
_{\kappa=1}^{\infty}\frac{(-1)^{\kappa}\,\mathcal{K}_{\varepsilon}%
^{\,2(\kappa-1)}}{(2\kappa+1)}\{\mathfrak{G}\mathcal{(}b+\pi/2,\kappa
,z+z^{\prime},\varrho)+\mathfrak{G}\mathcal{(}b+\pi/2,\kappa,z-z^{\prime
},\varrho)\}.\nonumber
\end{align}
The triple sum leads to terms too small to consider here and has been dropped.
\ In the case of the double sum, whenever the argument $\ z+z^{\prime}%
=2\pi(k-j)\lambda_{\varepsilon}(r)$ occurs we note that in the expressions for
$\overline{V}_{ne}$ where $1\leq j\leq\infty$ , and whenever $\ j$ has been
replaced by $\left|  k-j\right|  $ that $1\leq\left|  k-j\right|  \leq\infty$.
\ Since $j>k$ the region to be summed over is shown below.

\bigskip

\ The sum of interest taking into account the restriction \ $j>k$ is%
\begin{align}
\sum_{j=1}^{\infty}\sum_{k=1}^{\infty}(-1)^{\,j+k}\mathcal{V}_{1}^{\,}(\,j,k)
&  =\sum_{k=1}^{\infty}\sum_{j=k+1}^{\infty}(-1)^{\,j+k}\mathcal{V}_{1}%
^{\,}(\,j,k),\label{52}\\
&  =\sum_{j=2}^{\infty}\sum_{k=1}^{j-1}(-1)^{\,j+k}\mathcal{V}_{1}%
^{\,}(\,j,k).\nonumber
\end{align}
For the latter form of the double sum (52) (where the order of summation has
been reversed) we have shown the allowed region of summation in Fig. (9).%

\begin{figure}
\begin{center}
\includegraphics[
width=3.2258in
]%
{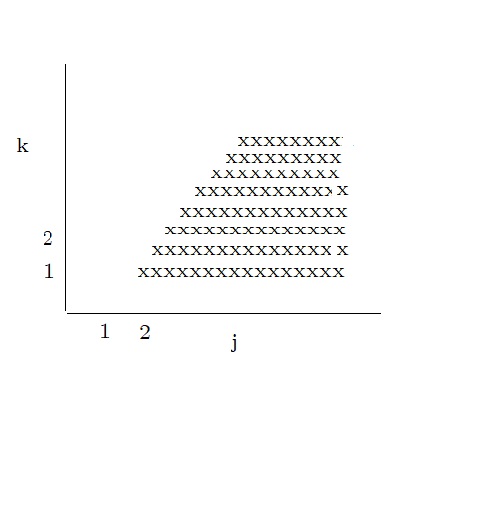}%
\caption{Allowed $k$ vs. $j$ region of summation}%
\end{center}
\end{figure}

We get%
\begin{equation}
\sum_{j=1}^{\infty}\sum_{k=1}^{\infty}(-1)^{\,j+k}\mathcal{V}_{1}%
^{\,}(\,j,k)=-\frac{Z}{2\pi^{3}}\frac{\lambda_{0}(r_{0})^{3/2}}{\sqrt{[\nu
_{0}^{\left(  1\right)  }-1]\,}\omega_{0}\,r_{0}^{2}}\sum_{j=2}^{\infty}%
\sum_{k=1}^{j-1}\frac{(-1)^{j+k}\sin(2\pi j\,[\nu_{0}^{\left(  1\right)
}-1]\lambda_{0}(r_{0})-\pi/4)}{\,\,j^{1/2}\,[j^{2}-k^{2}]}. \label{53}%
\end{equation}
Then the required expression for $V_{ne,\lambda,\nu}/Z$ \ becomes%
\begin{equation}
-\frac{V_{ne,\lambda,\nu}}{Z}=\tfrac{4}{\pi^{3}}\tfrac{\sqrt{2}\lambda
_{0}(r_{0})^{3/2}}{\sqrt{[\nu_{0}^{\left(  1\right)  }-1]\,}\omega_{0}%
\,r_{0}^{2}}\left\{  \operatorname{Re}\mathcal{S}(x)+\operatorname{Im}%
\mathcal{S}(x)\right\}  , \label{54}%
\end{equation}
with the complex quantity $x$ has been defined by
\[
x=\exp\{2\pi i\,\,[\nu_{0}^{\left(  1\right)  }-1]\lambda_{0}(r_{0})\},
\]
and%
\[
\mathcal{S}(x)=\sum_{j=2}^{\infty}\sum_{k=1}^{j-1}\frac{(-1)^{j+k}x^{\,j}%
}{\,\,j^{1/2}\,[j^{2}-k^{2}]}.
\]
The double sum $\mathcal{S}(x)$ can be reduced to the terms (cf. Appendix B)%
\begin{equation}
\mathcal{S}(x)=-\frac{1}{2}[\{1-\ln(2)\}x+\ln(2)Li_{3/2}(x)+Li_{5/2}%
(-x)]+\frac{1}{4}\sum_{j=2}^{\infty}\frac{x^{j}}{j^{3/2}}[\psi(j)-\psi
(j+1/2)], \label{55}%
\end{equation}
A closed-form expression for this expression is not known, however the
infinite sum converges rapidly and may be safely truncated to contain eight
terms to produce five figure accuracy.

Finally we have for $\overline{V}_{ne,\lambda}+\overline{V}_{ne,\lambda,\nu}$
the expression
\begin{align}
-\frac{\overline{V}_{ne,\lambda}+\overline{V}_{ne,\lambda,\nu}}{Z}  &
=\tfrac{\sqrt{2}}{\pi^{3}}\tfrac{\,\lambda_{0}(r_{0})^{3/2}}{\sqrt{[\nu
_{0}^{\left(  1\right)  }-1]}\omega_{0}\,r_{0}^{2}}\left\{  \operatorname{Re}%
[Li_{5/2}(-x)]+\operatorname{Im}[Li_{5/2}(-x)]\right\} \label{56}\\
&  +\tfrac{4\sqrt{2}}{\pi^{3}}\tfrac{\lambda_{0}(r_{0})^{3/2}}{\sqrt{[\nu
_{0}^{\left(  1\right)  }-1]\,}\omega_{0}\,r_{0}^{2}}\left\{
\operatorname{Re}\mathcal{S}(x)+\operatorname{Im}\mathcal{S}(x)\right\}
,\nonumber
\end{align}
with%
\[
\tfrac{\sqrt{2}}{\pi^{3}}\tfrac{\lambda_{0}(r_{0})^{3/2}}{\sqrt{[\nu
_{0}^{\left(  1\right)  }-1]}\omega_{0}\,r_{0}^{2}}=0.033375Z^{1/2}.
\]

\subsection{The Schwinger-Englert Energy}

As seen above the non-oscillating part of the chemical potential is zero, here
we wish to evaluate the derivative of the SE energy in order to complete the
calculation of the oscillating part of the chemical potential. \
\[
\mu_{osc}=\frac{dE(Z,Z)_{osc}}{dZ}-\frac{\overline{V}_{ne,osc}}{Z},
\]
In the work by SE the $j,k$ plane is divided into regions within which the
various contributions to the energy $E_{0}(Z,Z)$ have been computed. \ They
write the energy as
\begin{equation}
E_{osc}(Z,Z)=\sum_{k=1}^{\infty}E_{0,k}+\sum_{j=1}^{\infty}E_{j,0}+\sum
_{j=1}^{\infty}\sum_{k=1}^{\infty}E_{j,k}, \label{57}%
\end{equation}
with%
\begin{align*}
\sum_{k=1}^{\infty}E_{0,k}  &  =E_{\,\mathit{l}TF}(Z),\\
\sum_{j=1}^{\infty}E_{j,0}+\sum_{j=1}^{\infty}\sum_{k=1}^{\infty}E_{j,k}  &
=E_{\lambda-osc}(Z)+E_{\,\nu-osc}(Z)+E_{\nu,\lambda-osc}(Z).
\end{align*}

\bigskip

In that work the energy associated with the $\nu$ oscillation was small and
the $\lambda,\nu$ oscillations where found to be negligible compared to that
of the $\lambda$ oscillations. \ In the latter case the $E_{\nu,\lambda-osc}$
contribution is on the order of $0.000Z^{3/2}$ and has been dropped. \ As a
result we write%
\begin{equation}
E(Z,Z)=E_{\,\mathit{l}TF}(Z)+E_{\,\lambda-osc}(Z)+E_{\,\nu-osc}(Z). \label{58}%
\end{equation}

Schwinger and Englert have given $E_{\mathit{l}TF}\,(Z)$ as
\begin{equation}
E_{\,\mathit{l}\,TF}\,(Z)=\frac{\lambda_{0}}{r_{0}^{2}}[\frac{1}{2}%
\mathcal{K}_{1}C_{2}(\lambda_{0})-\lambda_{0}\nu_{0}^{^{(1)}}S_{1}(\lambda
_{0})\,], \label{59}%
\end{equation}
where the terms $C_{n}(\lambda_{0})$ and $S_{n}(\lambda_{0})$ are sums defined
below and $\mathcal{K}_{1}$ is a constant i.e.
\[
\mathcal{K}_{1}=-1+6\nu_{0}^{^{(1)}}-6(\nu_{0}^{(1)})^{3}+(\nu_{0}^{^{(1)}%
})^{5}=-5.709672
\]
Noting the general relations for the {sums}%
\begin{align*}
S_{n}(z)  &  =\sum_{k=1}^{\infty}\frac{(-1)^{k}\sin(2\pi kz)}{(\pi k)^{2n+1}%
},\\
C_{n}(z)  &  =\sum_{k=1}^{\infty}\frac{(-1)^{k}\cos(2\pi kz)}{(\pi k)^{2n}},\\
\frac{dS_{n}(z)}{dz}  &  =2C_{n}(z),\\
\frac{dC_{n+1}(z)}{dz}  &  =-2S_{n}(z),
\end{align*}
and%
\begin{align*}
\frac{d\lambda_{0}}{dZ}  &  =\frac{1}{3}\left(  \frac{\lambda_{0}}{Z}\right)
,\\
\frac{d}{dz}\left(  \frac{\lambda_{0}^{3}}{\omega_{0}r_{0}^{3}}\right)   &
=\frac{4}{3Z}\left(  \frac{\lambda_{0}^{3}}{\omega_{0}r_{0}^{3}}\right)  ,\\
\frac{d}{dz}\left(  \frac{\lambda_{0}}{r_{0}^{2}}\right)   &  =\frac{1}%
{Z}\left(  \frac{\lambda_{0}}{r_{0}^{2}}\right)  ,
\end{align*}
\ 

we have
\begin{equation}
\frac{dE_{\,\mathit{l}\,TF}\,(Z)}{dZ}=\frac{2\lambda_{0}}{3Zr_{0}^{2}%
}[\frac{3}{2}\mathcal{K}_{1}C_{2}(\lambda_{0})+\lambda_{0}\{\nu_{0}^{^{(1)}%
}-\mathcal{K}_{1}\}S_{1}(\lambda_{0})+\lambda_{0}^{2}\,\nu_{0}^{^{(1)}}%
C_{1}(\lambda_{0})], \label{60}%
\end{equation}
and%
\[
\frac{\lambda_{0}}{Zr_{0}^{2}}=0.26744
\]
The terms in the derivative above are seen to be of order $Z^{0}%
,Z^{1/3},Z^{2/3}$ respectively.

The SE energy of the $\lambda$ oscillations is given by%
\begin{equation}
-(r_{0}/\lambda_{0})^{2}E_{\,\lambda-osc}(Z)= \label{61}%
\end{equation}%
\begin{align*}
&  \mathbf{S}_{1}^{\prime}(\lambda_{0})-\frac{1}{4}S_{1}(2\lambda
_{0})+\mathcal{K}_{2}S_{0}(2\lambda_{0})-\lambda_{0}^{1/2}[\mathcal{K}%
_{3}\operatorname{Im}Li_{5/2}(-\exp(-4\pi i\lambda_{0}))]\\
&  +\frac{1}{Z^{1/3}}\left\{
\begin{array}
[c]{c}%
\mathcal{K}_{4}\left\{  \mathbf{C}_{1}^{\prime}(\lambda_{0})+\frac{1}%
{8}\widetilde{C}_{2}(2\lambda_{0})\right\}  -\mathcal{K}_{5}\mathbf{C}%
_{0}^{\prime}(\lambda_{0})+\mathcal{K}_{6}\widetilde{C}_{1}(2\lambda
_{0})+\mathcal{K}_{7}\widetilde{C}_{0}(2\lambda_{0})\\
+\lambda_{0}^{1/2}[\mathcal{K}_{8}\operatorname{Re}Li_{7/2}(-\exp(-4\pi
i\lambda_{0}))]+\lambda_{0}^{3/2}[\mathcal{K}_{9}\operatorname{Re}%
Li_{5/2}(-\exp(-4\pi i\lambda_{0}))]
\end{array}
\right\}  ,
\end{align*}
and%
\[
(\frac{\lambda_{0}}{r_{0}})^{2}=0.24818Z^{4/3},
\]
where the terms $\mathbf{S}_{n}^{\prime}$ $\mathbf{C}_{n}^{\prime}$
$\widetilde{S}_{n}$\ $\widetilde{C}_{n}$ \ are sums given below and the
numerical constants $\mathcal{K}_{i}$ are
\begin{align*}
\mathcal{K}_{2}  &  =[2-\nu_{0}^{^{(1)}}]/24\nu_{0}^{^{(1)}}=0.00134,\\
\mathcal{K}_{3}  &  =\frac{\sqrt{3}}{\sqrt{\nu_{0}^{^{(1)}}[(\nu_{0}^{^{(1)}%
})^{2}-1][-15+23(\nu_{0}^{^{(1)}})^{2}-5(\nu_{0}^{^{(1)}})^{4}]}\pi^{2}%
}=0.04905,\\
\mathcal{K}_{4}  &  =[3-4(\nu_{0}^{^{(1)}})^{2}]/(2\pi)^{4}=-0.00771,\\
\mathcal{K}_{5}  &  =[1-\nu_{0}^{^{(1)}}]/4(\nu_{0}^{^{(1)}})^{3}=-0.03222,\\
\mathcal{K}_{6}  &  =[34-48\nu_{0}^{^{(1)}}+31(\nu_{0}^{^{(1)}})^{2}%
]/576=0.09963,\\
\mathcal{K}_{7}  &  =7[92-162\nu_{0}^{^{(1)}}+150(\nu_{0}^{^{(1)}})^{2}%
-76(\nu_{0}^{^{(1)}})^{3}+15(\nu_{0}^{^{(1)}})^{4}]/46080=-0.000026,\\
\mathcal{K}_{8}  &  =\tfrac{[2(\nu_{0}^{^{(1)}})^{2}-1]}{(2\pi)^{3}}%
\sqrt{\tfrac{2}{\nu_{0}^{^{(1)}}-1}}=0.03832,\\
\mathcal{K}_{9}  &  =\left(  \tfrac{2}{\nu_{0}^{^{(1)}}-1}\right)
^{3/2}[2-6\nu_{0}^{^{(1)}}-6(\nu_{0}^{^{(1)}})^{2}+12(\nu_{0}^{^{(1)}}%
)^{3}-4(\nu_{0}^{^{(1)}})^{4}-2(\nu_{0}^{^{(1)}})^{5}-(\nu_{0}^{^{(1)}}%
)^{6}]/(4\pi)^{4}\\
&  =-0.01359
\end{align*}
The sums appearing above are given by \cite{SE2}%
\begin{align*}
\widetilde{S}_{0}(z)  &  =\sum_{k=1}^{\infty}\frac{\sin(2\pi k\,z)}{(\pi
k)}=-\left\langle z-1/2\right\rangle ,\\
\widetilde{C}_{0}(z)  &  =\sum_{k=1}^{\infty}\cos(2\pi k\,z)=-1/2,\\
\widetilde{S}_{1}(z)  &  =\sum_{k=1}^{\infty}\frac{\sin(2\pi k\,z)}{(\pi
k)^{3}}=\tfrac{2}{3}\left\langle z-1/2\right\rangle [\left\langle
z-1/2\right\rangle ^{2}-\tfrac{1}{4}],\\
\widetilde{C}_{1}(z)  &  =\sum_{k=1}^{\infty}\frac{\cos(2\pi k\,z)}{(\pi
k)^{2}}=\left\langle z-1/2\right\rangle ^{2}-\tfrac{1}{12},\\
\widetilde{S}_{2}(z)  &  =\sum_{k=1}^{\infty}\frac{\sin(2\pi k\,z)}{(\pi
k)^{5}}=-\tfrac{1}{3}\left\langle z-1/2\right\rangle [\tfrac{2}{5}\left\langle
z-1/2\right\rangle ^{4}-\tfrac{1}{3}\left\langle z-1/2\right\rangle
^{2}+\tfrac{7}{120}],\\
\widetilde{C}_{2}(z)  &  =\sum_{k=1}^{\infty}\frac{\cos(2\pi k\,z)}{(\pi
k)^{4}}=\tfrac{1}{90}-\tfrac{1}{3}[\left\langle z-1/2\right\rangle ^{2}%
-\tfrac{1}{4}]^{2}.
\end{align*}
and
\begin{align*}
\mathbf{S}_{0}^{\prime}(z)  &  =\sum_{k=0}^{\infty}(-1)^{k}\sin(2\pi
\lbrack2k+1]z)=0,\\
\mathbf{C}_{0}^{\prime}(z)  &  =\sum_{k=0}^{\infty}\frac{(-1)^{k}\cos
(2\pi\lbrack2k+1]z)}{[\pi(2k+1)]}=\tfrac{1}{4}(-1)^{\lfloor2z+1/2\rfloor},\\
\mathbf{S}_{1}^{\prime}(z)  &  =\sum_{k=0}^{\infty}\frac{(-1)^{k}\sin
(2\pi\lbrack2k+1]z)}{[\pi(2k+1)]^{2}}=\tfrac{1}{2}[<z+\tfrac{1}{4}%
>^{2}-<z-\tfrac{1}{4}>^{2}],\\
\mathbf{C}_{1}^{\prime}(z)  &  =\sum_{k=0}^{\infty}\frac{(-1)^{k}\cos
(2\pi\lbrack2k+1]z)}{[\pi(2k+1)]^{3}},\\
&  =\tfrac{1}{3}[<z+\tfrac{1}{4}>\{\tfrac{1}{4}-<z+\tfrac{1}{4}>^{2}%
\}-<z-\tfrac{1}{4}>\{\tfrac{1}{4}-<z-\tfrac{1}{4}>^{2}\}].
\end{align*}

\bigskip

The derivative of $E_{\,\lambda-osc}(Z)$ is given by%
\begin{equation}
-3Z\,(r_{0}/\lambda_{0})^{2}\frac{dE_{\,\lambda-osc}(Z)}{dZ}= \label{62}%
\end{equation}%
\begin{align*}
&  4\mathbf{S}_{1}^{\prime}(\lambda_{0})-S_{1}(2\lambda_{0})+4\mathcal{K}%
_{2}S_{0}(2\lambda_{0})-\lambda_{0}^{1/2}[\tfrac{9}{2}\mathcal{K}%
_{3}\operatorname{Im}Li_{5/2}(-\exp(-4\pi i\lambda_{0}))]\\
&  +\lambda_{0}[2\mathbf{C}_{0}^{\prime}(\lambda_{0})-C_{1}(2\lambda
_{0})]+\lambda_{0}^{3/2}[4\pi\,\mathcal{K}_{3}\operatorname{Re}Li_{3/2}%
(-\exp(-4\pi i\lambda_{0}))]\\
&  +\frac{1}{Z^{1/3}}\left\{
\begin{array}
[c]{c}%
\begin{array}
[c]{c}%
3\mathcal{K}_{4}[\mathbf{C}_{1}^{\prime}(\lambda_{0})+\frac{1}{8}\widetilde
{C}_{2}(2\lambda_{0})]-3\mathcal{K}_{5}\mathbf{C}_{0}^{\prime}(\lambda
_{0})+3\mathcal{K}_{6}\widetilde{C}_{1}(2\lambda_{0})+3\mathcal{K}%
_{7}\widetilde{C}_{0}(2\lambda_{0})\\
+\,\lambda_{0}^{1/2}[\frac{7}{2}\mathcal{K}_{8}\operatorname{Re}Li_{7/2}%
(-\exp(-4\pi i\lambda_{0}))]\\
-2\lambda_{0}[\mathcal{K}_{4}\{\mathbf{S}_{1}^{\prime}(\lambda_{0}%
)+\frac{1}{4}\widetilde{S}_{1}(2\lambda_{0})\}-\mathcal{K}_{5}\mathbf{S}%
_{0}^{\prime}(\lambda_{0})+2\mathcal{K}_{7}\widetilde{S}_{0}(2\lambda_{0})]\\
+\,\lambda_{0}^{3/2}[\frac{9}{2}\mathcal{K}_{9}\operatorname{Re}Li_{5/2}%
(-\exp(-4\pi i\lambda_{0}))+4\pi\,\mathcal{K}_{8}\operatorname{Im}%
Li_{5/2}(-\exp(-4\pi i\lambda_{0}))]
\end{array}
\\
+\lambda_{0}^{5/2}[4\pi\,\mathcal{K}_{9}\operatorname{Im}Li_{7/2}(-\exp(-4\pi
i\lambda_{0}))].
\end{array}
\right\}  ,
\end{align*}
with%
\[
\frac{1}{Z\,}(\frac{\lambda_{0}}{r_{0}})^{2}=0.24818Z^{\,1/3}.
\]
The $E_{\nu-osc}$ contribution to the energy is
\[
E_{\nu-osc}=2(\lambda_{0}/r_{0})^{2}\mathcal{K}_{10}\,\widetilde{S}_{1}%
([1+\nu_{0}^{\prime}]\lambda_{0}/2),
\]
where $\mathcal{K}_{10}$ is%
\[
\mathcal{K}_{10}=\frac{1}{[\left(  \nu_{0}^{\prime}\right)  ^{2}-1][1+\nu
_{0}^{\prime}-\left(  \nu_{0}^{\prime}\right)  ^{2}-\left(  \nu_{0}^{\prime
}\right)  ^{3}+\left(  \nu_{0}^{\prime}\right)  ^{4}]}=0.06046,
\]
with $E_{\nu-osc}$ being on the order of $0.00Z^{4/3\text{ \ }}$which is small
compared to the $\lambda$ oscillations. The derivative of $E_{\nu-osc}$ is%
\[
\frac{dE_{\nu-osc}}{dZ}=\frac{2}{3Z}(\frac{\lambda_{0}}{r_{0}})^{2}%
\mathcal{K}_{10}[4\widetilde{S}_{1}([1+\nu_{0}^{\prime}]\lambda_{0}%
/2)+[1+\nu_{0}^{\prime}]\lambda_{0}\widetilde{C}_{1}([1+\nu_{0}^{\prime
}]\lambda_{0}/2)]
\]
with $Z$ contributions to $dE_{\nu-osc}/dZ$ of order $Z^{2/3}$ and $\ Z^{1/3}$
and has been neglected. \ 

\subsection{Numerical Calculations}

\begin{center}
As a result of the analysis given above, the chemical potential is given by
\begin{equation}
\mu_{osc}=\frac{dE_{lTF}(Z)}{dZ}+\frac{dE_{\lambda-osc}(Z)}{dZ}%
-\frac{\overline{V}_{ne,lTF}}{Z}-\frac{\overline{V}_{ne,\lambda}+\overline
{V}_{ne,\lambda,\nu}}{Z}.\label{63}%
\end{equation}
The final calculation of the chemical potential can now proceed by combining
the eqations above.\ The results of these calculations are shown in Figs.
(10)
\begin{center}
\includegraphics[
width=285.8125pt
]%
{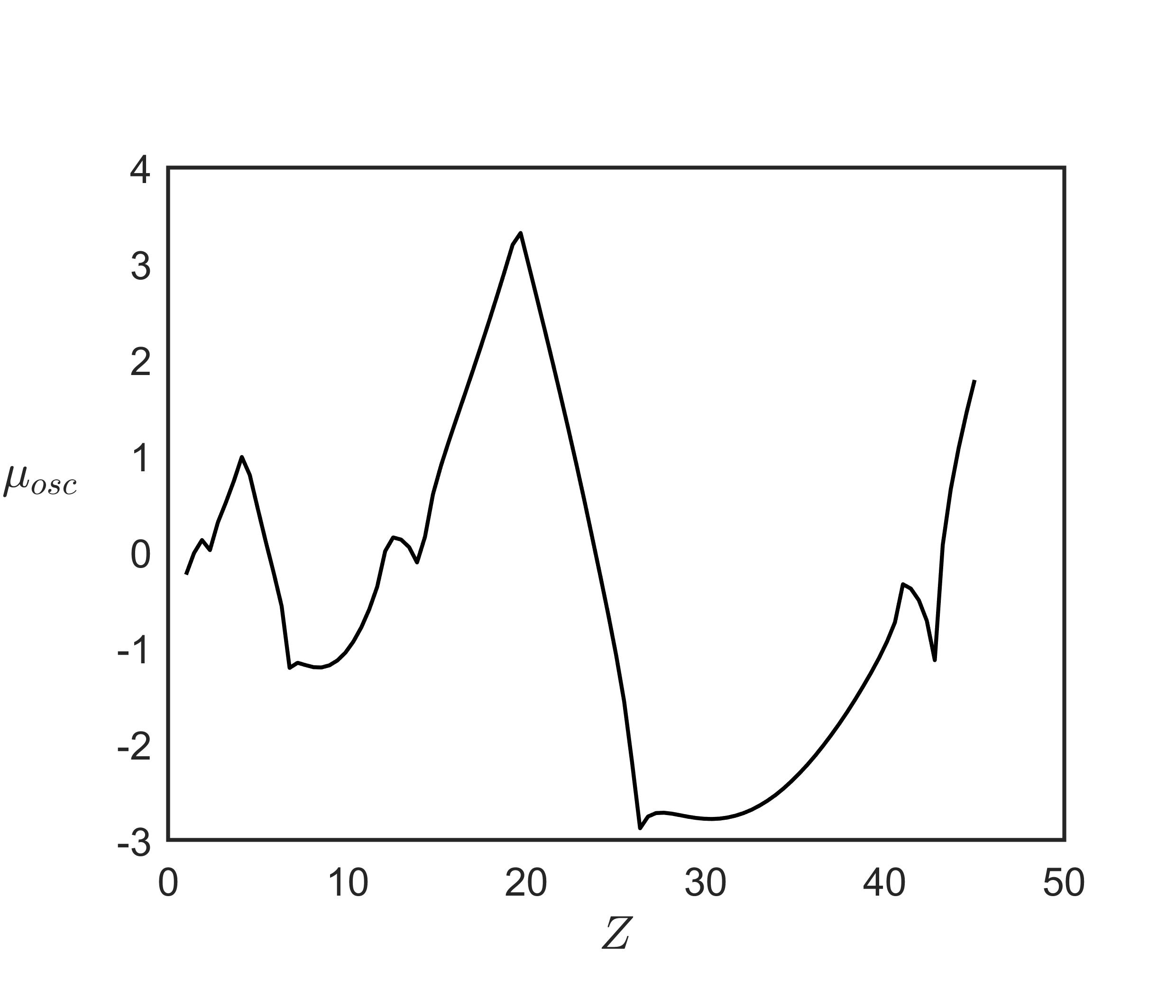}%
\end{center}
Figure 10 the oscillating part of the chemical potential vs. Z
\end{center}

The figure above shows the oscillating part of the chemical potential and
therefore cannot be directly compared with the chemical potential in Fig. 1.
\ Furthermore the results are only valid for large Z. We see that in the
semiclassical approximation to the potential energy that\ the fine scale
oscillations disappointingly have been smoothed out. \ In the case of very
large Z relativistic effects are important but these have\ not been included
in these calculations.

\newpage

\begin{center}
Appendix A\appendix                                                        
\end{center}

\subsubsection{Evaluation of the integrals $I_{1}(\nu,\lambda_{\varepsilon
},r)$ and $I_{2}(\nu,\lambda_{\varepsilon},r)$\bigskip\ }

Here we give details of the calculation of $I_{1}$ and $I_{2}$. \ In order for
$I_{1}$ to be evaluated, i.e.%
\[
I_{1}(\nu,\lambda_{\varepsilon},r)=\int\sqrt{2r^{2}\{\varepsilon_{\lambda,\nu
}-V\}-\lambda^{2}}d\nu,
\]
a relationship must be established between $\varepsilon_{\lambda,\nu}$ and
$\nu$ which will allow us to give an approximation for this integral. \ We
chose the Coulombic potential where this relationship is
\[
\varepsilon_{\lambda,\nu}=-\frac{Z^{2}}{2\left(  \lambda+\nu\right)  ^{2}}.
\]
Then the integral $I_{1}(\nu)$\ can be written as
\[
I_{1}(\nu,\lambda_{\varepsilon},r)=-\frac{Zr}{2}\int\sqrt{\frac{\varepsilon
_{\lambda,\nu}-V-\lambda^{2}/2r^{2}}{-\varepsilon_{\lambda,\nu}^{3}}%
}d\,\varepsilon_{\lambda,\nu},
\]
which immediately produces
\begin{equation}
I_{1}(\nu,\lambda_{\varepsilon},r)=Zr\left[  \sqrt{\frac{2r^{2}\left\{
\varepsilon_{\lambda,\nu}-V\right\}  -\lambda^{2}}{-2r^{2}\,\varepsilon
_{\lambda,\nu}}}-\arctan\sqrt{\frac{2r^{2}\left\{  \varepsilon_{\lambda,\nu
}-V\right\}  -\lambda^{2}}{-2r^{2}\,\varepsilon_{\lambda,\nu}}}\right]  .
\tag{A1}%
\end{equation}
The value of that integral along the curves of degeneracy is then
\[
I_{1}(\nu,\lambda_{\varepsilon},r)/Zr=\frac{\lambda_{\varepsilon}(r)}{Zr}%
\sqrt{\lambda_{\varepsilon}^{2}(r)-\lambda^{2}}-\arctan(\frac{\lambda
_{\varepsilon}(r)}{Zr}\sqrt{\lambda_{\varepsilon}^{2}(r)-\lambda^{2}}),
\]
where we have use the relations
\[
\varepsilon_{\lambda,\nu(\lambda|\varepsilon)}=\varepsilon_{\lambda
_{\varepsilon},0}=-Z^{2}/2\lambda_{\varepsilon}^{2},
\]
and
\[
\lambda_{\varepsilon}^{2}(r)=2r^{2}\left\{  \varepsilon_{\lambda_{\varepsilon
},\nu(\lambda|\varepsilon)}-V\right\}  ,
\]
or\ in terms of the angle $\theta$ \ (A1) becomes%
\[
\frac{I_{1}(\nu,\lambda_{\varepsilon},r)}{Zr}=\mathcal{K}_{\varepsilon}%
(r)\sin(\theta)-\arctan(\mathcal{K}_{\varepsilon}(r)\sin\theta),
\]
where $\mathcal{K}_{\varepsilon}(r)=\lambda_{\varepsilon}^{2}/rZ$. \ Since
$\mathcal{K}_{\varepsilon}(r)\sin\theta$ is less than unity we have the final
form of the integral%
\begin{equation}
\frac{I_{1}(\nu,\lambda_{\varepsilon},r)}{Zr}=\tfrac{1}{3}[\mathcal{K}%
_{\varepsilon}(r)\sin(\theta)]^{3}-\tfrac{1}{5}[\mathcal{K}_{\varepsilon
}(r)\sin(\theta)]^{5}+\cdots\tag{A2}%
\end{equation}
In a similar way, the integral for $I_{2}$ is%

\[
I_{2}(\nu,\lambda_{\varepsilon},r)=\sqrt{2}Z^{2}r\int\left[  \sqrt
{\tfrac{2r^{2}\left\{  \varepsilon_{\lambda,\nu}-V\right\}  -\lambda^{2}%
}{-2r^{2}\,\varepsilon_{\lambda,\nu}}}-\arctan\sqrt{\tfrac{2r^{2}\left\{
\varepsilon_{\lambda,\nu}-V\right\}  -\lambda^{2}}{-2r^{2}\,\varepsilon
_{\lambda,\nu}}}\right]  \frac{d\varepsilon_{\lambda,\nu}}{(-\varepsilon
_{\lambda,\nu})^{3/2}},
\]
which gives%
\[
I_{2}(\nu,\lambda,r)=
\]%
\[
-\sqrt{2}Z^{2}r\left[
\begin{array}
[c]{c}%
-\tfrac{\sqrt{2r^{2}\left\{  \varepsilon_{\lambda,\nu}-V\right\}  -\lambda
^{2}}}{\sqrt{2}r\varepsilon_{\lambda,\nu}}+\frac{2}{\sqrt{-\varepsilon
_{\lambda,\nu}}}\arctan\sqrt{\frac{2r^{2}\left\{  \varepsilon_{\lambda,\nu
}-V\right\}  -\lambda^{2}}{-2r^{2}\,\varepsilon_{\lambda,\nu}}}\\
-\,\frac{\sqrt{2}r}{\sqrt{2r^{2\{}\varepsilon_{\lambda,\nu}-V\}-\lambda
^{2}-2r^{2}\varepsilon_{\lambda,\nu}}}\ln(\sqrt{\frac{2r^{2}\left\{
\varepsilon_{\lambda,\nu}-V\right\}  -\lambda^{2}}{-2r^{2}\,\varepsilon
_{\lambda,\nu}}}-\sqrt{\tfrac{2r^{2}\{\varepsilon_{\lambda,\nu}-V\}-\lambda
^{2}-2r^{2}\varepsilon_{\lambda,\nu}}{-2r^{2}\varepsilon_{\lambda,\nu}}})
\end{array}
\right]  .
\]
Then
\begin{align*}
I_{2}(\nu,\lambda_{\varepsilon},r)  &  =2\lambda_{\varepsilon}^{2}%
\sqrt{\lambda_{\varepsilon}^{2}-\lambda^{2}}-4Zr\lambda_{\varepsilon}%
\arctan(\frac{\lambda_{\varepsilon}\sqrt{\lambda_{\varepsilon}^{2}-\lambda
^{2}}}{rZ})\\
&  +\frac{2r^{2}Z^{2}\lambda_{\varepsilon}}{\sqrt{r^{2}Z^{2}+\lambda
_{\varepsilon}^{2}(\lambda_{\varepsilon}^{2}-\lambda^{2})}}\ln(\frac{\lambda
_{\varepsilon}\sqrt{\lambda_{\varepsilon}^{2}-\lambda^{2}}}{rZ}-\frac{\sqrt
{\lambda_{\varepsilon}^{2}(\lambda_{\varepsilon}^{2}-\lambda^{2})+r^{2}Z^{2}}%
}{rZ})
\end{align*}
or with $\lambda=\lambda_{\varepsilon}\cos\theta$ we have%
\begin{align*}
I_{2}(\nu,\lambda_{\varepsilon},r)/2rZ\lambda_{\varepsilon}  &  =\frac{\lambda
_{\varepsilon}^{2}}{rZ}\sin\theta-2\arctan(\frac{\lambda_{\varepsilon}^{2}%
}{rZ}\sin\theta)\\
&  +\frac{1}{\sqrt{1+[\frac{\lambda_{\varepsilon}^{2}}{rZ}\sin\theta]^{2}}}%
\ln(\frac{\lambda_{\varepsilon}^{2}}{rZ}\sin\theta-\sqrt{[\frac{\lambda
_{\varepsilon}^{2}}{rZ}\sin\theta]^{2}+1]}\,),
\end{align*}
In the more compact form $I_{2}(\nu,\lambda_{\varepsilon},r)$ becomes%
\begin{equation}
\frac{I_{2}(\nu,\lambda_{\varepsilon},r)}{2rZ\lambda_{\varepsilon}%
}=\mathcal{K}_{\varepsilon}(r)\sin\theta-2\arctan(\mathcal{K}_{\varepsilon
}(r)\sin\theta)+\frac{\arcsin h\,(\mathcal{K}_{\varepsilon}(r)\sin\theta
)}{\sqrt{1+[\mathcal{K}_{\varepsilon}(r_{\varepsilon})\sin\theta]^{2}}}.
\tag{A3}%
\end{equation}
\ For small argument $\mathcal{K}_{\varepsilon}(r)\sin\theta$ we have
\begin{equation}
\frac{I_{2}(\nu,\lambda_{\varepsilon},r)}{2rZ\lambda_{\varepsilon}}=\tfrac
{2}{15}\left[  \mathcal{K}_{\varepsilon}(r)\sin\theta\right]  ^{5}-\tfrac
{6}{35}\left[  \mathcal{K}_{\varepsilon}(r)\sin\theta\right]  ^{7}+\cdots.
\tag{A4}%
\end{equation}
\pagebreak 

\begin{center}
\pagebreak \appendix       Appendix B
\end{center}

The double sum%
\[
\mathcal{S}(x)=\sum_{j=2}^{\infty}\sum_{k=1}^{j-1}\frac{(-1)^{j+k}x^{j}%
}{\,\,j^{1/2}\,[j^{2}-k^{2}]},
\]
with $x=\exp(i\theta)$ and $\theta=2\pi\lbrack\,[\nu_{0}^{\left(  1\right)
}-1]\lambda_{0}(r_{0})$ $]$ can be simplified as follows. \ The sum over $k$
can be written as
\[
\sum_{k=1}^{j-1}\frac{(-1)^{k}}{\,\,[j^{2}-k^{2}]}=-\frac{1}{2j^{2}%
}-\frac{(-1)^{j}}{j}\{\frac{1}{2}\ln(2)+\frac{1}{4}\psi(j+1/2)-\frac{1}{4}%
\psi(j)\},
\]
where $\psi$ is the digamma function. \ Then $\mathcal{S}(x)$is given by%
\begin{align*}
\mathcal{S}(x)  &  =-\frac{1}{2}[\{1-\ln(2)\}x+\ln(2)Li_{3/2}(x)+Li_{5/2}%
(-x)]+\frac{1}{4}\sum_{j=2}^{\infty}\frac{x^{j}}{j^{3/2}}[\psi(j)-\psi
(j+1/2)],\\
&  =-\frac{1}{2}[\{1-\ln(2)\}x+\ln(2)Li_{3/2}(x)-Li_{5/2}(x)+\frac{1}{2^{3/2}%
}Li_{5/2}(x^{2})]\\
&  +\frac{1}{4}\sum_{j=2}^{\infty}\frac{x^{j}}{j^{3/2}}[\psi(j)-\psi(j+1/2)],
\end{align*}
The remaining sum in \ the equation above is rapidly convergent, eight terms
being sufficient for six figure accuracy.

\pagebreak \ 

\begin{center}
\bigskip Appendix C
\appendix                                                        
\end{center}

\subsection{The Primitive Integrals $C_{ss},$ $S_{ss},$ $C_{cs},$ $S_{sc}$}

The \textit{key} primitive integrals $C_{ss},S_{ss},C_{cs},S_{sc}$ which
appear in the text above and where $z$ and $\varrho$ are $<0$ i.e.
\begin{align}
\mathcal{C}_{cc}\mathcal{(}\kappa,z,\varrho)  &  =\int_{0}^{\pi/2}%
\sin^{2\kappa+2\,}\theta\cos\theta\cos(z\cos\theta)\cos(\varrho\cos^{2}%
\theta)d\theta,\tag{C1}\\
\mathcal{S}_{ss}\mathcal{(}\kappa,z,\varrho)  &  =\int_{0}^{\pi/2}%
\sin^{2\kappa+2\,}\theta\cos\theta\sin(z\cos\theta)\sin(\varrho\cos^{2}%
\theta)d\theta,\nonumber\\
\mathcal{C}_{cs}\mathcal{(}\kappa,z,\varrho)  &  =\int_{0}^{\pi/2}%
\sin^{2\kappa+2\,}\theta\cos\theta\cos(z\cos\theta)\sin(\varrho\cos^{2}%
\theta)d\theta,\nonumber\\
\mathcal{S}_{sc}\mathcal{(}\kappa,z,\varrho)  &  =\int_{0}^{\pi/2}%
\sin^{2\kappa+2\,}\theta\cos\theta\sin(z\cos\theta)\cos(\varrho\cos^{2}%
\theta)d\theta.\nonumber
\end{align}
can be expressed in terms of the integrals$\ \mathcal{J}(\kappa,\,j,a,z),$ and
$\mathcal{H}(\kappa,\,j,a,z)$ which have closed forms and are defined as
\[
\mathcal{J}(\kappa,\,j,a,z)=\int_{0}^{\pi/2}\sin^{2\kappa+2\,}\theta
\,\cos^{4\,j+2\,a+1}\theta\,\sin(z\cos\theta)\,d\theta,
\]%
\[
\mathcal{H}(\kappa,\,j,a,z)=\int_{0}^{\pi/2}\sin^{2\kappa+2\,}\theta
\,\cos^{4\,j+2\,a+1}\theta\,\cos(z\cos\theta)\,d\theta.
\]
The latter integrals are given by Maple\thinspace as
\[
\mathcal{J}(\kappa,\,j,a,z)=\tfrac{1}{2}z\,B(2j+a+3/2,\kappa+3/2)\,_{1}%
F_{2}\tbinom{2\,j+a+3/2\,;\,-z^{2}/4}{3/2\,,\,\kappa+2\,j+a+3},
\]%
\[
\mathcal{H}(\kappa,\,j,a,z)=\tfrac{1}{2}B(2j+a+1,\kappa+3/2)\,_{1}F_{2}%
\tbinom{2\,j+a+1\,;\,-z^{2}/4}{1/2\,,\,\kappa+2\,j+a+5/2},
\]
where $B(\alpha,\beta)$ is the Beta function and $_{1}F_{2}\tbinom
{a_{1\,};\,z}{b_{1}\,,\,b_{2}}$ is a generalized hypergeometic function. \ As
will be seen below, these integrals are also expressible in terms of
quantities which involve only products of the Bessel functions $J_{0}(z)$, and
$J_{1}(z)$ or the Struve functions $\mathbf{H}_{0}(z),$ $\mathbf{H}_{1}(z)$,
and polynomials with arguments $1/z.$ That is to say%

\begin{subequations}
\begin{align}
&  \mathcal{J\,}(\kappa,\,j,a,z)\tag{C2}\\
&  =(-1)^{a+1}\tfrac{\pi}{2}[J_{0}(z)\,{\Large \wp}(\kappa,{\small 4j+2a+1}%
,z)+J_{1}(z)\,\mathcal{Q}(\kappa,{\small 4j+2a+1},z)],\nonumber\\
&  \mathcal{H}(\kappa,\,j,a,z)\tag{C3}\\
&  =({\small -1)}^{a}\tfrac{\pi}{2}[\mathbf{H}_{0}{\small (z)}{\Large \wp
}({\small \kappa,4j+2a+1,z}){\small +}\mathbf{H}_{1}{\small (z)\,}%
\mathcal{Q}({\small \kappa},{\small 4j+2a+1,z}){\small +}\tfrac{2}{\pi
}\mathcal{R(}{\small \kappa,4j+2a+1,z})],\nonumber
\end{align}
where the quantities ${\Large \wp,}\mathcal{Q}$, $\mathcal{R}$ are related to
the Lommel polynomials. \ In the latter forms the oscillatory behavior of
these integrals and those of the \textit{key} integrals mentioned above is
made manifest.

\subsubsection{Proof of The $\mathcal{J\,}(\kappa,\,j,a,z)$ and $\mathcal{H}%
(\kappa,\,j,a,z)$ expressions\bigskip}

Using the integral representations for Bessel functions of the first kind
$J_{\nu}(z)$ and the Struve functions $\mathbf{H}_{\nu}(z)$ \cite{NBS} i.e.
\end{subequations}
\[
\tfrac{\sqrt{\pi}\Gamma(\nu+1/2)}{2}\frac{J_{\nu}(z)}{\left(  z/2\right)
^{\nu}}=\int_{0}^{\pi/2}\sin^{2\nu}\theta\cos(z\cos\theta)\,d\theta,
\]%
\[
\tfrac{\sqrt{\pi}\Gamma(\nu+1/2)}{2}\frac{\mathbf{H}_{\nu}(z)}{(z/2)^{\nu}%
}=\int_{0}^{\pi/2}\sin^{2\nu}\theta\sin(z\cos\theta)\,d\theta,
\]
and differentiating these expressions with respect to $z,$ $4j+2a+1$ times
gives%
\begin{equation}
\mathcal{J}(\kappa,\,j,a,z)=(-1)^{a+1}\tfrac{\pi}{2}{\small (}2\kappa
+1{\small )!!}\frac{d^{\,\,4\,j+2\,a+1}[J_{\kappa+1}(z)/z^{\kappa+1}%
]}{d\,z^{\,4\,j+2\,a+1}}, \tag{C4}%
\end{equation}%
\begin{equation}
\mathcal{H}(\kappa,\,j,a,z)=(-1)^{a}\tfrac{\pi}{2}{\small (}2\kappa
+1{\small )!!}\frac{d^{\,\,4\,j+2\,a+1}[\mathbf{H}_{\kappa+1}(z)/z^{\kappa
+1}]}{d\,z^{\,4\,j+2\,a+1}}, \tag{C5}%
\end{equation}
where $(2\kappa+1)!!$ is the double factorial function i.e.%
\[
(2\kappa+1)!!=\prod_{i=1}^{\kappa}(2i+1)=\frac{2^{\kappa+1}}{\sqrt{\pi}}%
\Gamma(\kappa+3/2).
\]
We have by direct observation \cite{LAURENZI}that for $\nu\geq0,$
\begin{align}
(2\kappa+1)!!\frac{d^{\,\nu}}{d\,z^{\,\nu}}[\frac{J_{\kappa+1}(z)}%
{z^{\kappa+1}}]  &  ={\Large \wp}(\kappa,\nu,z)\,J_{0}(z)\,\,+\mathcal{Q}%
(\kappa,\nu,z)\,J_{1}(z)\,,\tag{C6}\\
(2\kappa+1)!!\frac{d^{\,\nu}}{d\,z^{\,\nu}}[\frac{\mathbf{H}_{\kappa+1}%
(z)}{z^{\kappa+1}}]  &  ={\Large \wp}(\kappa,\nu,z)\,\mathbf{H}_{0}%
(z)\,\,+\mathcal{Q}(\kappa,\nu,z)\,\mathbf{H}_{1}(z)+\tfrac{2}{\pi
}\,\mathcal{R(}\kappa,\nu,z), \tag{C7}%
\end{align}
Using (C4) and (C5) together with (C6) and (C7) we get (C2) and (C3).\ 

\subsection{Properties of the Polynomials ${\protect\LARGE \wp}(\kappa,\nu
,z)$, $\mathcal{Q}(\kappa,\nu,z)$, and $\mathcal{R(}\kappa,\nu,z)$}

The polynomials ${\Large \wp}(\kappa,\nu,z)$, $\mathcal{Q}(\kappa,\nu,z),$ and
$\mathcal{R(}\kappa,\nu,z)$ are interrelated by differential recurrence
relations (which follow from the expressions for $\frac{d^{\,\nu+1}%
}{d\,z^{\,\nu+1}}[\frac{J_{\kappa+1}(z)}{z^{\kappa+1}}]$ and $\frac{d^{\,\nu
+1}}{d\,z^{\,\nu+1}}[\frac{\mathbf{H}_{\kappa+1}(z)}{z^{\kappa+1}}]$ and the
linear independence of the Bessel $J_{0}$, $J_{1}$ and Struve functions
$\mathbf{H}_{0}$, $\mathbf{H}_{1}$), we have
\begin{align}
{\Large \wp}(\kappa,\nu+1,z)  &  =\frac{d{\Large \wp}(\kappa,\nu,z)}%
{dz}+\mathcal{Q}(\kappa,\nu,z),\tag{C8}\\
\mathcal{Q}(\kappa,\nu+1,z)  &  =\frac{d\mathcal{Q}(\kappa,\nu,z)}%
{dz}-{\Large \wp}(\kappa,\nu,z)-\frac{\mathcal{Q}(\kappa,\nu,z)}{z^{{}}%
},\nonumber\\
\mathcal{R(}\kappa,\nu+1,z)  &  =\frac{d\,\mathcal{R(}\kappa,\nu,z)}%
{dz}+{\Large \wp}(\kappa,\nu,z).\nonumber
\end{align}
with
\begin{align*}
{\Large \wp}(0,1,z)  &  =1/z,\\
\mathcal{Q}(0,1,z)  &  =-2/z^{2},\\
\mathcal{R(}0,1,z)  &  =0.
\end{align*}
In addition it follows from the differential equations defining $J_{\kappa
+1}(z)$ and $\mathbf{H}_{\kappa+1}(z)$ that the functions $J_{\kappa
+1}(z)/z^{\kappa+1}$ and $\mathbf{H}_{\kappa+1}(z)/z^{\kappa+1}$ satisfy the
differential equations\
\[
z\frac{d^{\,2}\left[  J_{\kappa+1}(z)/z^{\kappa+1}\right]  }{d\,z^{2}%
}+(2\kappa+3)\frac{d\,\left[  J_{\kappa+1}(z)/z^{\kappa+1}\right]  }%
{d\,z}+z[J_{\kappa+1}(z)/z^{\kappa+1}]=0,
\]%
\[
z\frac{d^{\,2}\left[  \mathbf{H}_{\kappa+1}(z)/z^{\kappa+1}\right]  }%
{d\,z^{2}}+(2\kappa+3)\frac{d\,\left[  \mathbf{H}_{\kappa+1}(z)/z^{\kappa
+1}\right]  }{d\,z}+z[\mathbf{H}_{\kappa+1}(z)/z^{\kappa+1}]=\frac{2}%
{\pi(2\kappa+1)!!\,}.
\]
Repeated differentiation of these relations gives for $\nu\geq3,$%
\[
z\frac{d^{\,\nu}\Im(\kappa,z)}{d\,z^{\nu}}+(2\kappa+\nu+1)\frac{d^{\,\nu-1}%
\Im(\kappa,z)}{d\,z^{\nu-1}}+z\frac{d^{\,\nu-2}\Im(\kappa,z)}{d\,z^{\nu-2}%
}+(\nu-2)\frac{d^{\,\nu-3}\Im(\kappa,z)}{d\,z^{\nu-3}}=0,
\]
where $\Im(\kappa,z)$ is either $J_{\kappa+1}(z)/z^{\kappa+1}$ or
$\mathbf{H}_{\kappa+1}(z)/z^{\kappa+1}.$ Using (C2) and (C3) it follows that
the coefficients of the functions $J_{0}(z)$, $\ J_{1}(z),$ $\mathbf{H}%
_{0}(z),$ and $\mathbf{H}_{1}(z)$ in the resulting relations vanish and we get
the pseudo 4th-order recurrence relations
\[
z\,F(\kappa,\nu,z)+(2\kappa+\nu+1)F(\kappa,\nu-1,z)+z\,F(\kappa,\nu
-2,z)+(\nu-2)F(\kappa,\nu-3,z)=0,\hspace{0.25in}\nu\geq3
\]
where $F(\kappa,\nu,z)$ stands for \textit{any} of the polynomials
${\Large \wp}(\kappa,\nu,z),$ $\mathcal{Q}(\kappa,\nu,z),$ $\mathcal{R(}%
\kappa,\nu,z)$. \ In the special case where $\nu<3$ the $\mathcal{R(}%
\kappa,\nu,z)$ polynomials are interrelated by%
\[
z\,\mathcal{R}(\kappa,2,z)+(2\kappa+3)\mathcal{R}(\kappa,1,z)+z\,\mathcal{R}%
(\kappa,0,z)=1.
\]

We also note that since%
\[
(2\kappa+3)!!\frac{d^{\,\nu}}{d\,z^{\,\nu}}\left[  \frac{J_{\kappa+2}%
(z)}{z^{\kappa+2}}\right]  =(2\kappa+1)!!\left\{  \frac{d^{\,\nu}}%
{d\,z^{\,\nu}}\left[  \frac{J_{\kappa+1}(z)}{z^{\kappa+1}}\right]
+\frac{d^{\,\nu+2}}{d\,z^{\,\nu+2}}\left[  \frac{J_{\kappa+1}(z)}{z^{\kappa
+1}}\right]  \right\}  ,
\]
(as well as the corresponding relation for the $\mathbf{H}_{\kappa
+1}(z)/z^{\kappa+1}$) it follows from the definition of the polynomial
${\Large \wp}(\kappa,\nu,z),$ $\mathcal{Q}(\kappa,\nu,z),$ $\mathcal{R(}%
\kappa,\nu,z)$, that they also satisfy partial difference equations in
$\kappa$ and $\nu,$ we have%
\[
F(\kappa+1,\nu,z)=F(\kappa,\nu,z)+F(\kappa,\nu+2,z).
\]

In summary, we will see that all of the integrals occurring above in the body
of the text can be expressed in terms of the Bessel and Struve functions of
orders \textit{zero} and \textit{one} together with the polynomials
${\Large \wp}(\kappa,\nu,z),$ $\mathcal{Q}(\kappa,\nu,z),$ $\mathcal{R(}%
\kappa,\nu,z)$ or the Lommel polynomials $R_{m,\nu}(z)$.

Below, the polynomials${\Large \wp,}$ $\mathcal{Q},$ and $\mathcal{R}$ and
their relation to the Lommel polynomials \cite{LOMMEL}$\ R_{m,\nu}(z)$ is
examined.$\medskip$

\subsection{${\protect\Large \wp}(\kappa,\nu,z),$ $\mathcal{Q}(\kappa,\nu,z),$
and $\mathcal{R(}\kappa,\nu,z)$ and the Lommel polynomials}

As will be seen below, the polynomials ${\Large \wp}(\kappa,\nu,z)$,
$\mathcal{Q}(\kappa,\nu,z),$ and $\mathcal{R(}\kappa,\nu,z)$ can be expressed
explicitly as sums of the Lommel polynomials $R_{m,\nu}(z)$ the latter being
given by
\begin{equation}
R_{m,\nu}(z)=\sum_{n=0}^{\lceil\frac{m}{2}\rceil}\frac{(-1)^{n}(m-n)!\Gamma
(\nu+m-n)}{n!(m-2n)!\Gamma(\nu+n)}(2/z)^{m-2n}. \tag{C9}%
\end{equation}
Using (C9), we see that the leading terms for the Lommel polynomials are
\begin{align*}
R_{2\kappa,\ \mu}(z)  &  \thicksim(-1)^{\kappa}[1-2\kappa(\kappa+1)(\kappa
+\mu-1)(\kappa+\mu)/z^{2}]+\cdots,\\
R_{2\kappa+1,\ \mu}(z)  &  \thicksim(-1)^{\kappa}2(\kappa+1)(\kappa
+\mu)/z+\cdots,
\end{align*}
expressions which will be useful in the sequel in obtaining the leading terms
of the ${\Large \wp}(\kappa,\nu,z)$, $\mathcal{Q}(\kappa,\nu,z),$ and
$\mathcal{R(}\kappa,\nu,z)$ polynomials. \ 

It is important to note that the $R_{m,\nu}(z)$ polynomials can also be
generated by the Bessel function relations \cite{Watson} first obtained by
Lommel i.e.%
\begin{equation}
J_{\mu+m}(z)=J_{\mu}(z)\ R_{m,\ \mu}(z)-J_{\mu-1}(z)\ R_{m-1,\ \mu+1}(z).
\tag{C10}%
\end{equation}
The corresponding relations involving the Struve functions while being more
complicated are given by \cite{BRYCHKOV}%
\begin{align}
\mathbf{H}_{\mu+m}(z)  &  =\mathbf{H}_{\mu}(z)\ R_{m,\ \mu}(z)-\mathbf{H}%
_{\mu-1}(z)\ R_{m-1,\ \mu+1}(z)\tag{C11}\\
&  +\frac{1}{\sqrt{\pi}}\left(  \frac{z}{2}\right)  ^{\mu+m-1}\sum_{j=0}%
^{m-1}\frac{R_{j,\ \mu+m-j}(z)}{\Gamma(\mu+m+1/2-j)}\left(  \frac{2}%
{z}\right)  ^{j}.\nonumber
\end{align}

Generalizing differential expressions due to Brychkov \cite{BRYCHKOV2} we have
the relations
\begin{equation}
\frac{d^{\,\nu}\,\left[  J_{\kappa+1}(z)/z^{\kappa+1}\right]  }{d\,z^{\,\nu}%
}=\frac{(-1)^{\nu}\nu!}{z^{\kappa+1}}\sum_{i=0}^{\lfloor\frac{\nu}{2}\rfloor
}\frac{(-1)^{i}J_{\kappa+1+\nu-i}(z)}{i!(\nu-2i)!(2z)^{i}} \tag{C12}%
\end{equation}%
\begin{align}
\frac{d^{\,\nu}\,\left[  \mathbf{H}_{\kappa+1}(z)/z^{\kappa+1}\right]
}{d\,z^{\,\nu}}  &  =\frac{(-1)^{\nu}\nu!}{z^{\kappa+1}}\sum_{i=0}%
^{\lfloor\frac{\nu}{2}\rfloor}\frac{(-1)^{i}\mathbf{H}_{\kappa+1+\nu-i}%
(z)}{i!(\nu-2i)!(2z)^{i}}\nonumber\\
&  +\frac{(-1)^{\nu+1}\nu!(z/2)^{\nu-1}}{\pi\ 2^{\kappa+1}}\sum_{i=0}%
^{\lfloor\frac{\nu}{2}\rfloor}\frac{(-1)^{i}}{i!(\nu-2i)!\ (z)^{2i}}\sum
_{j=0}^{\nu-1-i}\frac{\Gamma(j+1/2)}{\Gamma(\kappa+3/2+\nu-j-i)}(\frac{2}%
{z})^{2j}. \tag{C13}%
\end{align}

Using \ Eq. (C10) with $\mu=1$ and $\kappa$ replaced by $\kappa+\nu-i$ in Eqs.
(C12) and (C13) we get using Eqs. (C6) and (C7) the desired expressions for
the ${\Large \wp}(\kappa,\nu,z)$, $\mathcal{Q}(\kappa,\nu,z),$ and
$\mathcal{R(}\kappa,\nu,z)$ polynomials. (\textbf{computed in Lommelmarch4.mw}%
)
\begin{align*}
{\Large \wp}(\kappa,\nu,z)\,  &  =\frac{(-1)^{\nu+1}\nu!(2\kappa
+1)!!}{z^{\kappa+1}}\sum_{i=0}^{\lfloor\frac{\nu}{2}\rfloor}\frac{(-1)^{i}%
R_{\kappa+\nu-1-i},_{2}(z)}{i!(\nu-2i)!(2z)^{i}},\\
\mathcal{Q}(\kappa,\nu,z)  &  =\frac{(-1)^{\nu}\nu!(2\kappa+1)!!}{z^{\kappa
+1}}\sum_{i=0}^{\lfloor\frac{\nu}{2}\rfloor}\frac{(-1)^{i}R_{\kappa+\nu
-i},_{1}(z)}{i!(\nu-2i)!(2z)^{i}},\\
\mathcal{R(}\kappa,\nu,z)  &  =\frac{(-z/2)^{\nu-1}\,\nu!(2\kappa
+1)!!}{2^{\kappa+2}}[\sum_{m=0}^{\lfloor\frac{\nu}{2}\rfloor}\frac{c(\nu
,m,m)\,}{\Gamma(\kappa+\nu+3/2-m)}(\frac{2}{z})^{2m}\\
&  +\sum_{m=\,\lfloor\frac{\nu}{2}\rfloor+1}^{\nu-1}\frac{c(\nu,m,\lfloor
\frac{\nu}{2}\rfloor)\,}{\Gamma(\kappa+\nu+3/2-m)}(\frac{2}{z})^{2m}\\
&  -\sum_{i=0}^{\lfloor\frac{\nu}{2}\rfloor}\frac{(-1/z^{2})^{i}}{i!(\nu
-2i)!}\sum_{j=0}^{\kappa+\nu-1-i}\frac{\sqrt{\pi}R_{j,\,\kappa+\nu
+1-i-j}(z)\,}{\Gamma(\kappa+\nu+3/2-i-j)}(\frac{2}{z})^{\,j}],
\end{align*}
where the coefficients $c(\nu,m,N)$ are given by
\[
c(\nu,m,N)=\sum_{j=0}^{N}\frac{(-1/4)^{\,j}\Gamma(m+1/2-j)}{j!(\nu-2j)!}.
\]
\ \bigskip

In the case of ${\Large \wp}(\kappa,\nu,z)$, and $\mathcal{Q}(\kappa,\nu,z)$
it is interesting to note that only higher powers of $1/z$ occur. The leading
terms of these polynomials are then given by%
\begin{align*}
{\Large \wp}(2\kappa,2\mu+1,z)  &  \thicksim(-1)^{\kappa+\mu}(4\kappa
+1)!!/z^{2\kappa+1},\\
{\Large \wp}(2\kappa+1,2\mu+1,z)  &  \thicksim(-1)^{\kappa+\mu}(4\kappa
+3)!!\{2\kappa^{2}+2\kappa\lbrack2\mu+3]+7\mu+4\}/z^{2\kappa+3},\\
{\Large \wp}(2\kappa,2\mu,z)  &  \thicksim(-1)^{\kappa+\mu}(4\kappa
+1)!!\{2\kappa^{2}+2\kappa\lbrack2\mu+1]+3\mu\}/z^{2\kappa+2},\\
{\Large \wp}(2\kappa+1,2\mu,z)  &  \thicksim(-1)^{\kappa+\mu+1}(4\kappa
+3)!!/z^{2\kappa+2},
\end{align*}
and%
\begin{align*}
\mathcal{Q}(2\kappa,2\mu+1,z)  &  \thicksim(-1)^{\kappa+\mu+1}(4\kappa
+1)!!\{2\kappa^{2}+4\kappa\lbrack\mu+1]+3\mu+2\}/z^{2\kappa+2},\\
\mathcal{Q}(2\kappa+1,2\mu+1,z)  &  \thicksim(-1)^{\kappa+\mu}(4\kappa
+3)!!/z^{2\kappa+2},\\
\mathcal{Q}(2\kappa,2\mu,z)  &  \thicksim(-1)^{\kappa+\mu}(4\kappa
+1)!!/z^{2\kappa+1},\\
\mathcal{Q}(2\kappa+1,2\mu,z)  &  \thicksim(-1)^{\kappa+\mu}(4\kappa
+3)!!\{2\kappa^{2}+4\kappa\lbrack\mu+1]+5\mu+2\}/z^{2\kappa+3}.
\end{align*}
In the case of the polynomials $\mathcal{R(}\kappa,\mu,z)$ the leading terms
are more difficult to obtain. \ Using the differential difference equations
for the ${\Large \wp}(\kappa,\nu,z)$, $\mathcal{Q}(\kappa,\nu,z),$ and
$\mathcal{R(}\kappa,\nu,z)$ polynomials we get
\begin{align*}
\mathcal{R(}\kappa,\nu,z)  &  =\frac{d^{\,\nu-1}\mathcal{R(}\kappa
,1,z)}{d\,z^{\,\nu-1}}+\sum_{j=1}^{\nu-1}\frac{d^{\,\nu-1-j}{\Large \wp
}\mathcal{(}\kappa,j,z)}{d\,z^{\,\nu-1-j}},\\
\mathcal{R(}\kappa,\nu,z)  &  \thicksim(-1)^{\nu}\frac{\nu!}{z^{\nu+1}%
}+{\Large \wp}\mathcal{(}\kappa,\nu-1,z)+\frac{d^{\,}{\Large \wp}%
\mathcal{(}\kappa,\nu-2,z)}{d\,z^{\,}}+\cdots,\nu>2,\\
\mathcal{R(}\kappa,\nu,z)  &  \thicksim(-1)^{\nu}\frac{\nu!}{z^{\nu+1}%
}+2{\Large \wp}\mathcal{(}\kappa,\nu-1,z)-\mathcal{Q(}\kappa,\nu-2,z).
\end{align*}
Using the leading term expressions for the ${\Large \wp}(\kappa,\nu,z)$, and
$\mathcal{Q}(\kappa,\nu,z)$ polynomials we get
\[
\mathcal{R(}2\kappa+1,2\mu,z)\thicksim\left\{
\begin{tabular}
[c]{c}%
$(2\mu)!/z^{2\mu+1},\hspace{0.25in}\mu\leq\kappa,$\\
$(-1)^{\kappa+\mu+1}(4\kappa+3)!!\{2\kappa^{2}+4\kappa(\mu+1)+5\mu
+1\}/z^{2\kappa+3},\hspace{0.25in}\mu>\kappa,$%
\end{tabular}
\right.  ,
\]%
\[
\mathcal{R(}2\kappa+1,2\mu+1,z)\thicksim\left\{
\begin{tabular}
[c]{c}%
$-(2\mu+1)!/z^{2\mu+2},\hspace{0.25in}\mu<\kappa,$\\
$-[(4\kappa+3)!!+(2\kappa+1)!]/z^{2\kappa+2},\hspace{0.25in}\mu=\kappa,$\\
$(-1)^{\kappa+\mu+1}(4\kappa+3)!!/z^{2\kappa+2},\hspace{0.25in}\mu>\kappa,$%
\end{tabular}
\right.  ,
\]

\bigskip%

\[
\mathcal{R(}2\kappa,2\mu+1,z)\thicksim\left\{
\begin{tabular}
[c]{c}%
$-(2\mu+1)!/z^{2\mu+2},\hspace{0.25in}\mu<\kappa,$\\
$\lbrack(4\kappa+1)!!\{6\kappa^{2}+7\kappa+1\}-(2\kappa+1)!]/z^{2\kappa
+2},\hspace{0.25in}\mu=\kappa,$\\
$(-1)^{\kappa+\mu}(4\kappa+1)!!\{2\kappa^{2}+4\kappa(\mu+1)+3\mu
+1\}/z^{2\kappa+2},\hspace{0.25in}\mu>\kappa,$%
\end{tabular}
\right.  ,
\]

\bigskip%

\[
\mathcal{R(}2\kappa,2\mu,z)\thicksim\left\{
\begin{tabular}
[c]{c}%
$(2\mu)!/z^{2\mu+1},\hspace{0.25in}\mu<\kappa,$\\
$-[(4\kappa+1)!!-(2\kappa)!]/z^{2\kappa+1},\hspace{0.25in}\mu=\kappa,$\\
$(-1)^{\kappa+\mu+1}(4\kappa+1)!!/z^{2\kappa+1},\hspace{0.25in}\mu>\kappa,$%
\end{tabular}
\right.  .
\]

\ 

\subsubsection{The Integrals $\ \mathcal{C}_{cc},\mathcal{S}_{ss}%
,\mathcal{C}_{cs},\mathcal{S}_{sc}$ and $T_{cc},T_{ss},T_{sc},T_{cs}$}

\bigskip

Recalling the integrals $\mathcal{C}_{cc},\mathcal{S}_{ss},\mathcal{C}%
_{cs},\mathcal{S}_{sc}$ defined above .i.e.
\begin{align}
\mathcal{C}_{cc}\mathcal{(}\kappa,z,\varrho)  &  =\int_{0}^{\pi/2}%
\sin^{2\kappa+2\,}\theta\cos\theta\cos(z\cos\theta)\cos(\varrho\cos^{2}%
\theta)d\theta,\tag{C14}\\
\mathcal{S}_{ss}\mathcal{(}\kappa,z,\varrho)  &  =\int_{0}^{\pi/2}%
\sin^{2\kappa+2\,}\theta\cos\theta\sin(z\cos\theta)\sin(\varrho\cos^{2}%
\theta)d\theta,\nonumber\\
\mathcal{C}_{cs}\mathcal{(}\kappa,z,\varrho)  &  =\int_{0}^{\pi/2}%
\sin^{2\kappa+2\,}\theta\cos\theta\cos(z\cos\theta)\sin(\varrho\cos^{2}%
\theta)d\theta,\nonumber\\
\mathcal{S}_{sc}\mathcal{(}\kappa,z,\varrho)  &  =\int_{0}^{\pi/2}%
\sin^{2\kappa+2\,}\theta\cos\theta\sin(z\cos\theta)\cos(\varrho\cos^{2}%
\theta)d\theta,\nonumber
\end{align}
we note that the $\cos(\varrho\cos^{2}\theta)$ and $\sin(\varrho\cos^{2}%
\theta)$ terms appearing in these integrals can be expanded in terms of Bessel
functions \cite{BESSELEXPANSION} using the well known relations i.e.
\begin{align*}
\cos(\varrho\cos\phi)  &  =J_{0}(\varrho)+2\sum_{\kappa=1}^{\infty
}(-1)^{\kappa}J_{2\kappa}(\varrho)\cos(2\kappa\phi),\\
\sin(\varrho\cos\phi)  &  =2\sum_{\kappa=0}^{\infty}(-1)^{\kappa}J_{2\kappa
+1}(\varrho)\cos([2\kappa+1]\phi).
\end{align*}
The former can be written as powers of the $\cos^{2}\theta$ as
\begin{align*}
\cos(\varrho\cos^{2}\theta)  &  =J_{0}(\varrho)+2\sum_{\kappa=1}^{\infty
}(-1)^{\kappa}J_{2\kappa}(\varrho)T_{2\kappa}(\cos^{2}\theta),\\
\sin(\varrho\cos^{2}\theta)  &  =2\sum_{\kappa=0}^{\infty}(-1)^{\kappa
}J_{2\kappa+1}(\varrho)T_{2\kappa+1}(\cos^{2}\theta).
\end{align*}
where $T_{n}(x)$ are the Chebychev polynomials \cite{CHEBYCHEV} of the first
kind and are explicitly given by%

\begin{align*}
T_{0}(x)  &  =1,\text{ \ \ }\\
T_{n}(x)  &  =\frac{n}{2}\sum_{j=0}^{\lfloor\frac{n}{2}\rfloor}(-1)^{j}%
\frac{(n-j-1)!}{j!(n-2j)!}(2x)^{n-2j}\text{ \ for \ }n\geq1.
\end{align*}
Using these expressions and Eq. (26) and having interchanged the order of the
summations we get
\begin{align*}
\cos(\varrho\cos^{2}\theta)  &  =J_{0}(\varrho)+2\sum_{\kappa=1}^{\infty
}J_{2\kappa}(\varrho)\\
&  +\sum_{m=1}^{\infty}\frac{(-1)^{m}2^{2m}}{(2m)!}\cos^{4m}\theta\left\{
\sum_{\kappa=m}^{\infty}\frac{2\kappa(\kappa+m-1)!}{(\kappa-m)!}J_{2\kappa
}(\varrho)\right\}  ,\\
\sin(\varrho\cos^{2}\theta)  &  =\sum_{m=0}^{\infty}\frac{(-1)^{m}2^{2m+1}%
}{(2m+1)!}\cos^{4m+2}\theta\left\{  \sum_{\kappa=m}^{\infty}\frac{(2\kappa
+1)(\kappa+m)!}{(\kappa-m)!}J_{2\kappa+1}(\varrho)\right\}  .
\end{align*}
The sums containing the Bessel functions can be further reduced to terms
containing $J_{0}(\varrho)$ and $J_{1}(\varrho)$ and the Lommel polynomials
i.e.
\begin{align*}
\sum_{\kappa=m}^{\infty}\frac{2\kappa(\kappa+m-1)!}{(\kappa-m)!}J_{2\kappa
}(\varrho)  &  =J_{1}(\varrho)\,\sigma_{c}^{(1)}(m,\varrho)-J_{0}%
(\varrho)\,\sigma_{c}^{(2)}(m,\varrho),\\
\sum_{\kappa=m}^{\infty}\frac{(2\kappa+1)(\kappa+m)!}{(\kappa-m)!}%
J_{2\kappa+1}(\varrho)  &  =J_{1}(\varrho)\,\sigma_{s}^{(1)}(m,\varrho
)-J_{0}(\varrho)\,\sigma_{s}^{(2)}(m,\varrho),
\end{align*}
and in the sums below as $N$ approaches $\infty$ they converge rapidly and $N$
can safely be set to 4. \ The required $\sigma$ polynomials together with
their leading term expressions are then given by (for $N\geq m$ )
\begin{align*}
\sigma_{c}^{(1)}(m,\varrho)  &  =\sum_{\kappa=m}^{N}\tfrac{2\kappa
\,(\kappa+m-1)!}{(\kappa-m)!}R_{2\kappa-1,1}(\varrho)\thicksim\tfrac{2}{\zeta
}(-1)^{N+1}[N(N+1)-m]\tfrac{(N+m)!}{(N-m)!},\\
\sigma_{c}^{(2)}(m,\varrho)  &  =\sum_{\kappa=m}^{N}\tfrac{2\kappa
\,(\kappa+m-1)!}{(\kappa-m)!}R_{2\kappa-2,2}(\varrho)\thicksim(-1)^{N+1}%
\frac{(N+m)!}{(N-m)!},\\
\sigma_{s}^{(1)}(m,\varrho)  &  =\sum_{\kappa=m}^{N}\tfrac{(2\kappa
+1)\,(\kappa+m)!}{(\kappa-m)!}R_{2\kappa,1}(\varrho)\thicksim(-1)^{N}%
\frac{(N+m+1)!}{(N-m)!},\\
\sigma_{s}^{(2)}(m,\varrho)  &  =\sum_{\kappa=m}^{N}\tfrac{(2\kappa
+1)\,(\kappa+m)!}{(\kappa-m)!}R_{2\kappa-1,2}(\varrho)\thicksim\tfrac{2}%
{\zeta}(-1)^{N+1}[N(N+2)-m]\tfrac{(N+m+1)!}{(N-m)!},\\
\widehat{\sigma}_{c}^{(1)}(\varrho)  &  =\sum_{\kappa=1}^{N}R_{2\kappa
-1,1}(\varrho)\thicksim\tfrac{1}{\zeta}(-1)^{N+1}N(N+1),\\
\widehat{\sigma}_{c}^{(2)}(\varrho)  &  ={\small -1/2+}\sum_{\kappa=1}%
^{N}R_{2\kappa-2,2}(\varrho)\thicksim\frac{1}{2}(-1)^{N+1}.
\end{align*}

Combining the terms above we have
\begin{align*}
\cos(\varrho\cos^{2}\theta)  &  =J_{1}(\varrho)\left[  2\widehat{\sigma}%
_{c}^{(1)}(\varrho)+\sum_{m=1}^{\infty}\frac{(-4)^{m}}{(2m)!}\sigma_{c}%
^{(1)}(m,\varrho)\cos^{4m}\theta\right] \\
&  -J_{0}(\varrho)\left[  2\widehat{\sigma}_{c}^{(2)}(\varrho)+\sum
_{m=1}^{\infty}\frac{(-4)^{m}}{(2m)!}\sigma_{c}^{(2)}(m,\varrho)\cos
^{4m}\theta\right]  ,
\end{align*}
and%
\begin{align*}
\sin(\varrho\cos^{2}\theta)  &  =2\,J_{1}(\varrho)\left[  \sum_{m=0}^{\infty
}\frac{(-4)^{m}}{(2m+1)!}\sigma_{s}^{(1)}(m,\varrho)\cos^{4m+2}\,\theta\right]
\\
&  -2\,J_{0}(\varrho)\left[  \sum_{m=0}^{\infty}\frac{(-4)^{m}}{(2m+1)!}%
\sigma_{s}^{(2)}(m,\varrho)\cos^{4m+2}\,\theta\right]  .
\end{align*}
The integrals in Eq. (28) in the main text\ can then be rewritten in terms of
the $\mathcal{J}$ and $\mathcal{H}$ integrals as (The sums over $m$ and
$\kappa$ converge rapidly and the upper limits can be replaced with the first
four terms i.e. $0\leq m\leq4$ , and $m\leq\kappa\leq4$\ with accuracy in the
8th place.) \
\begin{align*}
\mathcal{C}_{cc}\mathcal{(}\kappa,z,\varrho)  &  =J_{1}(\varrho)\left[
2\widehat{\sigma}_{c}^{(1)}(\varrho)\mathcal{H}(\kappa,0,0,z)+\sum
_{m=1}^{\infty}\tfrac{(-4)^{m}}{(2m)!}\sigma_{c}^{(1)}(m,\varrho
)\,\mathcal{H}(\kappa,m,0,z)\right] \\
&  -J_{0}(\varrho)\left[  2\widehat{\sigma}_{c}^{(2)}(\varrho)\mathcal{H}%
(\kappa,0,0,z)+\sum_{m=1}^{\infty}\tfrac{(-4)^{m}}{(2m)!}\sigma_{c}%
^{(2)}(m,\varrho)\,\mathcal{H}(\kappa,m,0,z)\right]  ,\\
& \\
\mathcal{S}_{ss}\mathcal{(}\kappa,z,\varrho)  &  =2J_{1}(\varrho)\sum
_{m=0}^{\infty}\tfrac{(-4)^{m}}{(2m+1)!}\sigma_{s}^{(1)}(m,\varrho
)\mathcal{J}(\kappa,m,1,z)\\
&  -2J_{0}(\varrho)\sum_{m=0}^{\infty}\tfrac{(-4)^{m}}{(2m+1)!}\sigma
_{s}^{(2)}(m,\varrho)\mathcal{J}(\kappa,m,1,z),\\
& \\
\mathcal{C}_{cs}\mathcal{(}\kappa,z,\varrho)  &  =2J_{1}(\varrho)\sum
_{m=0}^{\infty}\tfrac{(-4)^{m}}{(2m+1)!}\sigma_{s}^{(1)}(m,\varrho
)\mathcal{H}(\kappa,m,1,z)\\
&  -2J_{0}(\varrho)\sum_{m=0}^{\infty}\tfrac{(-4)^{m}}{(2m+1)!}\sigma
_{s}^{(2)}(m,\varrho)\mathcal{H}(\kappa,m,1,z),\\
& \\
\mathcal{S}_{sc}\mathcal{(}\kappa,z,\varrho)  &  =J_{1}(\varrho)\left[
2\widehat{\sigma}_{c}^{(1)}(\varrho)\mathcal{J}(\kappa,0,0,z)+\sum
_{m=1}^{\infty}\tfrac{(-4)^{m}}{(2m)!}\sigma_{c}^{(1)}(m,\varrho
)\,\mathcal{J}(\kappa,m,0,z)\right] \\
&  -J_{0}(\varrho)\left[  2\widehat{\sigma}_{c}^{(2)}(\varrho)\mathcal{J}%
(\kappa,0,0,z)-\sum_{m=1}^{\infty}\tfrac{(-4)^{m}}{(2m)!}\sigma_{c}%
^{(2)}(m,\varrho)\,\mathcal{J}(\kappa,m,0,z)\right]  .
\end{align*}
where the integrals $\mathcal{J}(\kappa,\,j,a,z)$ and $\mathcal{H}%
(\kappa,\,j,a,z)$ are given by Eq. (23) \ 

The $\mathcal{C}_{cc}\mathcal{(}\kappa,z,\varrho),\mathcal{S}_{ss}%
\mathcal{(}\kappa,z,\varrho),\mathcal{C}_{cs}\mathcal{(}\kappa,z,\varrho
),\mathcal{S}_{sc}\mathcal{(}\kappa,z,\varrho)$ integrals can also be written
in terms of the Bessel and Struve functions using Eqs (23). \ We have,
\begin{subequations}
\begin{align}
\mathcal{C}_{cc}\mathcal{(}\kappa,z,\varrho)  &  =\pi J_{1}(\varrho
)[\mathbf{H}_{0}(z)\widehat{\,\mathbf{P}}_{c}^{(1)}\mathbf{(}\kappa
,0,z,\varrho\mathbf{)+H}_{1}(z)\,\widehat{\mathbf{Q}}_{c}^{(1)}\mathbf{(}%
\kappa,0,z,\varrho\mathbf{)+}\frac{2}{\pi}\widehat{\mathbf{R}}_{c}%
^{(1)}\mathbf{(}\kappa,0,z,\varrho\mathbf{)]}\tag{C15}\\
&  -\pi J_{0}(\varrho)[\mathbf{H}_{0}(z)\,\widehat{\mathbf{P}}_{c}%
^{(2)}{\small (}\kappa,0,z,\varrho{\small )}\mathbf{+H}_{1}(z)\,\widehat
{\mathbf{Q}}_{c}^{(2)}{\small (}\kappa,0,z,\varrho{\small )}\mathbf{+}%
\frac{2}{\pi}\widehat{\mathbf{R}}_{c}^{(2)}\mathbf{(}\kappa,0,z,\varrho
\mathbf{)]},\nonumber\\
& \nonumber\\
\mathcal{S}_{ss}\mathcal{(}\kappa,z,\varrho)  &  =\pi J_{1}(\varrho
)[J_{0}(z)\,\mathbf{P}_{s}^{(1)}\mathbf{(}\kappa,1,z,\varrho\mathbf{)+\,}%
J_{1}(z)\,\mathbf{Q}_{s}^{(1)}\mathbf{(}\kappa,1,z,\varrho\mathbf{)]}%
\nonumber\\
&  -\pi J_{0}(\varrho)\,[J_{0}(z)\,\mathbf{P}_{s}^{(2)}\mathbf{(}%
\kappa,1,z,\varrho\mathbf{)}+\mathbf{\,}J_{1}(z)\,\,\mathbf{Q}_{s}%
^{(2)}\mathbf{(}\kappa,1,z,\varrho\mathbf{)]},\tag{C16}\\
& \nonumber\\
\mathcal{C}_{cs}\mathcal{(}\kappa,z,\varrho)  &  =-\pi J_{1}(\varrho
)[\mathbf{H}_{0}(z)\,\mathbf{P}_{s}^{(1)}\mathbf{(}\kappa,1,z,\varrho
\mathbf{)+H}_{1}(z)\,\mathbf{Q}_{s}^{(1)}\mathbf{(}\kappa,1,z,\varrho
)\mathbf{+}\frac{2}{\pi}\mathbf{R}_{s}^{(1)}\mathbf{(}\kappa,1,z,\varrho
\mathbf{)]}\nonumber\\
&  +\pi J_{0}(\varrho)[\mathbf{H}_{0}(z)\,\mathbf{P}_{s}^{(2)}\mathbf{(}%
\kappa,1,z,\varrho\mathbf{)+\mathbf{H}_{1}}(z)\mathbf{\,\mathbf{Q}}_{s}%
^{(2)}(\kappa,1,z,\varrho)\mathbf{+}\frac{2}{\pi}\mathbf{R}_{s}^{(2)}%
\mathbf{(}\kappa,1,z,\varrho\mathbf{)],}\tag{C17}\\
& \nonumber\\
\mathcal{S}_{sc}\mathcal{(}\kappa,z,\varrho)  &  =-\pi J_{1}(\varrho
)[J_{0}(z)\,\widehat{\mathbf{P}}_{c}^{(1)}\mathbf{(}\kappa,0,z,\varrho
\mathbf{)+\,}J_{1}(z)\,\widehat{\mathbf{Q}}_{c}^{(1)}\mathbf{(}\kappa
,0\mathbf{,}z,\varrho\mathbf{)]}\nonumber\\
&  +\pi J_{0}(\varrho)[J_{0}(z)\,\widehat{\mathbf{P}}_{c}^{(2)}\mathbf{(}%
\kappa,0,z,\varrho\mathbf{)+\,}J_{1}(z)\,\widehat{\mathbf{Q}}_{c}%
^{(2)}\mathbf{(}\kappa,0\mathbf{,}z,\varrho\mathbf{)].} \tag{C18}%
\end{align}
where
\end{subequations}
\begin{align*}
\mathbf{P}_{s}^{(\eta)}\mathbf{(}\kappa,a,z,\varrho\mathbf{)}  &
\mathbf{=}\sum_{i=0}^{\infty}\frac{(-1)^{i}2^{2i}}{(2i+1)!}\sigma_{s}^{(\eta
)}(i,\varrho){\Large \wp}(\kappa,4i+2a+1,z),\\
\mathbf{Q}_{s}^{(\eta)}\mathbf{(}\kappa\mathbf{,}a\mathbf{,}z,\varrho
\mathbf{)}  &  \mathbf{=}\sum_{i=0}^{\infty}\frac{(-1)^{i}2^{2i}}%
{(2i+1)!}\sigma_{s}^{(\eta)}(i,\varrho)\,\mathcal{Q\,}(\kappa,4i+2a+1,z),\\
\mathbf{R}_{s}^{(\eta)}\mathbf{(}\kappa,a,z,\varrho\mathbf{)}  &
\mathbf{=}\sum_{i=0}^{\infty}\frac{(-1)^{i}2^{2i}}{(2i+1)!}\sigma_{s}^{(\eta
)}(i,\varrho)\,\mathcal{R}(\kappa,4i+2a+1,z),
\end{align*}
and%
\begin{align*}
\mathbf{P}_{c}^{(\eta)}\mathbf{(}\kappa,a,z,\varrho\mathbf{)}  &
\mathbf{=}\sum_{i=1}^{\infty}\frac{(-1)^{i}2^{2i}}{(2i)!}\sigma_{c}^{(\eta
)}(i,\varrho){\Large \wp}(\kappa,4i+2a+1,z),\\
\mathbf{Q}_{c}^{(\eta)}\mathbf{(}\kappa\mathbf{,}a\mathbf{,}z,\varrho
\mathbf{)}  &  \mathbf{=}\sum_{i=1}^{\infty}\frac{(-1)^{i}2^{2i}}{(2i)!}%
\sigma_{c}^{(\eta)}(i,\varrho)\,\mathcal{Q\,}(\kappa,4i+2a+1,z),\\
\mathbf{R}_{c}^{(\eta)}\mathbf{(}\kappa,a,z,\varrho\mathbf{)}  &
\mathbf{=}\sum_{i=1}^{\infty}\frac{(-1)^{i}2^{2i}}{(2i)!}\sigma_{c}^{(\eta
)}(i,\varrho)\,\mathcal{R}(\kappa,4i+2a+1,z),
\end{align*}%
\begin{align*}
\widehat{\mathbf{P}}_{c}^{(\eta)}{\small (}\kappa,0,z,\varrho{\small )}  &
{\small =}\tfrac{1}{2}\,\mathbf{P}_{c}^{(\eta)}\mathbf{(}\kappa,0,z,\varrho
\mathbf{)+}\widehat{\sigma}_{c}^{(\eta)}(\varrho){\Large \wp}(\kappa,1,z),\\
\,\widehat{\mathbf{Q}}_{c}^{(\eta)}{\small (}\kappa,0,z,\varrho{\small )}  &
=\tfrac{1}{2}\,\mathbf{Q}_{c}^{(\eta)}\mathbf{(}\kappa,0,z,\varrho
\mathbf{)+}\widehat{\sigma}_{c}^{(\eta)}\mathbf{(}\varrho)\mathcal{Q\,}%
(\kappa,1,z),\\
\widehat{\mathbf{R}}_{c}^{(\eta)}{\small (}\kappa,0,z,\varrho{\small )}  &
{\small =}\tfrac{1}{2}\mathbf{R}_{c}^{(\eta)}\mathbf{(}\kappa,0,z,\varrho
\mathbf{)+}\widehat{\sigma}_{c}^{(\eta)}(\varrho)\mathcal{R}(\kappa,1,z),
\end{align*}
and where\ $\eta$ is $1$ or $2$.

The leading terms for $\mathbf{P}_{s}^{(\eta)}\mathbf{(}\kappa,a,z,\varrho
\mathbf{)}$ , $\mathbf{P}_{c}^{(\eta)}\mathbf{(}\kappa,a,z,\varrho\mathbf{)}$
and $\mathbf{Q}_{s}^{(\eta)}\mathbf{(}\kappa\mathbf{,}a\mathbf{,}%
z,\varrho\mathbf{)}$ , $\mathbf{Q}_{c}^{(\eta)}\mathbf{(}\kappa\mathbf{,}%
a\mathbf{,}z,\varrho\mathbf{)}$\ are%
\begin{align*}
\mathbf{P}_{s}^{(1)}\mathbf{(}2\kappa,a,z,\varrho\mathbf{)}  &  \thicksim
\mathbf{\,}5\tfrac{(-1)^{\kappa+a}(4\kappa+1)!!}{z^{2\kappa+1}},\\
\mathbf{P}_{s}^{(1)}\mathbf{(}2\kappa+1,a,z,\varrho\mathbf{)}  &
\thicksim\tfrac{(-1)^{\kappa+a}(4\kappa+3)!!}{z^{2\kappa+3}}[10\kappa
^{2}+\kappa(20a+670)+25a+820],\\
\mathbf{P}_{s}^{(2)}\mathbf{(}2\kappa,a,z,\varrho\mathbf{)}  &  \thicksim
\mathbf{\,}80\tfrac{(-1)^{\kappa+a+1}(4\kappa+1)!!}{\varrho\,z^{2\kappa+1}},\\
\mathbf{P}_{s}^{(2)}\mathbf{(}2\kappa+1,a,z,\varrho\mathbf{)}  &
\thicksim2\tfrac{(-1)^{\kappa+a+1}(4\kappa+3)!!}{\varrho\,z^{2\kappa+3}%
}[10\kappa^{2}+\kappa(20a+670)+25a+820],
\end{align*}%
\begin{align*}
\mathbf{P}_{c}^{(1)}\mathbf{(}2\kappa,a,z,\varrho\mathbf{)}  &  \thicksim
\mathbf{\,}80\tfrac{(-1)^{\kappa+a+1}(4\kappa+1)!!}{\varrho\,z^{2\kappa+1}},\\
\mathbf{P}_{c}^{(1)}\mathbf{(}2\kappa+1,a,z,\varrho\mathbf{)}  &
\thicksim2\tfrac{(-1)^{\kappa+a+1}(4\kappa+3)!!}{\varrho\;z^{2\kappa+3}%
}[80\kappa^{2}+\kappa(160a+11568)+200a+14320],\\
\mathbf{P}_{c}^{(2)}\mathbf{(}2\kappa,a,z,\varrho\mathbf{)}  &  \thicksim
8\tfrac{(-1)^{\kappa+a+1}(4\kappa+1)!!}{\,z^{2\kappa+1}},\\
\mathbf{P}_{c}^{(2)}\mathbf{(}2\kappa+1,a,z,\varrho\mathbf{)}  &
\thicksim\tfrac{(-1)^{\kappa+a+1}(4\kappa+3)!!}{\,z^{2\kappa+3}}[16\kappa
^{2}+\kappa(32a+1008)+40a+1232],
\end{align*}%
\begin{align*}
\widehat{\mathbf{P}}_{c}^{(1)}\mathbf{(}2\kappa,0,z,\varrho\mathbf{)}  &
\thicksim\mathbf{\,}60\tfrac{(-1)^{\kappa+1}(4\kappa+1)!!}{\varrho
\,z^{2\kappa+1}},\\
\widehat{\mathbf{P}}_{c}^{(1)}\mathbf{(}2\kappa+1,0,z,\varrho\mathbf{)}  &
\thicksim\tfrac{(-1)^{\kappa+1}(4\kappa+3)!!}{\varrho\;z^{2\kappa+3}%
}[120\kappa^{2}+11574\kappa+14324],\\
\widehat{\mathbf{P}}_{c}^{(2)}\mathbf{(}2\kappa,0,z,\varrho\mathbf{)}  &
\thicksim\frac{9}{2}\tfrac{(-1)^{\kappa+1}(4\kappa+1)!!}{\,z^{2\kappa+1}},\\
\widehat{\mathbf{P}}_{c}^{(2)}\mathbf{(}2\kappa+1,0,z,\varrho\mathbf{)}  &
\thicksim\tfrac{(-1)^{\kappa+1}(4\kappa+3)!!}{\,z^{2\kappa+3}}[9\kappa
^{2}+511\kappa+620],
\end{align*}%

\begin{align*}
\mathbf{Q}_{s}^{(1)}\mathbf{(}2\kappa\mathbf{,}a\mathbf{,}z,\varrho\mathbf{)}
&  \thicksim\tfrac{(-1)^{\kappa+a+1}(4\kappa+1)!!}{\,z^{2\kappa+2}}%
[10\kappa^{2}+\kappa(20a+660)+15a+490],\\
\mathbf{Q}_{s}^{(1)}\mathbf{(}2\kappa+1\mathbf{,}a\mathbf{,}z,\varrho
\mathbf{)}  &  \thicksim5\tfrac{(-1)^{\kappa+a}(4\kappa+3)!!}{\,z^{2\kappa+2}%
},\\
\mathbf{Q}_{s}^{(2)}\mathbf{(}2\kappa\mathbf{,}a\mathbf{,}z,\varrho\mathbf{)}
&  \thicksim2\tfrac{(-1)^{\kappa+a}(4\kappa+1)!!}{\varrho\,\,z^{2\kappa+2}%
}[80\kappa^{2}+\kappa(160a+9504)+120a+7088],\\
\mathbf{Q}_{s}^{(2)}\mathbf{(}2\kappa+1\mathbf{,}a\mathbf{,}z,\varrho
\mathbf{)}  &  \thicksim80\tfrac{(-1)^{\kappa+a+1}(4\kappa+3)!!}%
{\varrho\,\,z^{2\kappa+2}},
\end{align*}%
\begin{align*}
\mathbf{Q}_{c}^{(1)}\mathbf{(}2\kappa\mathbf{,}a\mathbf{,}z,\varrho\mathbf{)}
&  \thicksim2\tfrac{(-1)^{\kappa+a}(4\kappa+1)!!}{\varrho\,\,z^{2\kappa+2}%
}[80\kappa^{2}+\kappa(160a+9464)+120a+8576],\\
\mathbf{Q}_{c}^{(1)}\mathbf{(}2\kappa+1\mathbf{,}a\mathbf{,}z,\varrho
\mathbf{)}  &  \thicksim80\tfrac{(-1)^{\kappa+a+1}(4\kappa+3)!!}%
{\varrho\,\,z^{2\kappa+2}},\\
\mathbf{Q}_{c}^{(2)}\mathbf{(}2\kappa\mathbf{,}a\mathbf{,}z,\varrho\mathbf{)}
&  \thicksim\tfrac{(-1)^{\kappa+a}(4\kappa+1)!!}{\,\,z^{2\kappa+2}}%
[16\kappa^{2}+\kappa(32a+992)+24a+736],\\
\mathbf{Q}_{c}^{(2)}\mathbf{(}2\kappa+1\mathbf{,}a\mathbf{,}z,\varrho
\mathbf{)}  &  \thicksim8\tfrac{(-1)^{\kappa+a+1}(4\kappa+3)!!}{\,z^{2\kappa
+2}},
\end{align*}%
\begin{align*}
\widehat{\mathbf{Q}}_{c}^{(1)}\mathbf{(}2\kappa\mathbf{,}0\mathbf{,}%
z,\varrho\mathbf{)}  &  \thicksim\tfrac{(-1)^{\kappa}(4\kappa+1)!!}%
{\varrho\,\,z^{2\kappa+2}}[120\kappa^{2}+4\kappa+2],\\
\widehat{\mathbf{Q}}_{c}^{(1)}\mathbf{(}2\kappa+1\mathbf{,}0\mathbf{,}%
z,\varrho\mathbf{)}  &  \thicksim60\tfrac{(-1)^{\kappa+1}(4\kappa
+3)!!}{\varrho\,\,z^{2\kappa+2}},\\
\widehat{\mathbf{Q}}_{c}^{(2)}\mathbf{(}2\kappa\mathbf{,}0\mathbf{,}%
z,\varrho\mathbf{)}  &  \thicksim\tfrac{(-1)^{\kappa}(4\kappa+1)!!}%
{\,z^{2\kappa+2}}[9\kappa^{2}+498\kappa+369],\\
\widehat{\mathbf{Q}}_{c}^{(2)}\mathbf{(}2\kappa+1\mathbf{,}0\mathbf{,}%
z,\varrho\mathbf{)}  &  \thicksim\frac{9}{2}\tfrac{(-1)^{\kappa+1}%
(4\kappa+3)!!}{\,z^{2\kappa+2}}.
\end{align*}

The leading terms for $\mathbf{R}_{s}^{(\eta)}\mathbf{(}\kappa,a,z,\varrho
\mathbf{)}$ and $\mathbf{R}_{c}^{(\eta)}\mathbf{(}\kappa,a,z,\varrho
\mathbf{)}$ for $a$ equal to $0$ or $1$ being
\begin{align*}
\left.
\begin{tabular}
[c]{c}%
$\mathbf{R}_{s}^{(1)}(2\kappa,0,z,\varrho\mathbf{)}$\\
$\mathbf{R}_{s}^{(1)}(2\kappa+1,0,z,\varrho\mathbf{)}$%
\end{tabular}
\right\}   &  \thicksim-\frac{5}{z^{2}}+O(\frac{1}{z^{2\kappa+2}}),\\
\left.
\begin{tabular}
[c]{c}%
$\mathbf{R}_{s}^{(1)}(2\kappa,1,z,\varrho\mathbf{)}$\\
$\mathbf{R}_{s}^{(1)}(2\kappa+1,1,z,\varrho\mathbf{)}$%
\end{tabular}
\right\}   &  \thicksim O(\frac{1}{z^{2\kappa+2}}),
\end{align*}%
\begin{align*}
\left.
\begin{tabular}
[c]{c}%
$\mathbf{R}_{s}^{(2)}(2\kappa,0,z,\varrho\mathbf{)}$\\
$\mathbf{R}_{s}^{(2)}(2\kappa+1,0,z,\varrho\mathbf{)}$%
\end{tabular}
\right\}   &  \thicksim\frac{1}{\varrho\,}\left\{  \frac{240}{\,z^{2}%
}+O(\frac{1}{z^{2\kappa+2}})\right\}  ,\\
\left.
\begin{tabular}
[c]{c}%
$\mathbf{R}_{s}^{(2)}(2\kappa,1,z,\varrho\mathbf{)}$\\
$\mathbf{R}_{s}^{(2)}(2\kappa+1,1,z,\varrho\mathbf{)}$%
\end{tabular}
\right\}   &  \thicksim\frac{1}{\varrho\,}O(\frac{1}{z^{2\kappa+2}}),
\end{align*}%
\[
\left.
\begin{tabular}
[c]{c}%
$\mathbf{R}_{c}^{(1)}(2\kappa,0,z,\varrho\mathbf{)}$\\
$\mathbf{R}_{c}^{(1)}(2\kappa+1,0,z,\varrho\mathbf{)}$\\
$\mathbf{R}_{c}^{(1)}(2\kappa,1,z,\varrho\mathbf{)}$\\
$\mathbf{R}_{c}^{(1)}(2\kappa+1,1,z,\varrho\mathbf{)}$%
\end{tabular}
\right\}  \thicksim\frac{1}{\varrho}O(\frac{1}{z^{2\kappa+2}}),
\]%
\[
\left.
\begin{tabular}
[c]{c}%
$\mathbf{R}_{c}^{(2)}(2\kappa,0,z,\varrho\mathbf{)}$\\
$\mathbf{R}_{c}^{(2)}(2\kappa+1,0,z,\varrho\mathbf{)}$\\
$\mathbf{R}_{c}^{(2)}(2\kappa,1,z,\varrho\mathbf{)}$\\
$\mathbf{R}_{c}^{(2)}(2\kappa+1,1,z,\varrho\mathbf{)}$%
\end{tabular}
\right\}  \thicksim O(\frac{1}{z^{2\kappa+2}}),
\]%
\begin{align*}
\left.
\begin{tabular}
[c]{c}%
$\widehat{\mathbf{R}}_{c}^{(1)}(2\kappa,0,z,\varrho\mathbf{)}$\\
$\widehat{\mathbf{R}}_{c}^{(1)}(2\kappa+1,0,z,\varrho\mathbf{)}$%
\end{tabular}
\right\}   &  \thicksim\frac{20}{\varrho z^{2}}+\frac{1}{\varrho}%
O(\frac{1}{z^{2k+2}}),\\
\left.
\begin{tabular}
[c]{c}%
$\widehat{\mathbf{R}}_{c}^{(2)}(2\kappa,0,z,\varrho\mathbf{)}$\\
$\widehat{\mathbf{R}}_{c}^{(2)}(2\kappa+1,0,z,\varrho\mathbf{)}$%
\end{tabular}
\right\}   &  \thicksim O(\frac{1}{z^{2\kappa+2}}),\\
\left.
\begin{tabular}
[c]{c}%
$\widehat{\mathbf{R}}_{c}^{(2)}(2\kappa,1,z,\varrho\mathbf{)}$\\
$\widehat{\mathbf{R}}_{c}^{(2)}(2\kappa+1,1,z,\varrho\mathbf{)}$%
\end{tabular}
\right\}   &  \thicksim\frac{1}{2z^{2}}+O(\frac{1}{z^{2k+2}}).
\end{align*}
Lastly the integrals $T_{cc},T_{ss},T_{sc},T_{cs}$ defined by%

\begin{align*}
T_{cc}(\mathcal{K}_{\varepsilon},z,\varrho)  &  =\int_{0}^{\pi/2}%
\mathcal{I}_{1}(\mathcal{K}_{\varepsilon}\sin\theta)\sin^{4}\theta\cos
\theta\cos(z\cos\theta)\cos(\varrho\cos^{2}\theta)d\theta,\\
T_{ss}(\mathcal{K}_{\varepsilon},z,\varrho)  &  =\int_{0}^{\pi/2}%
\mathcal{I}_{1}(\mathcal{K}_{\varepsilon}\sin\theta)\sin^{4}\theta\cos
\theta\sin(z\cos\theta)\sin(\varrho\cos^{2}\theta)d\theta,\\
T_{cs}(\mathcal{K}_{\varepsilon},z,\varrho)  &  =\int_{0}^{\pi/2}%
\mathcal{I}_{1}(\mathcal{K}_{\varepsilon}\sin\theta)\sin^{4}\theta\cos
\theta\cos(z\cos\theta)\sin(\varrho\cos^{2}\theta)d\theta,\\
T_{sc}(\mathcal{K}_{\varepsilon},z,\varrho)  &  =\int_{0}^{\pi/2}%
\mathcal{I}_{1}(\mathcal{K}_{\varepsilon}\sin\theta)\sin^{4}\theta\cos
\theta\sin(z\cos\theta)\cos(\varrho\cos^{2}\theta)d\theta,
\end{align*}
with
\[
\mathcal{I}_{1}(z)=\sum_{\kappa=0}^{\infty}\frac{(-1)^{k}\,z^{2\kappa}%
}{(2\kappa+3)},
\]
can be rewritten
\begin{align*}
T_{cc}(\mathcal{K}_{\varepsilon},z,\varrho)  &  =\sum_{\kappa=0}^{\infty
}\frac{(-1)^{\kappa}\mathcal{K}_{\varepsilon}^{\,2\,\kappa}}{(2\kappa
+3)}\mathcal{C}_{cc}\mathcal{(}\kappa+1,z,\varrho),\\
T_{ss}(\mathcal{K}_{\varepsilon},z,\varrho)  &  =\sum_{\kappa=0}^{\infty
}\frac{(-1)^{\kappa}\mathcal{K}_{\varepsilon}^{\,2\,\kappa}}{(2\kappa
+3)}\mathcal{S}_{ss}\mathcal{(}\kappa+1,z,\varrho),\\
T_{cs}(\mathcal{K}_{\varepsilon},z,\varrho)  &  =\sum_{\kappa=0}^{\infty
}\frac{(-1)^{\kappa}\mathcal{K}_{\varepsilon}^{\,2\,\kappa}}{(2\kappa
+3)}\mathcal{C}_{cs}\mathcal{(}\kappa+1,z,\varrho),\\
T_{sc}(\mathcal{K}_{\varepsilon},z,\varrho)  &  =\sum_{\kappa=0}^{\infty
}\frac{(-1)^{\kappa}\mathcal{K}_{\varepsilon}^{\,2\,\kappa}}{(2\kappa
+3)}\mathcal{S}_{sc}\mathcal{(}\kappa+1,z,\varrho),
\end{align*}

\begin{center}
\bigskip\pagebreak 

\bigskip

\textbf{Table \ 3}

$\mathbf{\bigskip}$\textbf{The }$\mathbf{{\Large \wp}(}k,0,z\mathbf{)}$ \textbf{polynomials}%

\begin{tabular}
[c]{|c|c|}\hline
$k$ & ${\Large \wp}(k,0,z)$\\\hline
$1$ & $-3/z^{2}$\\\hline
$2$ & $-60/z^{4}$\\\hline
$3$ & $105/z^{4}-2520/z^{6}$\\\hline
$4$ & $11340/z^{6}-181440/z^{8}$\\\hline
$5$ & $-10395/z^{6}+1496880/z^{8}-19958400/z^{10}$\\\hline
$6$ & $-3243240/z^{8}+259459200/z^{10}-3113510400/z^{12}$\\\hline
$7$ & $2027025/z^{8}-972972000/z^{10}+58378320000/z^{12}-653837184000/z^{14}%
$\\\hline
$8$ & \\\hline
\end{tabular}

\bigskip

\textbf{Table \ 4}

$\mathbf{\bigskip}$\textbf{The }$\mathbf{\mathcal{Q}(}k,0,z\mathbf{)}$ \textbf{polynomials}%

\begin{tabular}
[c]{|c|c|}\hline
$k$ & $\mathcal{Q}(k,0,z)$\\\hline
$1$ & $6/z^{3}$\\\hline
$2$ & $-15/z^{3}+120/z^{5}$\\\hline
$3$ & $-840/z^{5}+5040/z^{7}$\\\hline
$4$ & $945/z^{5}-68040/z^{7}+362880/z^{9}$\\\hline
$5$ & $187110/z^{7}-7983360/z^{9}+39916800/z^{11}$\\\hline
$6$ & $-135135/z^{7}+38918880/z^{9}-1297296000/z^{11}+6227020800/z^{13}%
$\\\hline
$7$ & $-64864800/z^{9}+9729720000/z^{11}-280215936000/z^{13}%
+1307674368000/z^{15}$\\\hline
\end{tabular}

\bigskip

\textbf{Table \ 5}

$\mathbf{\bigskip}$\textbf{The }$\mathbf{\mathcal{R}(}k,0,z\mathbf{)}$ \textbf{polynomials}%

\begin{tabular}
[c]{|c|c|}\hline
$k$ & $\mathcal{R(}k,0,z)$\\\hline
$1$ & $1/z$\\\hline
$2$ & $1/z+20/z^{3}$\\\hline
$3$ & $1/z+7/z^{3}+840/z^{5}$\\\hline
$4$ & $1/z+9/z^{3}-756/z^{5}+60480/z^{7}$\\\hline
$5$ & $1/z+11/z^{3}+297/z^{5}-166320/z^{7}+6652800/z^{9}$\\\hline
$6$ & $1/z+13/z^{3}+429/z^{5}+154440/z^{7}-34594560/z^{9}+1037836800/z^{11}%
$\\\hline
$7$ & $1/z+15/z^{3}+585/z^{5}+32175/z^{7}+64864800/z^{9}-8562153600/z^{11}%
+217945728000/z^{13}$\\\hline
\end{tabular}

\bigskip

\textbf{Table \ 6}

$\mathbf{\bigskip}$\textbf{The }${\Large \wp}(1,\nu,z)$ \textbf{polynomials}%

\begin{tabular}
[c]{|c|c|}\hline
$\nu$ & ${\Large \wp}(1,\nu,z)$\\\hline
$0$ & $-3/z^{2}$\\\hline
$1$ & $12/z^{3}$\\\hline
$2$ & $3/z^{2}-60/z^{4}$\\\hline
$3$ & $-27/z^{3}+360/z^{5}$\\\hline
$4$ & $-3/z^{2}+225/z^{4}-2520/z^{6}$\\\hline
\end{tabular}

\bigskip

\textbf{Table \ 7}

$\mathbf{\bigskip}$\textbf{The }$\mathcal{Q}(1,\nu,z)$ \textbf{polynomials}%

\begin{tabular}
[c]{|c|c|}\hline
$\nu$ & $\mathcal{Q}(1,\nu,z)$\\\hline
$0$ & $6/z^{3}$\\\hline
$1$ & $3/z^{2}-24/z^{4}$\\\hline
$2$ & $-21/z^{3}+120/z^{5}$\\\hline
$3$ & $-3/z^{2}+144/z^{4}-720/z^{6}$\\\hline
$4$ & $36/z^{3}-1080/z^{5}+5040/z^{7}$\\\hline
\end{tabular}

\bigskip

\textbf{Table \ 8}

$\mathbf{\bigskip}$\textbf{The }$\mathbf{\mathcal{R}(}1,\nu,z\mathbf{)}$ \textbf{polynomials}%

\begin{tabular}
[c]{|c|c|}\hline
$\nu$ & $\mathbf{\mathcal{R}(}1,\nu,z\mathbf{)}$\\\hline
$0$ & $1/z$\\\hline
$1$ & $-4/z^{2}$\\\hline
$2$ & $20/z^{3}$\\\hline
$3$ & $3/z^{2}-120/z^{4}$\\\hline
$4$ & $-33/z^{3}+840/z^{5}$\\\hline
\end{tabular}
\pagebreak 
\end{center}


\begin{thebibliography}{9}
\bibitem {PHASEFACTOR}The phase factor $\cos^{2}(\int_{r_{l}}^{r}%
\sqrt{2\{E_{l,n_{r}}-V(r)-{\small (z+}1/2{\small )}^{2}{\small /}2r^{2}%
\}}\,\,dr-\pi/4)$ in $\left|  u_{l,n_{r}}(r)\right|  ^{2}$ has been replaced
by its average value of 1/2 (the factor of 1/$\sqrt{2}$ in Eq. (4)).
\bibitem {PARAMETERNOTATION}The notation $\nu(\lambda|\varepsilon)$ indicates
that the energy $\varepsilon$ is regarded as a parameter upon which $\nu$ as a
function of $\lambda$ depends.
\end{thebibliography}
\end{document}